\documentclass[fleqn,usenatbib]{mnras}
\usepackage{newtxtext,newtxmath}
\usepackage[T1]{fontenc}
\usepackage{ae,aecompl}
\usepackage{float}
\usepackage{graphicx}
\usepackage{amssymb}
\usepackage{amsmath}
\usepackage{subfig}
\usepackage{natbib}

\title[Magnetic obliquity of accreting T Tauri stars]{The magnetic obliquity of accreting T Tauri stars 
} 
\author[McGinnis et al.]{
Pauline McGinnis$^{1,2}$\thanks{E-mail: pmcginnis@cp.dias.ie}, J\'er\^ome Bouvier$^2$ and 
Florian Gallet$^2$ 
\\ 
$^1$Dublin Institute for Advanced Studies, School of Cosmic Physics, Astronomy \& Astrophysics 
Section, 31 Fitzwilliam Place, Dublin, Ireland \\
$^2$Univ. Grenoble Alpes, CNRS, IPAG, F-38000 Grenoble, France
} 

\date{Accepted 2020 July 08. Received 2020 July 08; in original form 2020 March 15.}

\pubyear{2020}

\begin{document}
\label{firstpage}
\pagerange{\pageref{firstpage}--\pageref{lastpage}}
\maketitle

\begin{abstract}
Classical T Tauri stars (CTTS) accrete material from their discs through their magnetospheres. 
The geometry of the accretion flow strongly depends on the magnetic obliquity, i.e., the angle 
between the rotational and magnetic axes. 
We aim at deriving the distribution of magnetic obliquities in a sample of 10 CTTSs. 
For this, we monitored the radial velocity variations of the HeI$\lambda$5876\AA \space line 
in these stars' spectra along their rotational cycle. He~I is produced in the accretion shock, 
close to the magnetic pole. 
When the magnetic and rotational axes are not aligned, the radial velocity of this 
line is modulated by stellar rotation. The amplitude of modulation is related to 
the star's projected rotational velocity, $v\sin i$, and the latitude of the hotspot. By 
deriving $v\sin i$ and HeI$\lambda$5876 radial velocity curves from our spectra we thus 
obtain an estimate of the magnetic obliquities. 
We find an average obliquity in our sample of 11.4$^{\circ}$ with an rms dispersion of 
5.4$^{\circ}$. The magnetic axis thus seems nearly, but not exactly aligned with the rotational 
axis in these accreting T Tauri stars, somewhat in disagreement with studies of spectropolarimetry, 
which have found a significant misalignment ($\gtrsim 20^{\circ}$) for several CTTSs. This 
could simply be an effect of low number statistics, or it may be due to a selection bias of 
our sample. 
We discuss possible biases that our sample may be subject to. 
We also find tentative evidence that the magnetic obliquity may vary according to the stellar 
interior and that there may be a significant difference between fully convective and partly 
radiative stars. 
\end{abstract}


\section{Introduction}

The accretion of circumstellar disc material onto young, low-mass stars known as T Tauri stars is 
believed to be strongly mediated by the stellar magnetic field. These actively accreting T Tauri 
stars are often referred to as classical T Tauri stars (CTTS), as opposed to the non-accreting 
weak-line T Tauri stars (WTTS). Magnetospheric accretion models predict that the stellar 
magnetosphere of a CTTS interacts with the inner accretion disc at a few stellar radii, truncating 
the inner disc \citep{shu94, romanova02}. At this region, circumstellar disc material is lifted 
above the disc mid-plane and falls onto the star following the magnetic field lines, forming what 
are known as accretion funnel flows or accretion columns \citep{bessolaz08}. 
When the material traveling at free-fall velocities collides with the stellar surface, accretion 
shocks are formed near the stellar surface. These shocks produce an excess continuum flux 
\citep{calvet98} and narrow components in emission lines such as He~I and the Ca~II triplet 
\citep[e.g.][]{dodin12}, which are observed in T~Tauri spectra 
\citep{joy45, appenzeller86, hamman92}. The excess emission flux often veils a T~Tauri star's 
spectrum, making the photospheric absorption lines appear shallower 
\citep{joy49, rydgren76, hartigan89}. 

Magnetohydrodynamics (MHD) simulations have shown that both the strength of the large-scale 
stellar magnetic field and the magnetic obliquity, i.e. the tilt between the axis of the 
stellar magnetic field and the stellar rotation axis, affect the geometry of the accretion 
flow \citep[see, e.g.,][]{kurosawa13}.
The magnetic obliquity therefore influences the star-inner disc interaction and has implications 
on the formation of inner disc warps \citep{romanova13}, as well as on the formation 
and migration of planets in the inner disc region (few 0.1 au from the star).
To better understand its role in the star-disc interaction, it is important to measure the magnetic 
obliquity of a number of CTTS. However, to do this normally requires mapping stellar magnetic field 
geometries using spectropolarimetry and Zeeman Doppler Imaging \citep[ZDI,][]{donati97}, a 
technique that is time-consuming and requires the use of large telescopes. It has therefore only 
been performed on a small number of CTTSs, making a statistical analysis impracticable. 

\begin{figure}%[t]
  \centering
  \includegraphics[width=0.45\textwidth, angle=0]{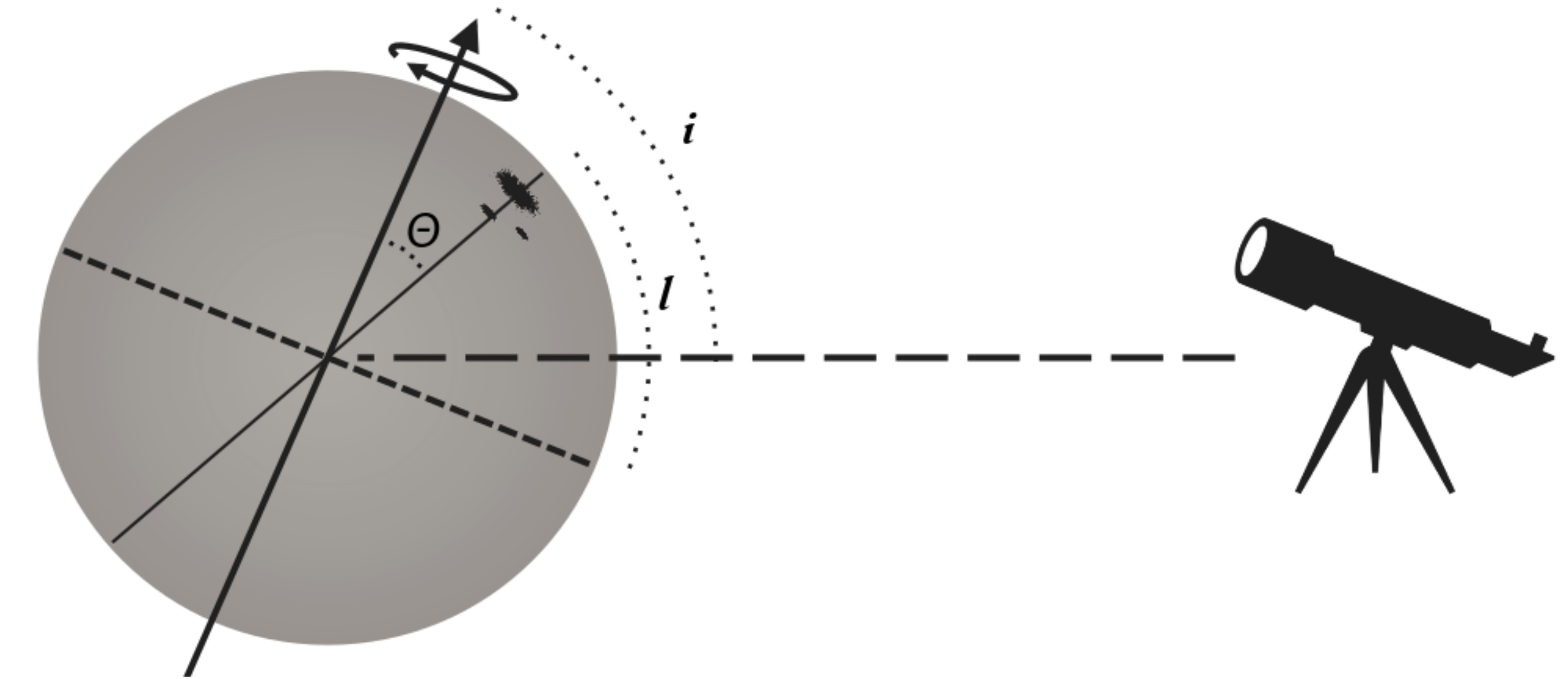}
  \caption{Illustration of a hotspot at latitude $l$ on the surface of a star viewed 
  at inclination $i$ (inclination between the rotation axis and the observer's line of sight). 
  The magnetic obliquity is identified as the angle $\Theta$. }
  \label{fig:sketch} 
\end{figure}

If we are interested in studying only the magnetic obliquity, and not the strength or complexity 
of the stellar magnetic field, there is a more cost-effective way to do this without the need to 
derive full stellar magnetic field configurations. 
Several CTTSs show narrow HeI$\lambda5876$\AA \space emission with redshifts of a few km~s$^{-1}$, 
which is believed to originate from the post-shock region, where the gas has been considerably 
decelerated. This line often shows radial velocity variations which may be periodically modulated by 
the star's rotation. In these cases, the amplitude of this variability $\Delta V_{rad}(\mathrm{HeI})$ 
is directly related to the cosine of the latitude $l$ of the accretion shock through the simple 
formula \citep{bouvier07}:

\begin{equation}\label{eq:cosl} 
\Delta V_{rad} (\mathrm{HeI}) = 2 \cdot v \sin i \cdot \cos l ,
\end{equation} 

\noindent
where $v\sin i$ is the projected stellar rotational velocity, which can be measured directly in an 
observed spectrum. 
Observations of CTTS magnetic fields using spectropolarimetry have shown that these accretion shocks 
form within a few degrees of the magnetic poles \citep[e.g,][]{donati08,donati10b,donati11a}.
Therefore by measuring the latitude of these shocks in CTTSs, their magnetic obliquities can be 
inferred without the need of direct magnetic field measurements.
Figure \ref{fig:sketch} illustrates this scenario, where a hotspot at latitude $l$ on 
the surface of a star (observed at an inclination $i$) traces the magnetic obliquity (represented 
by the angle $\Theta$).

As many other emission lines associated with accretion onto a young star, the 
HeI$\lambda$5876 line is often composed of more than one component. \citet{beristain01} studied 
a number of He~I profiles in CTTSs and found that they often present a narrow component (NC) and 
a broad component (BC). It is the NC that is believed to arise in the post-shock region, near the 
stellar surface, while the BC seems to have a composite origin. \citet{beristain01} found that 
the BC is sometimes observed blueshifted, in which case they attribute it to emission from a hot 
wind. When it is redshifted, they attribute it to emission from the accretion columns. 
Since all of these phenomena tend to be dynamic, these components will generally all be variable 
in their own way. Therefore in order to study one phenomenon or the other, it is important to 
either study systems that are dominated by one component, or to properly disentangle them. In 
the present study we attempt to do both. 

We propose here to estimate the magnetic obliquity of a sample of well-known accreting T Tauri 
stars in the Taurus-Auriga star forming region, using the radial velocity variations of the NC of 
the HeI$\lambda$5876 line. We describe our observations in Sect. \ref{sec:obs} and the method we 
use to derive hotspot latitudes in Sect. \ref{sec:analysis}. The results obtained for each object 
are detailed in Sect. \ref{sec:results}. We discuss our main findings in Sect. \ref{sec:discuss} 
and summarize our conclusions in Sect. \ref{sec:conc}. 

\section{Observations}\label{sec:obs}

\setlength\tabcolsep{0.125cm}

\begin{table*}%[t]
 \centering                                  
 \caption{Journal of Observations}           
 \label{obs}                                 
 \begin{tabular}{l c c c c c c c c c c}          
  \hline                                    
Name & Dates & N$_{spec}$ & S/N$_{600nm}$ & <$V_{rad}$> & $\sigma V_{rad}$ & <$v\sin i$> & $\sigma v\sin i$ & Classification & Spectral type & Mass \\ 
     & (Nov.2011) &       &              & (km~s$^{-1}$) &  (km~s$^{-1}$) & (km~s$^{-1}$) & (km~s$^{-1}$)   &  &  & M$_{\odot}$ \\ 
  \hline  
  DE Tau   & 21-28 & 13 & 26-33 & 14.66  & 0.61  & 8.38  & 0.54 & CTTS & M2.3 & 0.38 \\ 
  DF Tau   & 20-28 & 15 & 16-35 & 17.42  & 2.42  & 7.27  & 1.77 & CTTS & M2.7 & 0.32 \\ 
  DK Tau   & 21-28 & 15 & 21-35 & 15.82  & 2.38  & 12.69 & 2.18 & CTTS & K8.5 & 0.68 \\ 
  DN Tau   & 20-28 & 16 & 23-37 & 16.79  & 0.44  & 8.69  & 0.30 & CTTS & M0.3 & 0.55 \\ 
  GI Tau   & 21-28 & 11 & 24-34 & 17.06  & 0.53  & 9.23  & 0.51 & CTTS & M0.4 & 0.58 \\ 
  GK Tau   & 21-28 & 15 & 24-35 & 16.56  & 1.94  & 19.06 & 3.36 & CTTS & K6.5 & 0.76 \\ 
  GM Aur   & 20-27 & 14 & 21-36 & 14.95  & 0.98  & 12.59 & 1.02 & CTTS & K6.0 & 0.88 \\ 
  IP Tau   & 21-28 & 10 & 19-33 & 15.98  & 1.73  & 9.73  & 0.86 & CTTS & M0.6 & 0.59 \\ 
  IW Tau   & 20-28 & 15 & 27-46 & 16.03  & 0.30  & 8.53  & 0.30 & WTTS & M0.9 & 0.49 \\ 
  T Tau    & 18-28 & 26 & 14-39 & 19.47  & 1.28  & 23.54 & 1.61 & CTTS & K0   & 1.99 \\ 
  V826 Tau & 18-22 & 5  & 21-47 & 12.95  & 14.42 & 5.59  & 2.83 & WTTS & K7   & 0.74 \\ 
  V836 Tau & 23-28 & 7  & 30-33 & 17.73  & 0.86  & 10.51 & 0.77 & CTTS & M0.8 & 0.58 \\ 
  \hline  
 \end{tabular}
 
 \textbf{Notes:} Spectral types and masses are from \citet{herczeg14}. 
\end{table*}

Observations were carried out from November 18 to 28, 2011, at Observatoire de Haute-Provence 
using the SOPHIE spectrograph \citep{perruchot08} in High-Efficiency mode, which delivers a 
spectral resolution of $R\sim40,000$. The sample consists of ten accreting T Tauri 
stars (CTTS) and a control sample of two non-accreting T Tauri stars (WTTS). A total of 162 
spectra were obtained, comprising between 10 and 16 spectra for each source in the CTTS sample 
over eight nights (Nov. 21-28), with up to 26 spectra for T Tauri itself over 11 nights (Nov. 
18-28), and 5-7 spectra for each of the two WTTS of the control sample over the same time frame. 
The Journal of Observations is given in Table~\ref{obs}.  Depending on the source brightness, 
exposure times varied between 137~s (e.g., T~Tau) and 4410~s (e.g., IP~Tau). The resulting 
signal-to-noise ratio (S/N) at the continuum level around 600~nm ranges from 15 to 50, with 2/3 
of the spectra having S/N$\geq$30. 

The raw spectra were fully reduced at the telescope by the SOPHIE real-time pipeline 
\citep{bouchy09}. The data products include a re-sampled 1D spectrum with a constant wavelength 
step of 0.01 \AA, corrected for barycentric radial velocity, an order-by-order estimate of the 
signal-to-noise ratio, and a measurement of the source radial velocity, $V_{rad}$, and projected 
rotational velocity, $v\sin i$. These were derived from a cross-correlation analysis between the 
spectrum and a spectral mask template, either a G2 or a K5 mask, depending on the source's 
spectral type \citep[e.g.,][]{melo01,boisse10}. We list the values of these parameters in 
Table~\ref{obs} and use them for the subsequent analysis. The error quoted on $V_{rad}$ and 
$v\sin i$ in Table~\ref{obs} is the standard deviation of individual measurements. While the 
latter are usually accurate to within a fraction of a km~s$^{-1}$, the photospheric line profile 
variability induced by surface spots and/or the accretion flow in young stars yields much larger 
uncertainties \citep[e.g.,][]{petrov01}. The cross-correlation profiles themselves are presented 
in the online material and their night-to-night variations discussed there. 

\begin{figure*}%[t]
  \centering
  \includegraphics[width=\textwidth, angle=0]{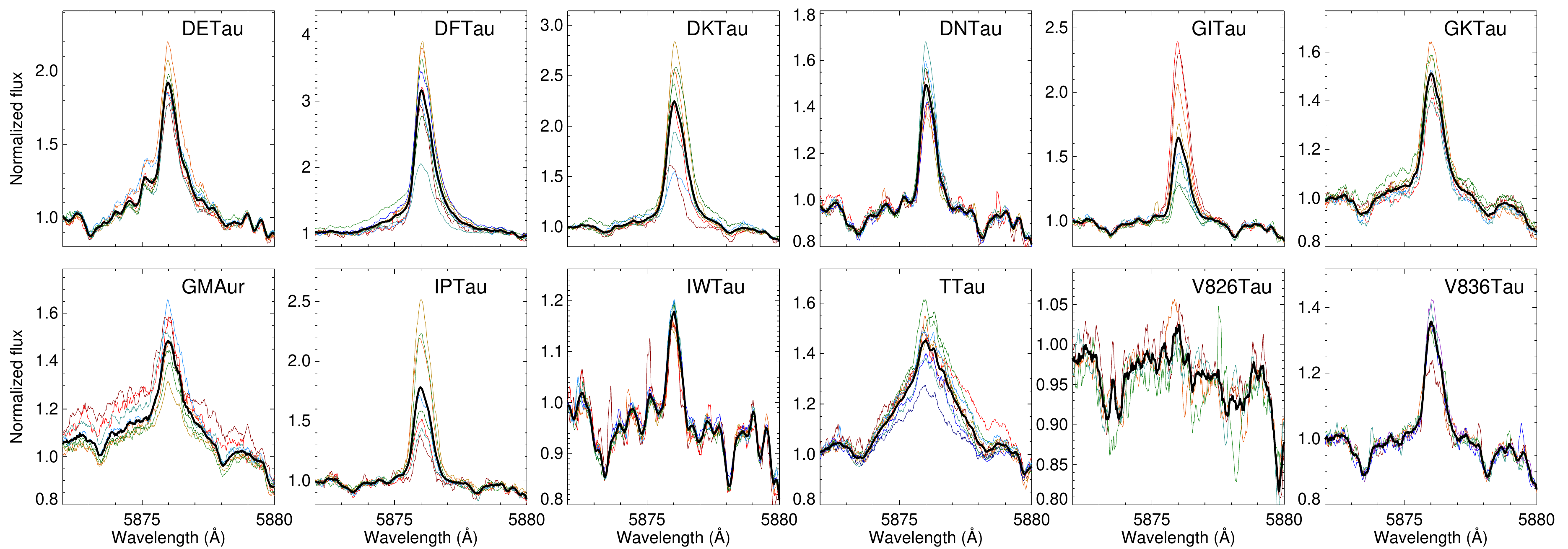}
  \caption{The HeI$\lambda5876$ line profiles of CTTS and WTTS observed in this study. 
  {\it Thin lines:} individual line profiles. {\it Thick black line:} averaged He~I line profile. 
  For clarity, the profiles shown here have been boxcar smoothed over 15 pixels (0.15 \AA).  }
  \label{fig:hei_profiles}
\end{figure*}

All 1D spectra were normalized to a continuum level of unity in the spectral region of 
interest. This was done using an IDL routine that identifies the continuum of a portion of the 
spectrum and fits a polynomial (of fourth or fifth order, depending on the curvature of the 
spectrum) through it. The continuum is identified by first manually excluding regions with emission 
lines, then taking only the points with the 20\% highest intensities in order to exclude absorption 
lines. The portion of the spectrum of interest is then divided by the function that fits the 
continuum, resulting in normalized spectra. The final normalized spectra are shown in the online 
supplementary material. 

\section{Spectral analysis: line profile Doppler shifts}\label{sec:analysis}

We aim at measuring Doppler shifts on the HeI$\lambda5876$ line as a function of rotational 
phase. The line profiles for our stellar sample are shown in Fig.~\ref{fig:hei_profiles}. 
With the exception of V826~Tau, which has very weak He~I emission that is only measurable 
in one observation, we see that the He~I line is observed in emission on every night for every 
star in this sample. This may seem unexpected if this emission comes from a rotationally 
modulated hotspot, since we would expect the spot to be behind the star in at least some epochs. 
We do note, however, that several stars show a significant decrease in the He~I line flux during 
certain epochs (this can be seen in Fig.~\ref{fig:hei_profiles}). This could occur if the spot 
passes behind the star, but a small portion of it always remains visible, because of the system 
inclination and spot latitude. Studies of ZDI have shown that it is common for the hotspot to be 
found at high latitudes \citep[e.g.][]{donati08,donati10b,donati11a}, which would indeed lead 
them to be observable during most of the rotational cycle. 

\begin{figure*}%[t]
   \centering
   \includegraphics[height=5cm]{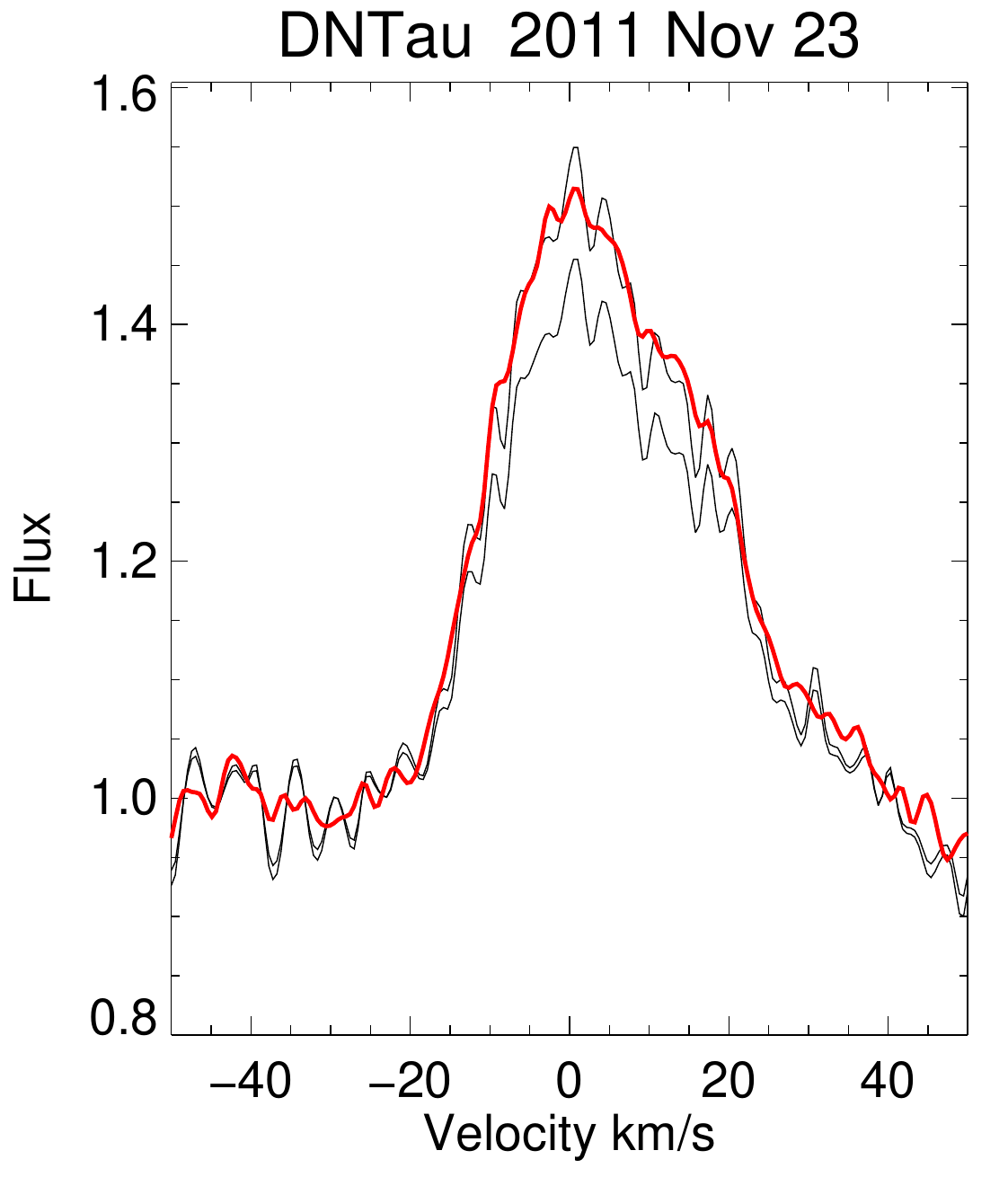}
   \includegraphics[height=5cm]{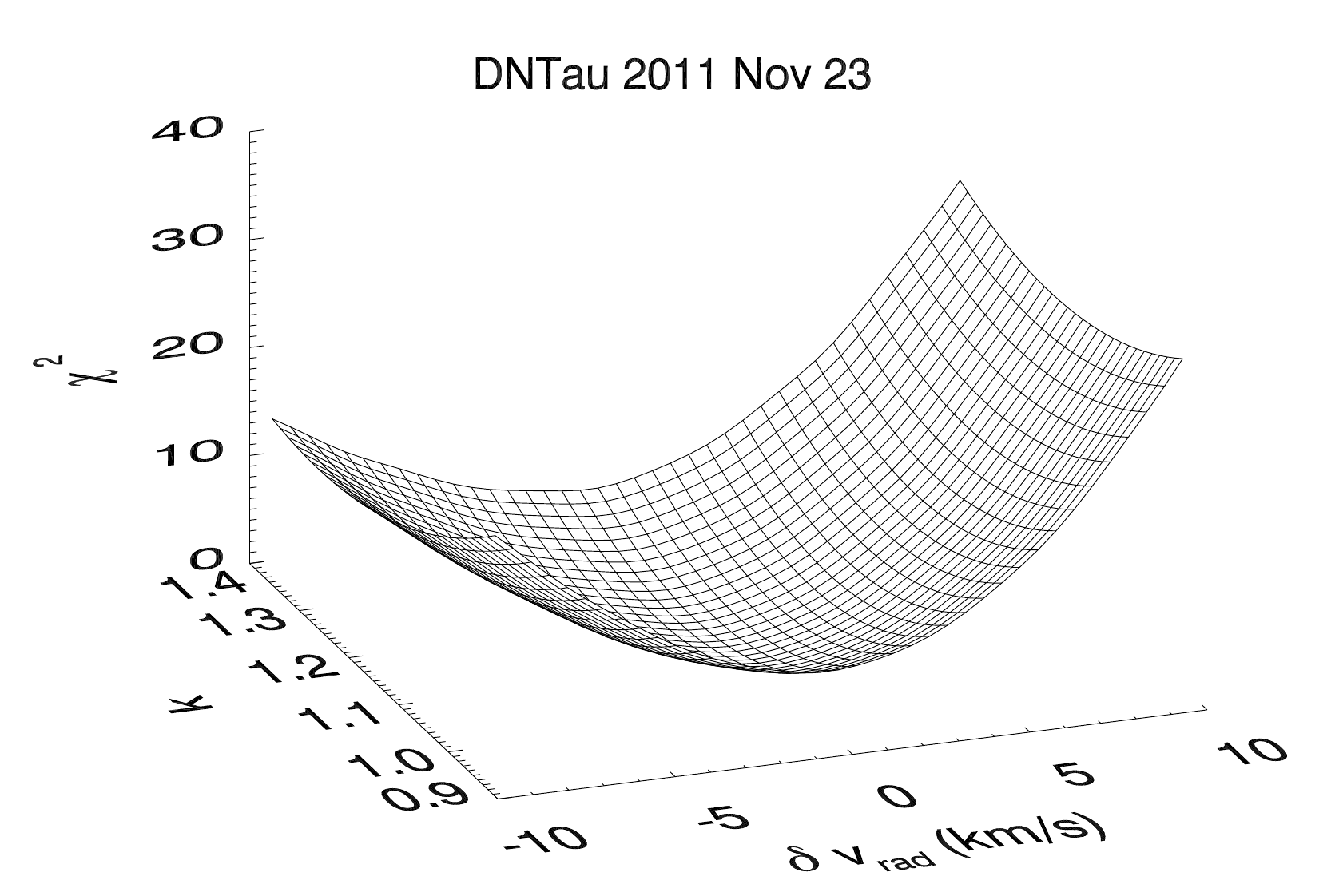}
   \includegraphics[height=5cm]{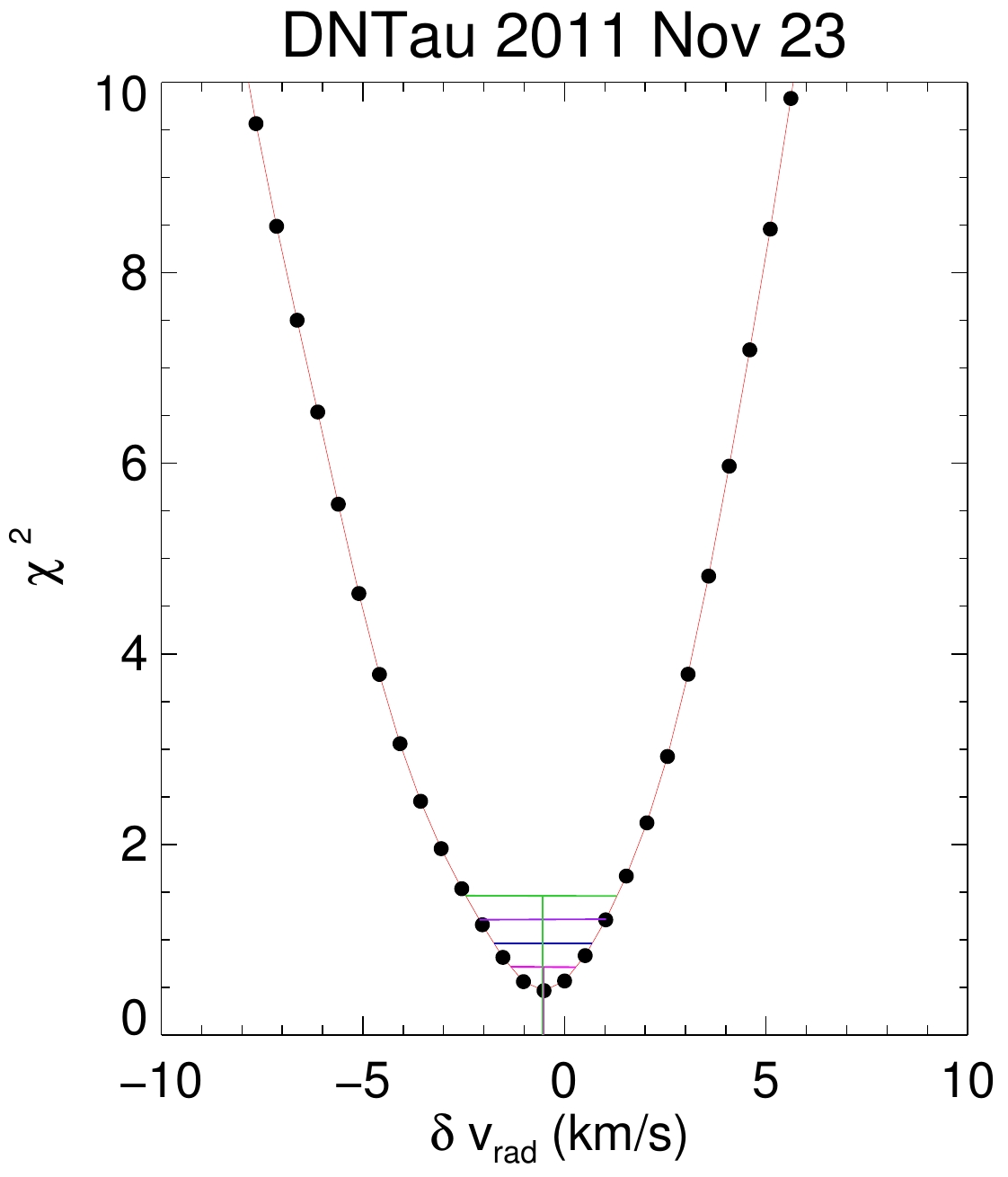}
   \caption{\textit{Left panel:} the He~I line profile of DN Tau obtained on Nov.~23 (lower 
   solid line) is wavelength shifted and intensity scaled (upper solid line) to fit the average 
   He~I profile (thick red solid line) of the spectral series. The best fit (lowest $\chi_\nu^2$)  
   solution is shown here. 
   \textit{Middle panel:} The $\chi_\nu^2 (k, \delta V_{rad})$ surface 
   is a shallow valley with a much steeper curvature along the radial velocity shift ($\delta 
   V_{rad}$) axis  than along the intensity scaling axis ($k$). 
   \textit{Right panel:} A cut in the 
   $\chi_\nu^2$ surface around the $\chi_\nu^2$ minimum along the $\delta V_{rad}$ axis is taken 
   (black dots) and linearly re-sampled (red curve). A sub-pixel estimate of the velocity shift is 
   obtained by computing the centroid of the core of the $\chi_\nu^2 (\delta V_{rad})$ curve over 
   4 velocity ranges (coloured horizontal lines). The adopted velocity shift $\delta V_{rad}$ is 
   computed as the mean of the 4 centroid estimates and the 1$\sigma$ limits on $\delta V_{rad}$ 
   are derived at the corresponding $\chi^2_{1\sigma} = \chi^2_{min} + 1$ value. 
   }
   \label{chi2}%
\end{figure*}

To measure the Doppler shift on each night, we first define a reference profile for each star, 
to which the profiles from each night are compared. We initially used as reference the line 
profile with the best S/N and performed the analysis described in this section to find the shifts 
between this and all other profiles obtained from the spectral series of that star. We then 
shifted each of these $N$ line profiles in velocity space so they would coincide as closely as 
possible with each other. We take the mean of these $N$ shifted profiles as the final reference. 
This is done so as to obtain a mean line profile without artificially broadening it by combining 
lines at slightly different radial velocities. 

The line profile of each night is intensity scaled by a factor $k$ and Doppler shifted by an 
amount of $\delta\lambda$ in order to fit the reference profile (see Fig.~\ref{chi2}). For 
each spectrum in the series, the best fit parameters, $k$ and $\delta\lambda$, are obtained by 
minimizing the quantity

\begin{equation}\label{eq:chi2} 
\chi_\nu^2 (k, \delta\lambda) = \sum_{i=1}^{n} w_i^2 \frac{\left[k\cdot (I(\lambda_i + \delta\lambda) - 1) - (\left< I(\lambda_i)\right> -1) \right]^2}{(n-2) \sigma_{cont}^2} ,
\end{equation}

\noindent where $n$ is the number of spectral pixels over the line profile, $w_i$ is the weight 
applied to each pixel, $k$ is a scaling factor that accounts for line intensity variations, 
$I(\lambda)$ the intensity of the individual profile, $\delta\lambda$ the Doppler shift, 
$\left< I(\lambda)\right>$ is the intensity of the reference profile, and 
$\sigma_{cont}$ is the rms noise of the spectrum at the continuum level next to the line. The 
summation runs over $\pm$2.5$\sigma$ from the line centre, where $\sigma$ is the standard 
deviation of a Gaussian fit to the average line profile.

A first guess to the scaling factor $k$ is provided by the ratio between the maximum intensity of 
the average and individual profiles. The scaling parameter is allowed to vary within 20\% of their 
ratio in 30 incremental steps. Doppler shifts are explored over a range of  $\pm$25~km~s$^{-1}$, 
with a spectral pixel step of 0.01 \AA, corresponding to $\sim$0.5~km~s$^{-1}$. Weights, $w_i$, 
which are chosen to scale as a Gaussian function (determined via a Gaussian fit to the 
reference profile), are applied to the $\chi_\nu^2$ calculation of each profile, in order to 
effectively increase the importance of the line core relative to the wings during the fitting 
procedure\footnote{For all objects considered here, the Doppler shifts derived using Gaussian 
or uniform weights differed by no more than 1 spectral pixel ($\sim$0.5~km~s$^{-1}$) with no 
systematics.}. For each profile, maps of $\chi_\nu^2 (k, \delta\lambda)$ are computed and the 
location of the deepest $\chi_\nu^2$ minimum is found. As the two model parameters are 
independent, the $\chi_\nu^2$ surface is a shallow valley, whose main axis lies along the $k$ 
dimension (see Fig.~\ref{chi2}). We therefore proceed to obtain a sub-pixel estimate of the 
location of the $\chi_\nu^2$ minimum in the $(k, \delta\lambda)$ grid by extracting a cut of 
the $\chi_\nu^2$ surface perpendicular to the $k$ direction around the deepest minimum. The 
$\chi_\nu^2 (\delta\lambda)$ curve thus obtained is re-sampled and a sub-pixel estimate of its 
centroid is derived to provide the final estimate on $\delta V_{rad}$, the Doppler shift between 
the individual and average He~I line profiles (see Fig.~\ref{chi2}). Finally, the 1$\sigma$ 
uncertainty on $\delta V_{rad}$ is derived along the $\chi_\nu^2 (\delta\lambda)$ curve at the 
location where $\chi_\nu^2 = \chi_{\nu,min}^2 + 1$ \citep{press92}. The whole procedure is 
illustrated in Fig.~\ref{chi2}.

\begin{figure}
   \centering
   \includegraphics[width=0.23\textwidth]{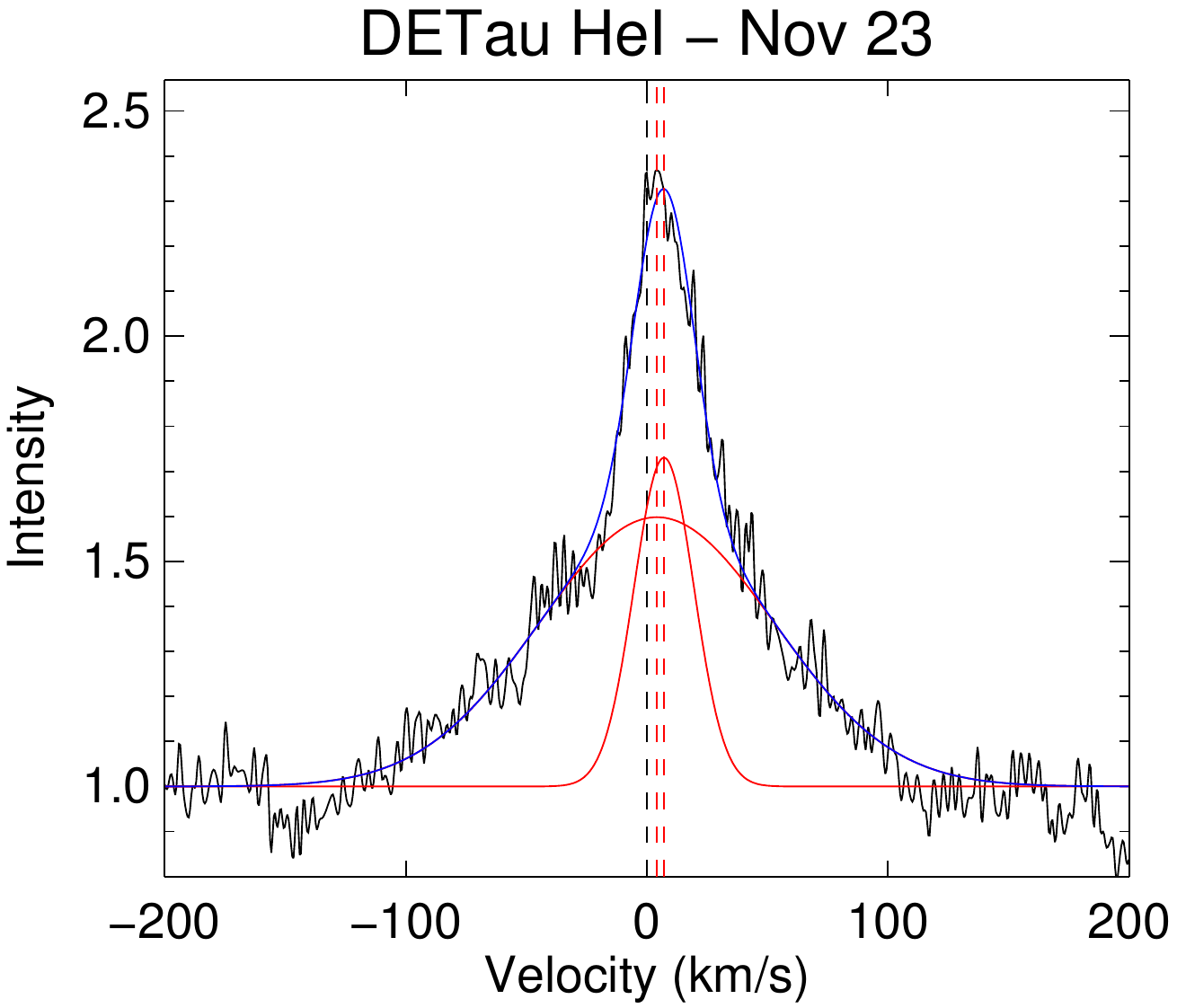}
   \includegraphics[width=0.23\textwidth]{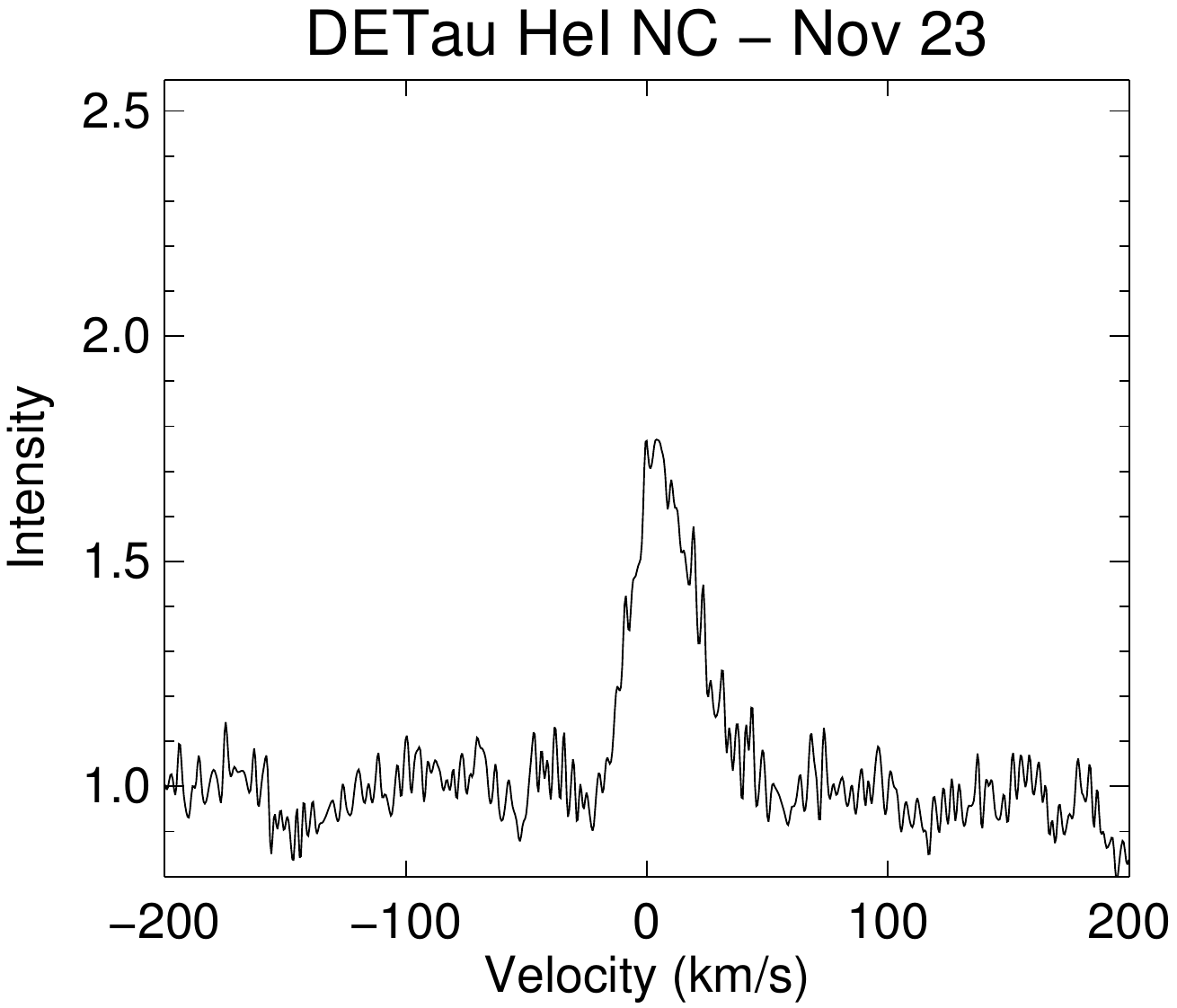}
   \includegraphics[width=0.23\textwidth]{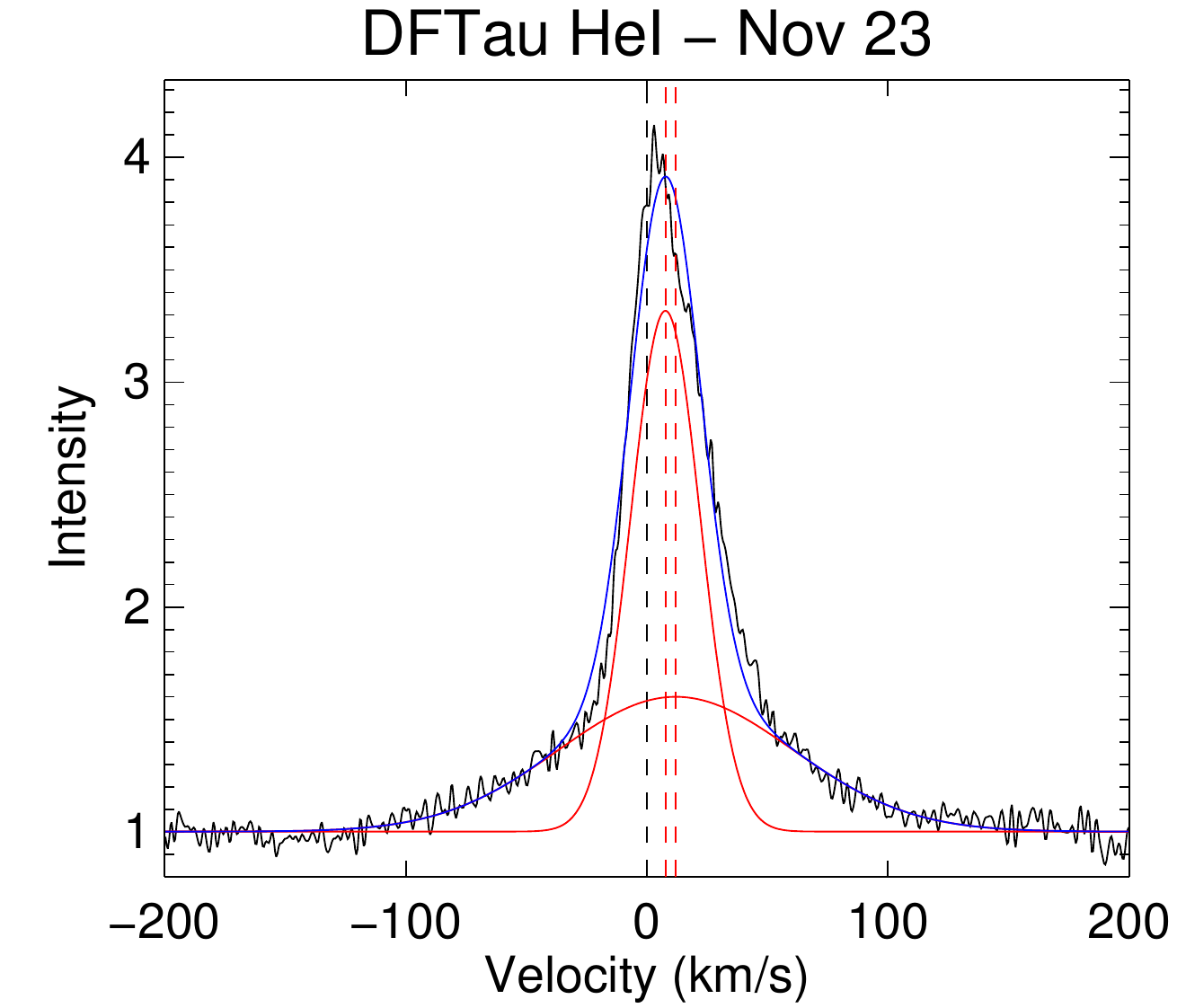}
   \includegraphics[width=0.23\textwidth]{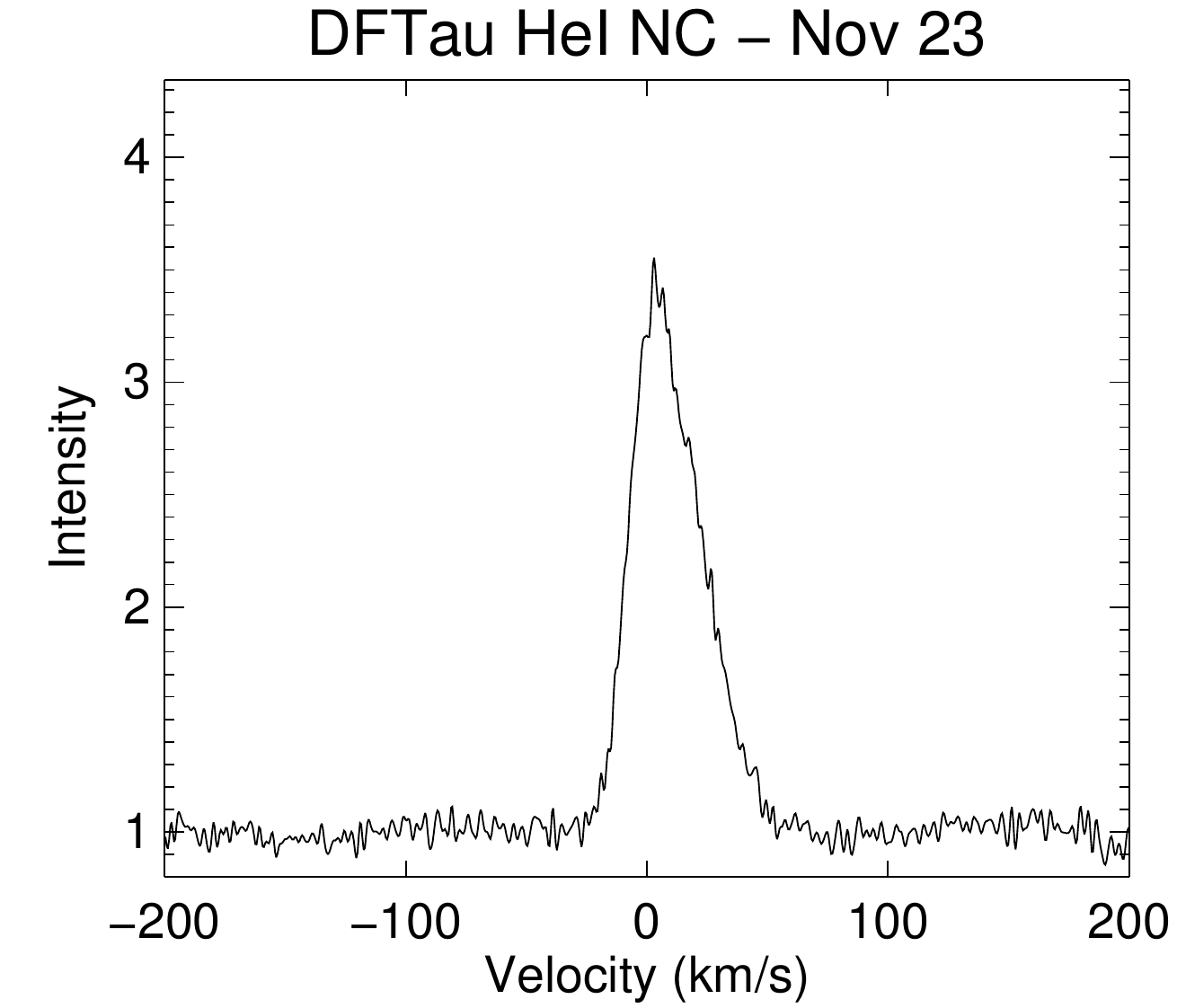}
   \includegraphics[width=0.23\textwidth]{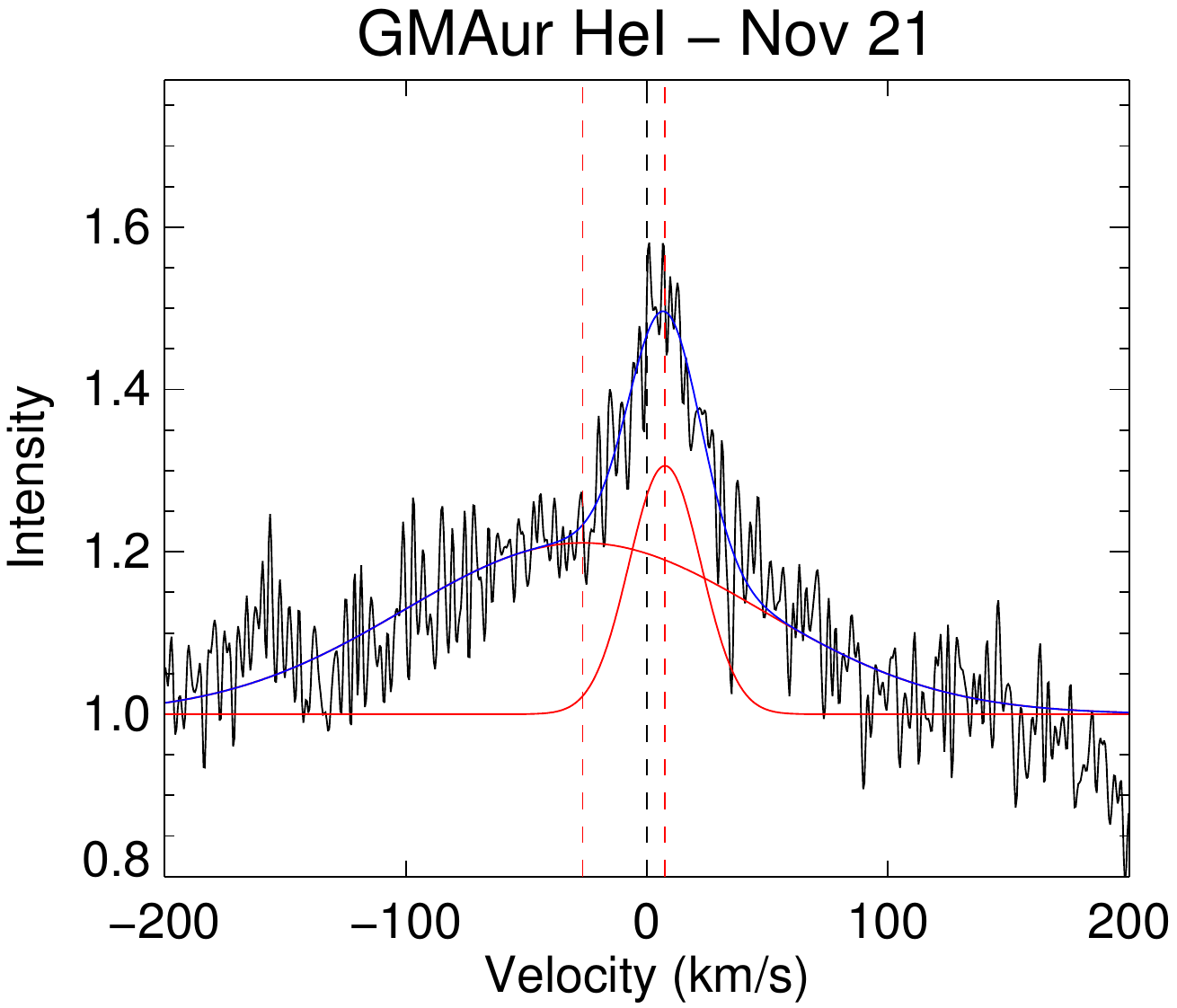}
   \includegraphics[width=0.23\textwidth]{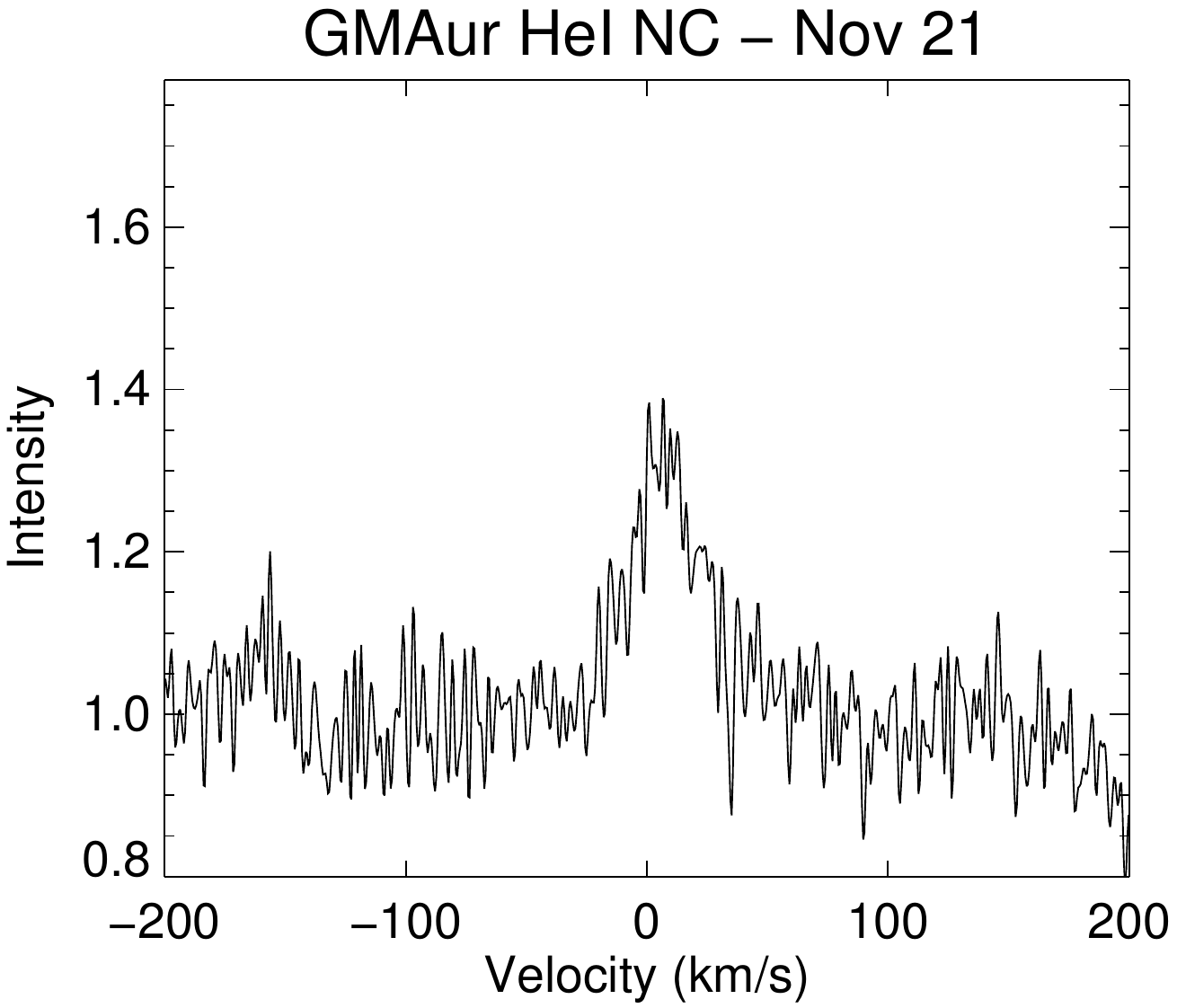}
   \includegraphics[width=0.23\textwidth]{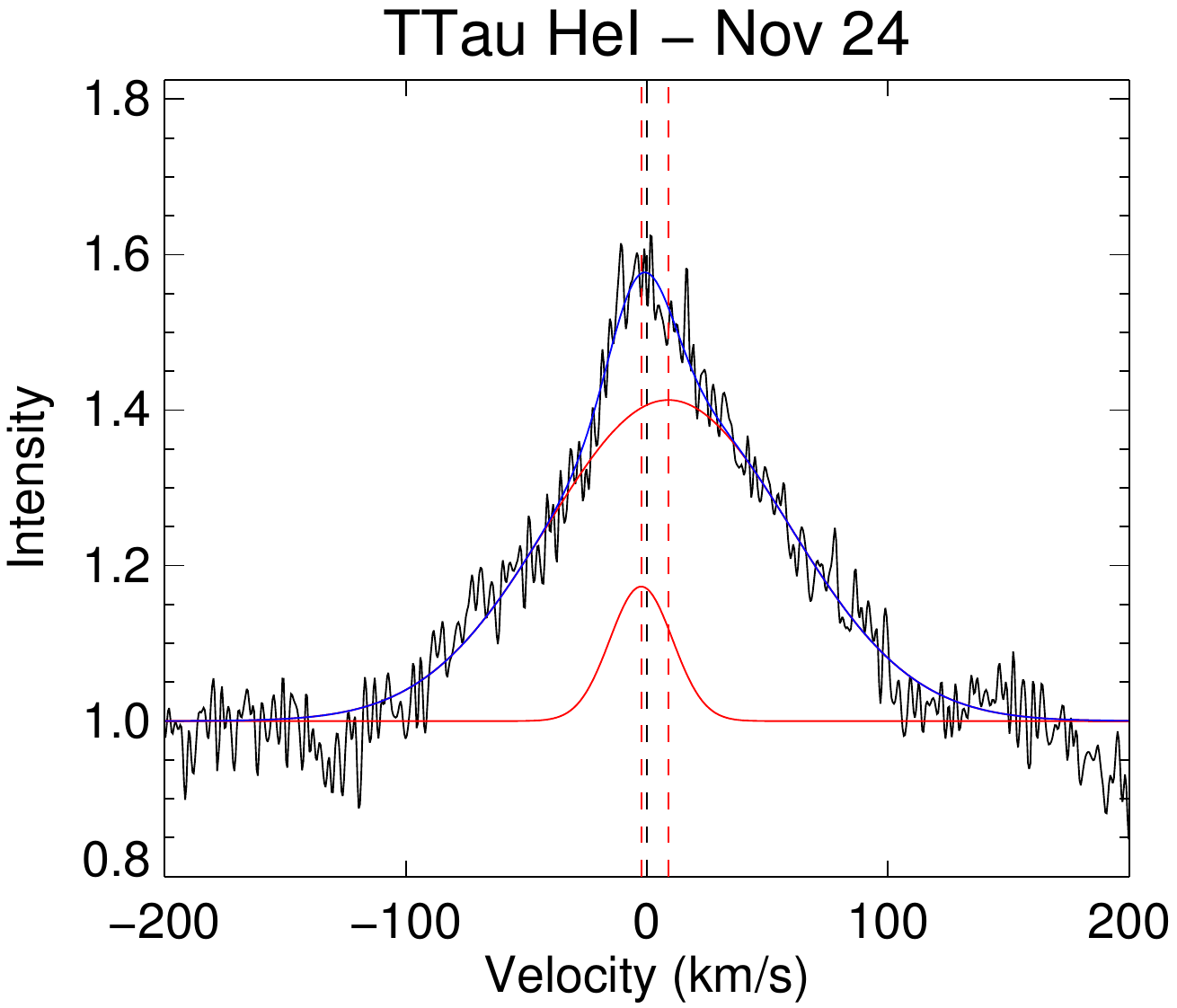}
   \includegraphics[width=0.23\textwidth]{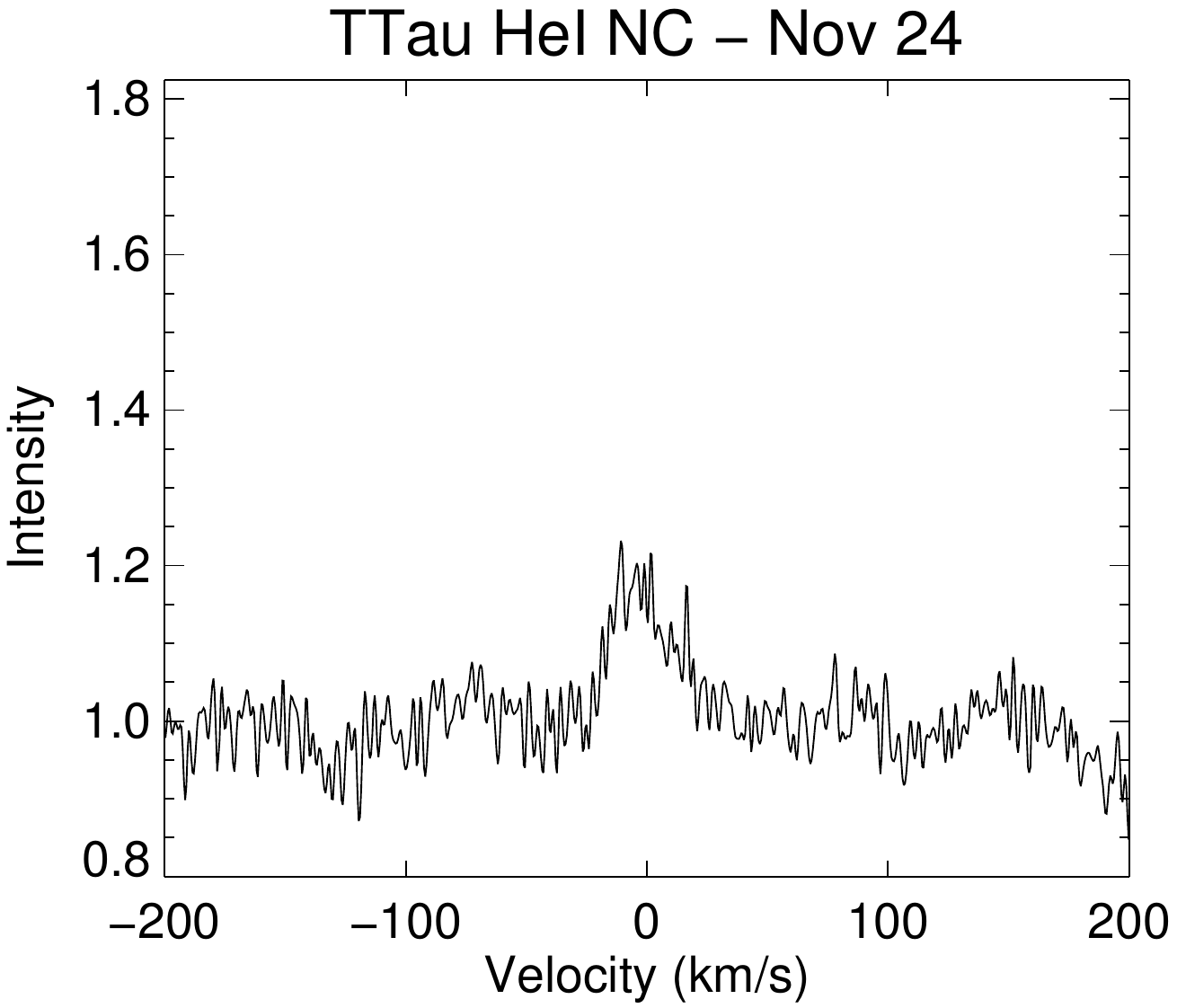}
   \caption{\textit{Left panels:} Examples of the Gaussian decomposition performed on the 
   HeI$\lambda5876$ line profiles that presented two components. \textit{Right panels:} residual 
   profiles obtained after subtracting the broad component from the observed spectrum. 
   }\label{fig:gausfit}%
\end{figure}

Care must be taken when using this procedure on the stars whose HeI$\lambda5876$ line profiles 
show more than one component, which is the case for DE Tau, DF Tau, GM Aur and T Tau. 
These components are believed to have different origins and therefore should present different 
variabilities \citep{beristain01}. The magnetic obliquity is related to the variability of the 
narrow component (NC), which is believed to arise in accretion shocks close to the stellar 
surface. Therefore, in the cases where a broad component (BC) is also present, we first fit the 
individual profiles from each night with two Gaussians in order to identify the two components, 
then analyse only the variability of the narrowest of the two components. 
However, the profiles that present only one component show that the NC is not symmetric. 
Thus, taking the centroid of the narrowest Gaussian component may introduce small uncertainties 
to the radial velocity measurements. On the other hand, a visual inspection of the profiles 
studied by \citet[][see their Fig. 1\textit{a}]{beristain01} 
shows that the BC often seems to be nearly symmetric and reasonably well reproduced by a 
Gaussian (this is not always the case, particularly for the strongest accretors, such as 
AS~353A, DG~Tau, DR~Tau and RW~Aur, all of which show more complex profiles, but it 
seems to be the case for the four stars in our sample). Therefore, to isolate the 
NC, we subtract the broadest Gaussian component from the composite profiles to remove the 
contribution of the BC. We then perform the cross-correlation analysis described above on 
the residual profile, which should contain only the contribution from the NC. We note that, 
in our sample, a typical NC has a full width at half maximum (FWHM) of $\sim 35$~km~s$^{-1}$, 
while a typical BC has FWHM $\sim 120$~km~s$^{-1}$. 
Figure \ref{fig:gausfit} shows an example of the Gaussian fit and residual profile of each 
of these stars. 

We find that the radial velocity measurements of the residual profiles agree, within the 
uncertainties, with the centroid velocities of the narrowest component of the Gaussian 
decomposition. In the case of DE Tau, DF Tau and GM Aur, the results also agree with the 
ones obtained from the cross-correlation of the full HeI$\lambda5876$ line profile (without 
subtracting the BC), though the error bars are larger when the BC is not removed. 
We can see in Fig. \ref{fig:gausfit} that the BC mostly affects the wings of these profiles, 
while the centre of the line is dominated by the NC. Therefore, by allowing the computation 
of $\chi_\nu^2$ to run over $\pm2.5\sigma$ of a Gaussian approximation to the centre of the line, 
the broad wings are ignored and as a result the BC does not strongly interfere with the 
analysis. For the star T~Tau, however, the HeI$\lambda5876$ line profile is clearly dominated 
by the BC. In this case, the cross-correlation analysis of the full profile yields very different 
results than when the contribution of the BC is removed. For this star, removing the 
contribution of the BC is essential to properly identify the latitude of the accretion shocks. 
The Gaussian decomposition of all observations of the star T~Tau can be seen in the online material 
(except for two observations, which have been excluded due to low S/N). 

\vspace{-0.1cm}

\section{Results}\label{sec:results}

\subsection{Variability of the He~I line} 

We present in this section the results obtained for the radial velocity variations of the
HeI$\lambda5876$ line profile of a sample of accreting T Tauri stars. 
The hotspot latitudes derived from the radial velocity variations of each star and Eq. 
\ref{eq:cosl} are shown in Table \ref{tab:results}. 
The variability of the narrow component of the He~I line and the veiling variability (with 
respect to the mean spectrum) for each star in our sample can be verified in Figs. 
\ref{fig:he1veil1} and \ref{fig:he1veil2} of the Appendix\footnote{When more than one spectrum 
of a given star were taken on the same night, we averaged these profiles in order to obtain a 
better signal to noise ratio. Therefore the plots shown in Figs. \ref{fig:he1veil1}, 
\ref{fig:he1veil2} and \ref{fig:li_he_ew} show only one measurement per night.}.
Veiling was measured in a region near the He~I line where several photospheric absorption 
lines are present. The spectrum of each night was compared with the mean spectrum of the same 
star\footnote{Because the spectra were compared with their own mean spectrum, rather than 
with the spectrum of a non-accreting star of the same spectral type (as is usually the case 
when determining veiling in a CTTS spectrum), both positive and negative values were found, 
representing the variation of veiling, rather than the value of veiling itself.}, 
in order to find the amount of excess continuum emission that minimized $\chi^2$. 
The uncertainties (derived at the corresponding $\chi^2_{1\sigma} = \chi^2_{min} + 1$ value) 
are large, but some of the trends seen in veiling appear to be confirmed by the variability 
of the equivalent width (EW) of the LiI$\lambda$6708 line (Fig. \ref{fig:li_he_ew}). The 
EW of this line (or any deep photospheric absorption line) gives an indirect measurement of 
veiling, since the excess emission causes the photospheric absorption lines to appear shallower 
in the normalized spectrum, reducing their EW. 

Our analysis also partly rests upon the pattern of spectral variability observed in other 
lines, such as H$\alpha$ and H$\beta$ (whose profiles are also shown in Figs. \ref{fig:he1veil1} 
and \ref{fig:he1veil2} and in more detail in the online material), as well as LiI$\lambda$6708 
(whose radial velocity curves and equivalent width variability can be seen in Fig. 
\ref{fig:li_he_ew}, alongside the variability of $V_{rad}$ and EW of the He~I line), and K2 
light curves (shown in Fig. \ref{fig:k2lcs}). 
The Li~I line profiles can be seen in the online material. 
Due to the specific behaviour of line profile variability in each object, we discuss
each sample source in turn\footnote{We do not discuss the star V826~Tau, because it is clear 
from Fig. \ref{fig:hei_profiles} that this is a spectroscopic binary (SB2) which presents 
almost no He~I emission.}. 

\begin{table}
 \caption{Hotspot latitudes and magnetic obliquities} 
 \label{tab:results} 
 \centering
 \begin{tabular}{l c c c c c}
  \hline
  Star & $\Delta V_{rad}$(HeI) & $v \sin i$     &  $\cos l$    &  $l$                    &  $\Theta$               \\
       &    (km~s$^{-1}$)      &  (km~s$^{-1}$) &              &  ($^{\circ}$)           &  ($^{\circ}$)           \\
  \hline                                                                                       
  DE Tau   &   1.8$\pm$3.2   &   8.4$\pm$0.5  &  0.11$\pm$0.19  & 84$\substack{+ 6\\-11}$ &  6$\substack{+11\\- 6}$ \\
  DF Tau   &   3.7$\pm$2.2   &   7.3$\pm$1.8  &  0.25$\pm$0.16  & 75$\pm10$               & 15$\pm10$               \\ 
  DK Tau   &   7.8$\pm$2.8   &  12.7$\pm$2.2  &  0.31$\pm$0.12  & 72$\substack{+ 7\\- 8}$ & 18$\substack{+ 8\\- 7}$ \\ 
  DN Tau   &   3.8$\pm$3.0   &   8.7$\pm$0.3  &  0.22$\pm$0.17  & 77$\substack{+10\\-11}$ & 13$\substack{+11\\-10}$ \\ 
  GI Tau   &   3.8$\pm$2.8   &   9.2$\pm$0.5  &  0.21$\pm$0.15  & 78$\pm9$                & 12$\pm9$                \\ 
  GK Tau   &   3.0$\pm$4.6   &  19.1$\pm$3.4  &  0.08$\pm$0.12  & 85$\substack{+ 5\\- 7}$ &  5$\substack{+ 7\\- 5}$ \\ 
  GM Aur   &   5.9$\pm$6.6   &  12.6$\pm$1.0  &  0.23$\pm$0.26  & 77$\substack{+13\\-16}$ & 13$\substack{+16\\-13}$ \\ 
  IP Tau   &   2.2$\pm$3.9   &   9.7$\pm$0.9  &  0.11$\pm$0.20  & 83$\substack{+ 7\\-12}$ &  7$\substack{+12\\- 7}$ \\ 
  IW Tau   &   1.9$\pm$4.1   &   8.5$\pm$0.3  &  0.11$\pm$0.24  & 84$\substack{+ 6\\-14}$ &  6$\substack{+14\\- 6}$ \\ 
  T Tau    &  18.4$\pm$7.9   &  23.5$\pm$1.6  &  0.39$\pm$0.17  & 67$\substack{+10\\-11}$ & 23$\substack{+11\\-10}$ \\ 
  V836 Tau &   3.7$\pm$4.3   &  10.5$\pm$0.8  &  0.17$\pm$0.21  & 80$\substack{+10\\-12}$ & 10$\substack{+12\\-10}$ \\ 
  \hline 
 \end{tabular}
\end{table}

\subsubsection{DE Tau}

This star's spectrum presents veiling that is clearly variable in the time-scale of the 
observations. The veiling increases steadily to reach a maximum on Nov. 26, then drops again, 
consistent with the rotation period of 7.6 days taken from the literature \citep{bouvier93}. 
The intensity of the He~I line reaches its maximum two days before the veiling maximum, on 
Nov. 24. The H$\alpha$ and H$\beta$ line profiles are more intense than average on both 
these nights, but show no clear trend with the 7.6 day period. 

\subsubsection{DF Tau}

The intensity of the He~I line profile, as well as its radial velocity variability, appear 
to be modulated on a time-scale of around 6 days, though the time-sampling of our observations 
is not enough to determine an accurate rotation period. 
This value is different from the photometric periods given in the literature, of 8.5 days 
\citep{bouvier95} and 7.18 days \citep{artemenko12}. However, neither of these two periods 
are found in this star's K2 light curve (N. Roggero private communication). 
The He~I line shows the lowest intensities just before it is most blueshifted, 
which is consistent with an origin in a stellar spot modulated by rotation. 
The veiling variability has too small an amplitude to derive any conclusions from it. 
However, the H$\alpha$ and H$\beta$ line profiles also show a trend with the $\sim$6 day 
period, where a blueshifted absorption component appears at phase $\phi \sim 0.6$, just after 
the accretion shock has passed in front of our line of sight (the He~I line profile is more 
redshifted than average). Since this component is associated with outflows intersecting our 
line of sight, this points to a spatial association between the accretion shocks and a wind. 
This could be a stellar wind or a non-axisymmetric disc wind that is launched close to the 
co-rotation radius, in order to show periodic behaviour on the same time-scale as the stellar 
rotation. 

\subsubsection{DK Tau}

The intensity and radial velocity of the He~I line, as well as the veiling, show a variability 
that is modulated on the time-scale of the observations (the variability of veiling is just 
within the error bars, but the trend is confirmed by the EW of the LiI line in Fig. 
\ref{fig:li_he_ew}). 
As the He~I line passes in our line of sight, its radial velocity going from blueshifted to 
redshifted with an amplitude of $\sim 10$~km~s$^{-1}$, the veiling and He~I intensity increase 
to reach a maximum on Nov. 25 and Nov. 27, before decreasing again. However, something occurs 
on Nov. 26 which results in a less intense, more centred He~I line, less veiling and much less 
intense H$\alpha$ and H$\beta$ line profiles. If not for this event, the behaviour of He~I and 
veiling in this star would be consistent with the rotational modulation of a hotspot on the 
stellar surface with the reported period of 8.18 days \citep{percy10, artemenko12}. It is 
possible that on this date, circumstellar material passed in front of the line of sight, 
partially obscuring the hotspot and resulting in the less intense emission lines and veiling. 
Its K2 light curve appears to be dominated by flux dips (see Fig. \ref{fig:k2lcs}), which is 
consistent with circumstellar material obscuring the star from time to time. 
The H$\beta$ line shows an inverse P Cygni profile on Nov. 24 and 25, just before this event. 
This type of profile, characterized by a redshifted absorption, occurs when the accretion 
funnel flows intersect our line of sight. 

\subsubsection{DN Tau}

The intensity and radial velocity of the He~I and Li~I lines are modulated with a period 
that is consistent with the value of 6.32 days given in the literature for this star's 
rotation period \citep{donati13,artemenko12}. 
The radial velocity variability of the Li~I line appears to be mirrored with respect to the 
He~I line (Fig. \ref{fig:li_he_ew}). This is consistent with rotational modulation of an 
accretion shock. While the He~I emission originating in the hotspot will be more blueshifted 
as this spot comes into view and more redshifted as it recedes, photospheric absorption lines 
such as Li~I appear more redshifted as a hot (or cold) spot comes into view and more blueshifted 
as it recedes, since the line is deformed due to the presence of spots on the stellar surface 
\citep[see e.g.][]{gahm13}. The amplitude of the veiling variability is too small to draw 
any conclusions from it. 

This star's H$\alpha$ and H$\beta$ line profiles present a redshifted absorption component 
in some observations, which is clearest on Nov. 27, when the He~I line is strongest and 
more centred, meaning that the accretion shock is in full view. As this redshifted 
absorption originates in accretion funnel flows crossing our line of sight, this points to 
a spatial association between the accretion shocks and funnel flows.

\subsubsection{GI Tau}

The intensity of the main emission line profiles (H$\alpha$, H$\beta$, He~I) is modulated 
on the time-scale of the observations. The intensity of the residual line profiles steadily 
decreases from Nov. 22-23 to reach a minimum around Nov. 27, then increases again up to Nov. 
29, a behaviour which is consistent with the modulation of the line flux by a bright spot 
rotating with a period of $\sim$7 or 8 days. This is in agreement with the past reported 
photometric period of 7.1 days \citep{percy10,artemenko12}. 
The veiling appears to decrease over a similar time period, however the variability amplitude 
is within the uncertainties and therefore cannot be confirmed. 

\subsubsection{GK Tau}

The variability of the radial velocity of the He~I line is consistent with a period of 4.6 
days, found in the literature \citep{percy10,artemenko12} and in this star's K2 light curve 
\citep{rebull20}. There is no clear trend in veiling or in the H$\alpha$ and H$\beta$ line 
profiles with this period, even though the variability of the line profiles is very strong. 

\subsubsection{GM Aur}

The intensity of the He~I line is modulated on the time-scale of the observations, 
decreasing until Nov. 24, then increasing again. The variability of this line's radial 
velocity is consistent with rotational modulation of a spot with the period of 6 days 
given in the literature \citep{percy10,artemenko12}. The veiling variability is too small 
to draw any conclusions from it. 
As with DN~Tau, the radial velocity variability of the Li~I line is mirrored with 
respect to the He~I line, in support of rotational modulation of a hotspot. 
There is no clear trend in the H$\alpha$ and H$\beta$ line profiles with this period.

\subsubsection{IP Tau}

The intensity and radial velocity of the He~I and Li~I lines, as well 
as the veiling, all show a variability that is consistent with a period of 5.6 days. Veiling 
increases along with the intensity of He~I, both reaching a maximum on Nov. 25, while the 
He~I line profile went from  more blueshifted than average on Nov. 23 and 24, to 
more redshifted on Nov. 25-28, and the Li~I line went from more redshifted to more blueshifted 
in this same time. This behaviour is consistent with the veiling and He~I line being modulated 
by a hotspot rotating at this period, which is different from the photometric period of 
3.25 days found by \citet{bouvier93}. The light curve from that study was therefore likely 
dominated by something other than rotational modulation from spots on the stellar surface. 

The H$\alpha$ and H$\beta$ line profiles show a redshifted absorption component on one night 
(Nov. 29), corresponding to phase $\phi \sim 0.25$, when the He~I line is more 
blueshifted and veiling is beginning to increase, meaning that the accretion shock is appearing 
in our line of sight. As was the case with DN~Tau, this points to a spatial association between 
the accretion funnel flows and accretion shocks.

\subsubsection{IW Tau} 

This star was one of the two WTTSs (non-accreting T Tauri stars) in our control sample. It 
shows a relatively narrow H$\alpha$ line profile, with a width at 10\% of the maximum intensity 
of only 180 km~s$^{-1}$, which is much lower than the threshold of 270 km~s$^{-1}$ usually used 
to classify a T Tauri star as actively accreting \citep{white03}. Therefore it would not be 
classified as a classical T Tauri star based on its H$\alpha$ line profile. However, this star 
does present a weak He~I emission line with a redshift of 4.3 $\pm$ 0.8 km~s$^{-1}$, which does 
not seem to be variable in our observations. \citet{beristain01} also observed weak 
HeI$\lambda5876$ emission in the spectra of three non-accreting T Tauri stars, however they were 
all on average centred at the stellar rest velocity, which led to the conclusion that they could 
be the result of very active chromospheres. 
The He~I emission we observe in IW~Tau is redshifted at above 3$\sigma$, which means it must 
originate in material falling onto the star. 

It is possible that this star is still undergoing accretion at a very reduced rate. In order 
to test this possibility, we estimated a mass accretion rate for this star from several 
emission lines in its spectra, following the relations given by \citet{alcala17}. We find an 
average value for the logarithmic mass accretion rate of 
$\log \dot{M}_{acc} = -9.5 (\pm 0.3)~\mathrm{M}_{\odot}~\mathrm{yr}^{-1}$ 
(corresponding to $\dot{M}_{acc} = 3 \times 10^{-10}~\mathrm{M}_{\odot}~\mathrm{yr}^{-1}$), 
where the uncertainty in $\log \dot{M}_{acc}$ represents the standard deviation across the ten 
emission lines used to calculate accretion luminosity. It is worth noting that this likely 
represents an upper limit to this star's mass accretion rate, since at these low levels of 
emission, these lines may be affected by chromospheric activity in a non-negligible way. At 
the same time, the strong agreement between the mass accretion rates derived from all these 
different lines (made clear by the relatively small dispersion in the values found) is a good 
indication of the reliability of this result. This star does not seem to be surrounded 
by a circumstellar disk \citep[besides possibly a thin debris-disk, typical of WTTSs in 
general; ][]{beckwith90, wolk96, furlan05}. 

This star's Li~I line is modulated on a 5.5 day time-scale, which could be caused by the 
rotational modulation of cold spots on the stellar surface, or by a binary companion 
\citep[since this star has been reported to be a binary system by][]{richichi94}. 
This star's K2 light curve clearly shows a spot-like behaviour, with two distinct periods, 
one of which agrees with this period found in the Li~I line 
\citep[see Fig. \ref{fig:k2lcs} and][]{rebull20}. 
It is likely that one of these periods represents the star's rotation while the other may be 
due to a companion.  

\subsubsection{T Tau}

This is a multiple system, consisting of three stars \citep[known as T~Tau~N, Sa and 
Sb;][]{koresko00, kohler00}, although the secondary component T~Tau~S(a+b) is not detectable in 
optical wavelengths, likely because of strong extinction from circumstellar or circumbinary 
material \citep{duchene05}. Thus the spectra presented here are entirely dominated by the 
brightest component, T~Tau~N. 

\begin{figure}%[p] 
   \centering
   \includegraphics[width=0.47\textwidth]{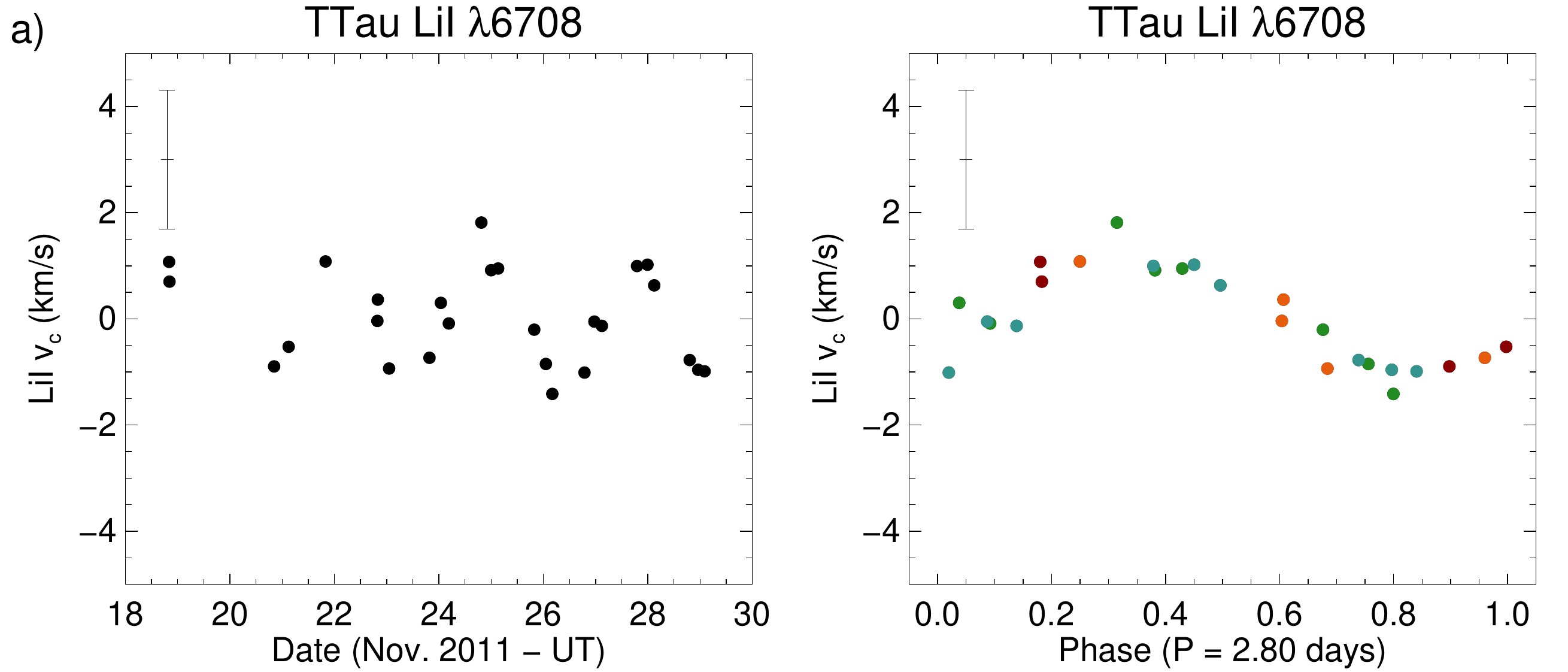}
   \includegraphics[width=0.47\textwidth]{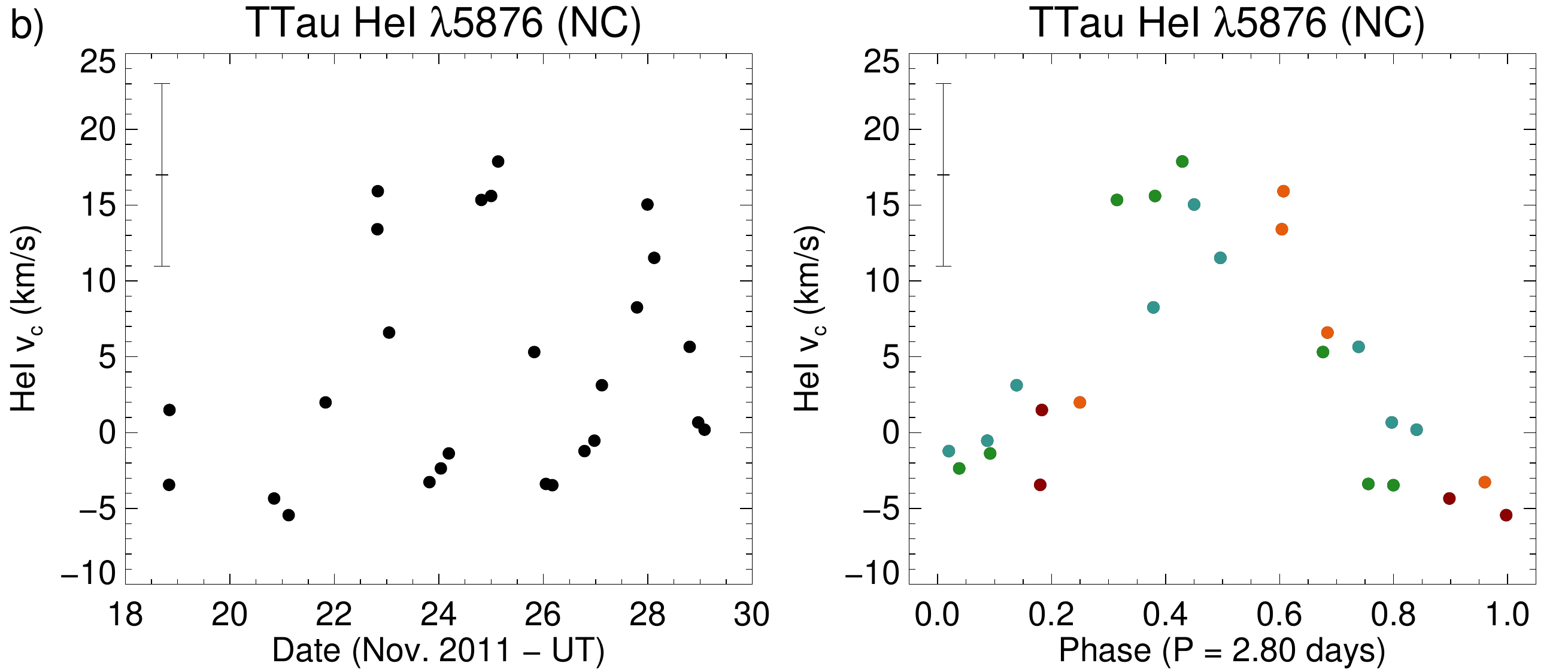}
   \includegraphics[width=0.47\textwidth]{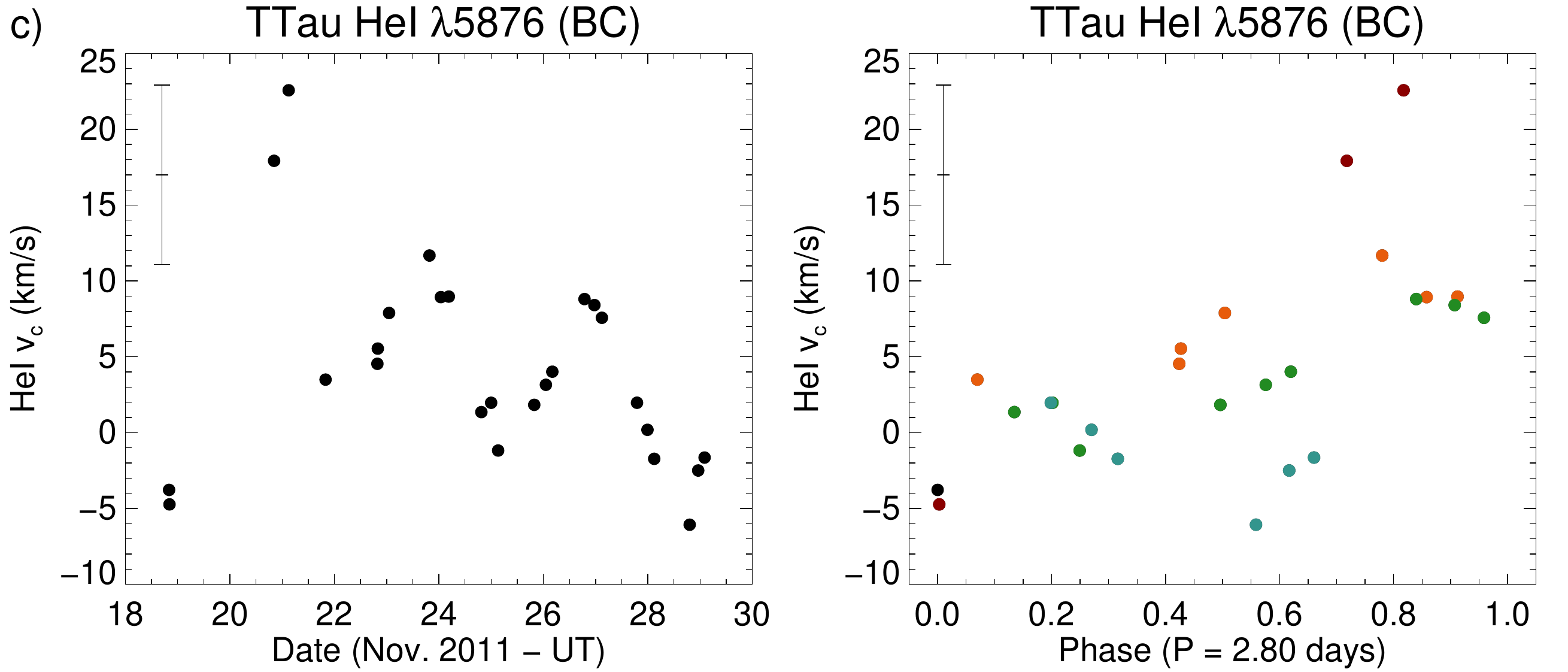}
   \includegraphics[width=0.47\textwidth]{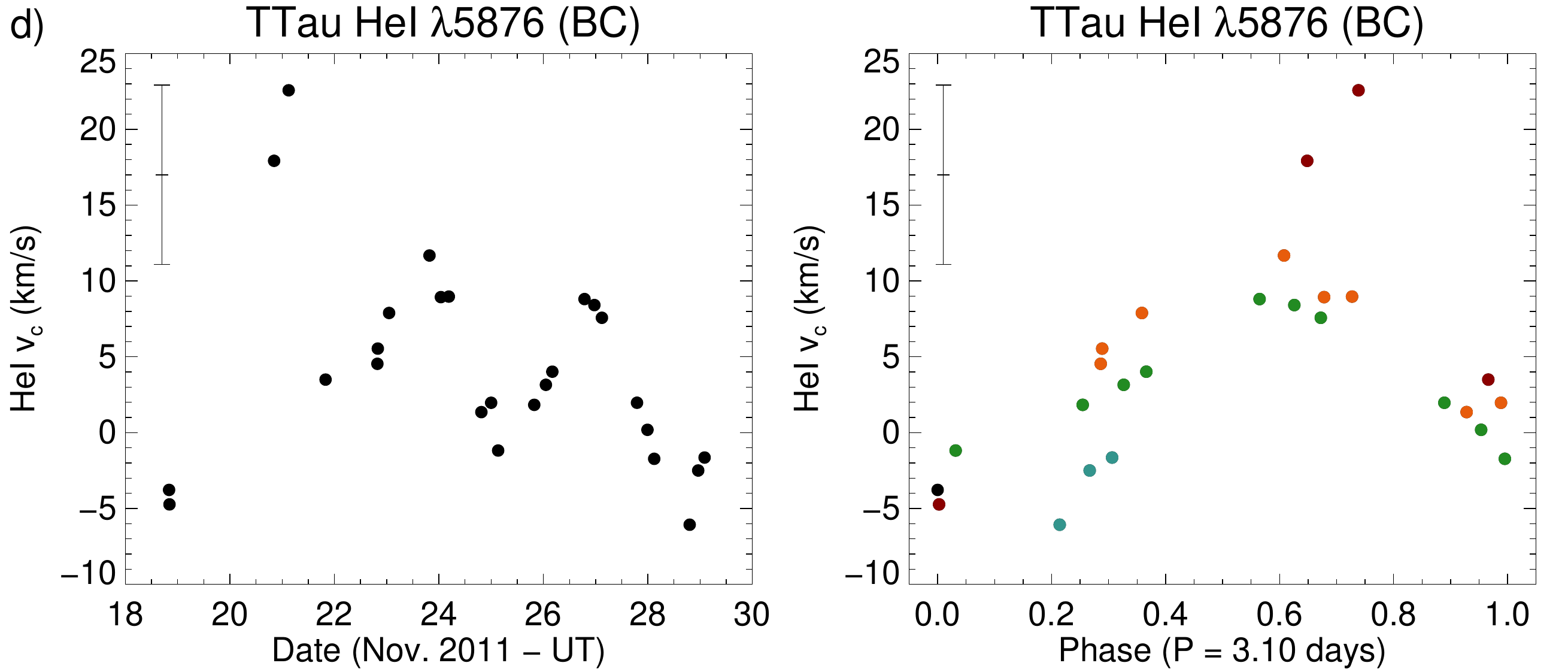}
   \caption{Radial velocity variability of \textit{a)} the Li~I line; \textit{b)} the narrow 
   component of the He~I line; \textit{c)} and \textit{d)} the broad component of the He~I 
   line, for the star T~Tau. The plots are shown in full in the \textit{left} panels and in 
   phase in the \textit{right} panels, using the period of $P=2.8$~days (panels \textit{a, b} 
   and \textit{c}) and $P=3.1$~days (panel \textit{d}). 
   Different colors represent different rotation cycles. 
   The median error bar on the radial velocity is shown on the top left corner of each plot.
   In these plots, we do not show the average measurement in each night, but rather include 
   all observations. 
   }
   \label{fig:ttaubc}
\end{figure}

The radial velocity of this star's He~I NC is very well modulated with a period of 2.7$\pm$0.4 
days, while the radial velocity of the Li~I line gives a similar period of 2.86 days. These 
periods coincide well with the photometric period of 2.81 days measured by \citet{artemenko12} 
and in its K2 light curve \citep{rebull20}. The top two panels of Fig. \ref{fig:ttaubc} show 
the Li~I line and the NC of the He~I line folded in phase with this period of $P=2.8$~days. 
The variability amplitude of the veiling is small, and there are no clear trends in the 
H$\alpha$ or H$\beta$ lines with the stellar rotation period. 
We can see that the modulus of the equivalent width of the He~I line's NC is largest when it 
is most redshifted (on Nov. 25 and Nov. 28), which may be due to a favorable viewing geometry. 
We have found that the magnetic obliquity of this star is $\theta = (23 \pm 10)^{\circ}$, which 
agrees, within the uncertainties, with its system inclination of $i = (28 \pm 1)^{\circ}$ 
\citep{manara19}. This means that when the accretion shock passes in front of us, we observe the 
flow of matter onto the star projected almost directly in our line of sight. 

\begin{figure}%[p] 
   \centering
   \includegraphics[width=0.47\textwidth]{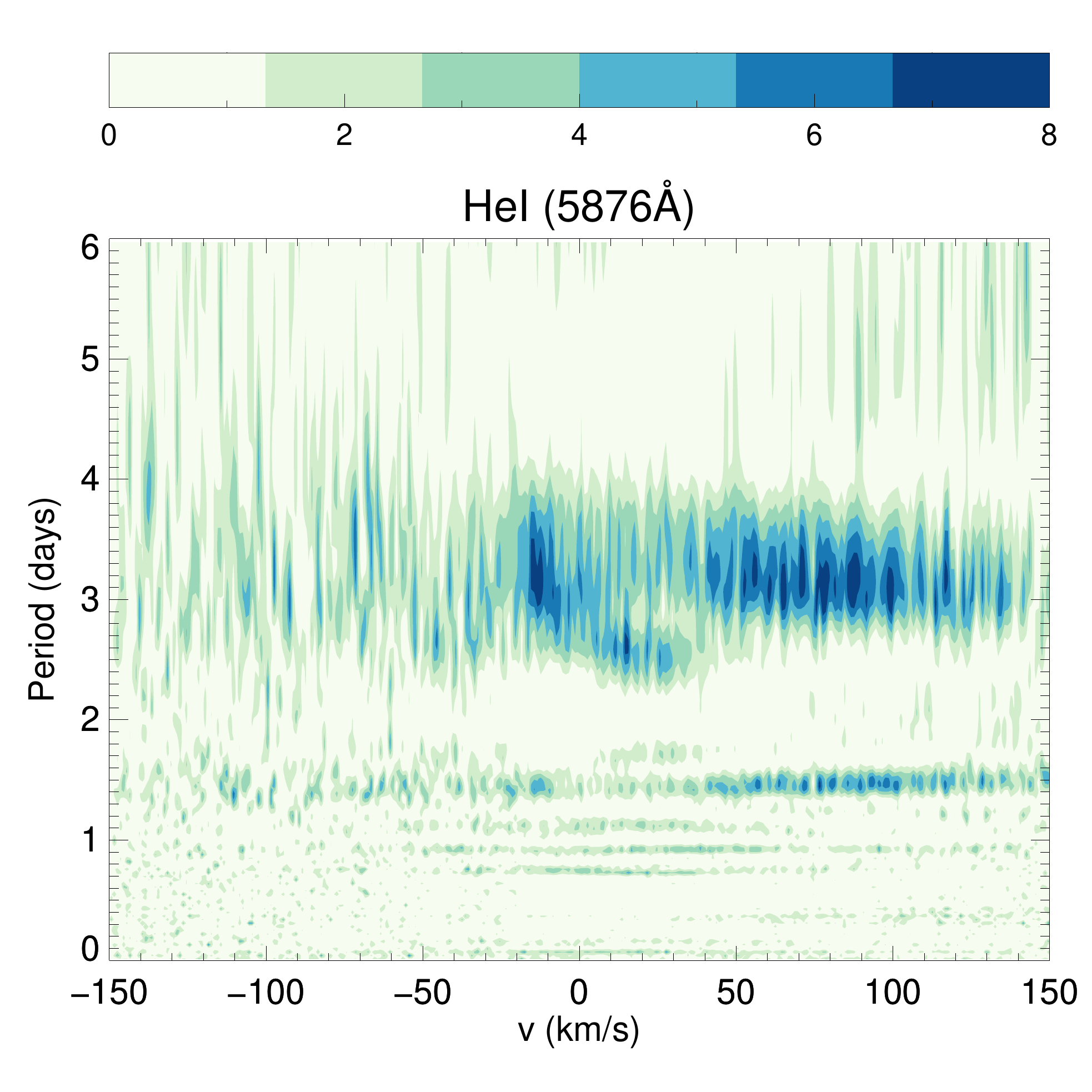}
   \caption{2-dimensional periodogram of the He~I line of T~Tau. The abscissa shows the velocity 
   bins (centred at stellar rest velocity) for which periodograms were measured, the ordinate shows 
   the range of periods spanned, and the colour gradient represents the periodograms' intensities. 
   }
   \label{fig:2dper}
\end{figure}

Another interesting aspect of this star's He~I line profile is that its BC\footnote{The Gaussian 
decomposition of each observation of this star can be verified in the online material.} 
also shows periodic radial velocity variability, but with a slightly longer period of 
3.1$\pm$0.3~days. The bottom two panels of Fig. \ref{fig:ttaubc} show the radial velocity 
variability of this component folded in phase with the period of the NC ($P=2.8$~days) and with 
the longer period of $P=3.1$~days. It is clear that this component folds much better 
in phase with the longer period. 
In order to confirm whether these two slightly different periods are real and not an artefact 
of the Gaussian decomposition, we performed a 2-dimensional periodogram analysis on the full 
line (Fig. \ref{fig:2dper}). From this figure it appears that the dominant period is the 3.1~day 
period of the BC, which is clear on the wings of the profile (especially the redshifted wing, 
between $\sim$40km~s$^{-1}$ and 100km~s$^{-1}$), where the BC dominates\footnote{Another 
peak can also be seen at approximately 1.5 days, but this is simply an artefact, equal to half 
of the 3.1 day period.}. However, near the centre of the line (around 5~-~10km~s$^{-1}$, where 
the NC appears), the periodogram seems to split and shows a small peak at 2.7 days. 
If the full line originated in the hotspot, then this apparent quasi-periodicity could occur, 
for instance, if the hotspot were to move on the stellar surface, or to split into more than one 
spot. However, because the wings of the profile come from the BC, which is believed to have a 
different origin than the NC, it seems more likely that the two components of the He~I line 
indeed have two slightly different periods\footnote{Because we are analysing the normalized 
spectra, variations of the continuum caused, for instance, by variability of veiling, would 
also affect the periodogram. Nevertheless, one can see in Fig. \ref{fig:he1veil2} that the 
veiling variability in this star is small and not periodic, so any period measured in Fig. 
\ref{fig:2dper} should not have any influence from this effect.}. 
Since the period from the NC agrees well with the rotation period of 2.8 days found from 
photometry in the literature and also from the Li~I line in this study, this likely represents 
the stellar rotation, while the slightly larger period of 3.1 days from the BC should be caused 
by a different phenomenon. 

According to \citet{beristain01}, the broad component of the HeI$\lambda5876$ line may come from 
a hot wind or from the accretion funnel flows. Since in T Tau this component is redshifted on 
average, it must be coming mostly from the accretion funnel flows at polar angles less than 
54.7$^{\circ}$ \citep[see Fig. 9 of][]{beristain01}. If this is indeed the case, then a 
slightly larger period measured from this component when compared to the stellar rotation period 
may be an indication that there is differential rotation throughout the accretion 
funnel flow. This could be expected if a star's magnetosphere were to truncate the inner disc at a 
radius larger than the co-rotation radius, resulting in the part of the accretion funnel flow that 
connects to the inner disc rotating more slowly than the stellar surface. The star would then 
accrete via the propeller regime, though this could be a temporary situation, not necessarily 
reflecting the steady state of the system. Its K2 light curve, for instance (which was observed 
in 2015 - four years after the spectroscopic observations in this paper), is periodic, which seems 
to indicate that this star was in a stable accretion regime at the time of the K2 observations 
(see Fig. \ref{fig:k2lcs}). 

We cannot, however, conclude with certainty that there is differential rotation in the 
accretion columns of T~Tau, since the two periods found in fact agree within their error bars. 
A more in-depth study, with more accurate periods, would be necessary to confirm this hypothesis. 
Additionally, current radiative transfer models are incapable of reproducing the main 
characteristics of the broad component of the HeI$\lambda5876$ line with standard accretion 
models \citep[see, e.g.][]{kurosawa11}, so we cannot say for certain what is the origin of this 
broad emission. 

\subsubsection{V836 Tau}

This star shows some variability in the radial velocity of its He~I line, but no clear period 
can be distinguished in the 6-day observation, which is shorter than the period of 7.0 days 
given in the literature \citep{rydgren84}. The He~I line intensity and veiling show little 
variability and no clear trend is seen in the H$\alpha$ or H$\beta$ line profiles. 

\subsection{Velocity of the flow of matter traced by He~I}

We estimate the velocity ($V_{flow}$) of the accretion flow that is traced by the narrow 
component of the HeI$\lambda5876$ line for each star in our sample, by taking the median He~I 
radial velocity of each star and de-projecting it from the line of sight. 
Therefore, \mbox{$V_{flow} =~V_{rad,med}/\cos\alpha$}, where $\alpha$ is the angle between the 
line of sight and the direction of the flow of matter onto the star. This angle is related to 
the system inclination ($i$ - given in Table \ref{tab:vflow}) and the latitude of the accretion 
shock ($l$ - given in Table \ref{tab:results}) through the relation $\alpha = |i-l|$ (see Fig. 
\ref{fig:sketch}). Table \ref{tab:vflow} shows the derived flow velocities $V_{flow}$, as well 
as the mean ($<V_{rad}>$), median ($V_{rad,med}$) and standard deviation ($\sigma$) of the He~I 
radial velocities measured in each star (with regard to the star's rest velocity), the mean 
error in $V_{rad}$ measurements for each star, and the system inclinations $i$.

\begin{table*}
 \caption{Average radial velocities of the HeI$\lambda5876$ line and resulting flow velocities} 
 \label{tab:vflow} 
 \centering
 \begin{tabular}{l c c c c c c c}
  \hline
  Star    & $<V_{rad}>$ & $V_{rad,med}$ & $<err V_{rad}>$ & $\sigma V_{rad}$ & Inclination  & Ref  & $V_{flow}$      \\ 
         & (km~s$^{-1}$) & (km~s$^{-1}$) & (km~s$^{-1}$)  & (km~s$^{-1}$)    & ($^{\circ}$) &      & (km~s$^{-1}$)   \\  
  \hline                                                                                                               
  DE~Tau   &   5.84      &   5.96           &   1.19          &   0.81           &  $70\pm7$    & P14  &   14$\substack{  +38\\-7}$   \\  
  DF~Tau   &   6.54      &   6.69           &   1.01          &   1.61           &  $19\pm8$    &  *   &  6.7$\substack{ +1.6\\-1.0}$ \\  
  DK~Tau   &   9.36      &   9.54           &   1.86          &   3.96           &  $13\pm3$    & M19  &  9.6$\substack{ +2.3\\-1.9}$ \\  
  DN~Tau   &   6.68      &   7.36           &   0.99          &   1.40           &  $35\pm 1$   & L18  &  7.9$\substack{ +2.1\\-1.5}$ \\  
  GI~Tau   &   6.72      &   6.32           &   1.25          &   1.65           &  $29\pm9$    &  *   &  7.9$\substack{ +9.4\\-2.8}$ \\  
  GK~Tau   &   7.21      &   7.40           &   1.95          &   1.18           &  $71\pm5$    & S17  &   18$\substack{  +27\\-9}$   \\  
  GM~Aur   &   7.10      &   7.90           &   2.72          &   2.56           &  $55\pm 1$   & T17  & 10.6$\substack{+10.0\\-4.9}$ \\  
  IP~Tau   &   5.99      &   6.14           &   1.08          &   0.87           &  $45\pm1$    & L18  &  7.8$\substack{ +3.7\\-2.2}$ \\  
  IW~Tau   &   4.33      &   4.14           &   0.77          &   0.77           &  $33\pm9$    &  *   &  4.6$\substack{ +3.0\\-1.3}$ \\  
  T~Tau    &   4.20      &   2.00           &   5.19          &   7.55           &  $28\pm1$    & M19  &  2.0$\substack{ +5.5\\-5.2}$ \\  
  V836~Tau &   6.01      &   6.22           &   1.30          &   1.12           &  $61\pm10$   & T17  &   10$\substack{  +16\\-4}$   \\  
  \hline 
 \end{tabular}
 
 \textbf{Notes:} The second and third columns show the mean and median radial velocity of He~I 
 for each star, column 4 shows the mean error on the radial velocity measurements, columns 5 and 
 6 give the inclination to the system and its reference, and the final column shows the flow 
 velocity derived for each star.
 References for inclinations are \citet{pietu14} (P14), \citet{manara19} (M19), \citet{long18} (L18), 
 \citet{simon17} (S17), \citet{tripathi17} (T17), and those marked with an asterisk have no direct 
 measurement of the disc in the literature, so inclinations were estimated using the relation 
 $v\sin i = (2\pi R_* / P_{rot}) \sin i$, where $P_{rot}$ and $v\sin i$ are given in this paper, 
 and $R_*$ was taken from \citet{herczeg14}. 

 \begin{tabular}{c c c c}
   \hline
    Average $<V_{rad}>$ & $\sigma <V_{rad}>$  & Average $V_{flow}$ & $\sigma V_{flow}$  \\
    (km~s$^{-1}$)       &  (km~s$^{-1}$)      &    (km~s$^{-1}$)   &  (km~s$^{-1}$)     \\    
   \hline                                                          
     6.4                &  1.4                &    9.0             & 4.3                \\
   \hline 
 \end{tabular}
\end{table*}

\section{Discussion}\label{sec:discuss}

\subsection{Comparison with other observational studies} 

As can be seen in Table \ref{tab:results}, the magnetic obliquities we find for the stars in
our sample are between 5$^{\circ}$ and 23$^{\circ}$, with most (80\%) below 15$^{\circ}$.
Therefore in our sample the magnetic fields seem to be generally well aligned with the stellar
rotation axis, which is somewhat in disagreement with what has been found from direct magnetic 
field measurements in T Tauri stars, many of which have a misalignment larger than 20$^{\circ}$
\citep[see Table \ref{tab:magob_otr} and, e.g.,][]{johnstone14}.
It is possible that our sample may be influenced by a selection bias. The stars in this sample
were selected based mainly on two criteria: having shown periodic photometric behaviour in a
previous study and presenting a HeI$\lambda5876$ line profile in emission that is dominated by 
the NC. The only exception to the latter criterion is T Tau, which was observed mainly
because of its brightness and is the only star in our sample for which the He~I line is
dominated by the broad component. It is also the star that shows the largest magnetic obliquity
in this study, of 23$^{\circ}$. It is possible that there may be a connection between the
processes that originate the BC of the HeI$\lambda5876$ line and large magnetic obliquities. 
This would lead to an observational bias in our sample, where by excluding stars that present 
very broad HeI$\lambda5876$ emission, we exclude the stars with large magnetic obliquities. 
In order to confirm this hypothesis, we would need to study a much larger sample of CTTSs with 
more diverse He~I line profiles.

The other selection bias that our sample could possibly be subject to is with periodicity,
since the stars in our sample all have a measured period in the literature. However, it is 
difficult to find an explanation in which this leads to a bias towards magnetic obliquities 
smaller than $\sim 15^{\circ}$, since MHD models of \citet{kurosawa13} and \citet{blinova16} 
predict that larger magnetic obliquities should lead to stars accreting 
more often in a stable accretion regime, which is believed to produce periodic signatures in a 
star's photometry. Therefore, based on these models, choosing a sample of stars with known 
periodicity should not lead us to exclude stars with large magnetic obliquities. 

While studying the spectroscopic variability of the very active star EX~Lupi, 
\citet{sicilia-aguilar15} found a periodic signature in the radial velocity variation of 
different He~I lines (at $\lambda = 4713$\AA, 5876\AA, and  6678\AA), as well as the 
HeII$\lambda4686$ line, which were consistent with the rotational modulation of a hotspot on 
the stellar surface. They found, however, that the amplitudes of the radial velocity variations 
differ for the different lines, with lines with higher excitation potentials having larger 
amplitudes and a clearer modulation. 
In general, lines with different excitation potentials may originate in regions with different 
temperatures, which could be at different latitudes, longitudes (a spot may have the highest 
temperature at its core and be surrounded by regions of lower temperature), or height above the 
stellar surface \citep[there may be a temperature structure along the accretion column, above 
the stellar surface, as has been detected by][]{dupree12,sicilia-aguilar17}. This could explain 
the different amplitudes found by \citet{sicilia-aguilar15} for the different lines. 

If there is a difference in the temperature structure on the stellar surface along latitude, 
then this may help to explain our apparently low magnetic obliquities. By choosing the 
HeI$\lambda5876$ line, which has a lower excitation potential than HeI$\lambda4713$ and 
HeII$\lambda4686$, we may be tracing a region that is at a slightly higher latitude than the 
central part of the accretion shock, which may be nearer to the magnetic pole. If true, this 
effect would probably bias the estimates of magnetic obliquity using this line to lower values, 
which may explain not only our sample, but also the fact that the three other stars from the 
literature whose obliquities were derived using the same method as the one described in our 
paper also present such low values of $\Theta$ \citep[EX~Lup, RU~Lup and DR~Tau\footnote{These 
are three very active stars and all present both a NC and BC in their HeI$\lambda$5876 line 
profiles. In all three cases, the two components were deconvolved using Gaussian fits and the 
radial velocity of the NC was used to derive the hotspot latitude, in accordance with our 
study.}, see Table \ref{tab:magob_otr} and][]{sicilia-aguilar15, gahm13, petrov11}. 
However, this does not seem to be the case. \citet{petrov01} found the same amplitude (within 
error bars) for the radial velocity variability of He~I and He~II lines for the star RW~Aur. 
Also, several studies of spectropolarimetry have found misalignements between the main 
component of the magnetic field and the stellar rotation axis that are consistent with the 
HeI$\lambda5876$ variability amplitudes found in the same observations \citep[e.g. 
BP~Tau, AA~Tau, LkCa15, and CI~Tau; ][]{donati08, donati10b, alencar18, donati20}. 

We can also verify how the variability of photospheric lines compares with that of the He~I 
lines. \citet{crockett12} studied the radial velocity variability of photospheric lines in 
optical and near-infrared spectra of several CTTSs, in order to study the effect of stellar 
spots on these lines. Their sample included four of the same stars as our sample. The amplitudes 
they found for the radial velocity variability of photospheric lines in optical wavelengths 
agree, within the uncertainties, with our amplitudes for the HeI$\lambda$5876 line, though the 
values for the photospheric lines are systematically lower, as expected for large cool spots. 

\subsection{Photometric variability} 

\citet{blinova16} predict that the location of the magnetospheric radius $R_m$ (the disc's 
truncation radius) with respect to the disc's co-rotation radius $R_{co}$ has a strong 
influence on the accretion regime, stronger than the magnetic obliquity ($\Theta$). 
None the less, for stars with magnetospheres of similar size, they predict 
that an increase of the magnetic obliquity leads to a decrease of the amplitude of the light 
curve oscillations that are associated with instabilities and an increase of the amplitude of 
the variability associated with the rotation of the star. 
According to \citet{blinova16}, for small values of $\Theta$ ($\lesssim 5^{\circ}$ for 
$R_{co} \sim 1 - 1.4 R_{m}$, or up to $\sim 10^{\circ}$ for smaller magnetospheres), the main 
source of variability should be unstable ordered hotspots and the light curve should be very 
irregular, though the stellar rotation period may still be found in the light curve's Fourier 
spectrum. For slightly larger values of $\Theta$ ($\sim 15^{\circ}$), the stellar rotation period 
should become more evident, but many short-scale oscillations from instabilities could still be 
seen in the light curve. When $\Theta$ is around $\sim 20^{\circ} - 30^{\circ}$, the oscillations 
from instabilities would have smaller amplitudes, leading to more regular light curves, but with 
the stellar rotation period becoming slightly less well defined in the Fourier spectrum. 

Among the 11 stars in our sample with He~I in emission, 6 were observed recently by the K2 
mission \citep[their light curves are shown in Fig. \ref{fig:k2lcs} and a more detailed 
analysis is explored in][and in Roggero et al. in prep.]{rebull20}. Even with such 
a small number, this sample shows a large diversity of light curve morphologies. 
IW~Tau presents a very regular light curve, consistent with modulation by 
the rotation of stable spots on the stellar surface \citep[as the spot-like light curves 
of][]{alencar10}, in combination with a second periodic event (possibly due to its binarity). 
T~Tau shows a periodic but somewhat irregular behaviour (altering between what look like flux 
bursts and flux dips) reminiscent of the class of light curves attributed to stochastic 
accretion by \citet{stauffer16}. The K2 light curve of DF~Tau is not periodic and appears to 
be dominated by flux bursts \citep[as those described in][]{stauffer14}. Finally, the remaining 
3 light curves (those of DK~Tau, GI~Tau and GK~Tau) seem to be dominated by flux dips \citep[as 
those described in][]{mcginnis15}. However, 
it is important to note that GI~Tau and GK~Tau consist of a binary system with a separation of 
13.2" \citep{akeson19}, so the light curve of GI~Tau may be slightly contaminated by that of 
the primary, GK~Tau (this is evident from the fact that a peak at 4.6 days, the rotation period 
of GK~Tau, is detected in a periodogram analysis of GI~Tau's light curve). 

We can compare our findings with the predictions of \citet{blinova16} regarding the relationship 
between the magnetic obliquity and the star's light curve, stated above. 
The most regular light curve in our sample belongs to IW~Tau. This star's He~I line did not 
show a detectable variation in radial velocity, therefore its magnetic obliquity must be small. 
This seems not to be in line with the theoretical prediction that with smaller magnetic 
obliquities, light curves are more irregular. However this particular star has a very low mass 
accretion rate (if it is in fact accreting and the He~I emission we observe does not come from 
a very active chromosphere), therefore it should have a large magnetosphere, which is also 
predicted to lead to stable configurations (so long as the magnetic radius $R_m$ does not exceed 
the co-rotation radius $R_{co}$). 
Excluding IW~Tau, the light curves of T~Tau and GK~Tau appear to have the least amount of 
irregular oscillations among the stars in this sample, with both showing a clear periodic signal. 
T~Tau presents the largest magnetic obliquity in the sample, of $23^{\circ} \pm 10 ^{\circ}$, 
consistent with the prediction that stars with a magnetic obliquity larger than $\sim 20^{\circ}$ 
would tend to have more regular light curves. However, the magnetic obliquity found for GK~Tau 
is low, of only $5^{\circ} (\substack{ +7^{\circ}\\-5^{\circ}})$, while the values found for 
DF~Tau, DK~Tau and GI~Tau (all of which present light curves that appear to be more irregular 
than GK~Tau's) are larger, between $12^{\circ}$ and $18^{\circ}$. It is clear that with this 
small sample we cannot distinguish any effects the magnetic obliquity may have on a star's light 
curve morphology. 

\subsection{Correlations with stellar properties}

In an attempt to shed some light on the origin of the magnetic obliquity, we joined our sample 
with other values from the literature and searched for possible correlations with stellar 
parameters. The magnetic obliquities taken from the literature are given in Table 
\ref{tab:magob_otr}, along with their references. They were derived mostly using ZDI, but a few 
cases were also found in which they were derived in a similar fashion as in our study (or in 
which a value of $\Delta V_{rad} (He I)$ and $v \sin i$ are given, allowing us to derive 
$\Theta$). 

We find a tentative correlation between the magnetic obliquity and stellar mass. 
A Kendal $\tau$ analysis gives a false alarm probability of $<1\%$, 
but it is important to note that there are few stars in this sample, so this result should be 
taken with care. It is, however, an interesting indication that the inner structure of the star 
may have an important role in determining its magnetic obliquity since, for stars of a similar 
age (as should be more or less the case in this sample), a more massive star is expected to 
have developed a radiative core while less massive stars remain fully convective. 

\begin{figure}%[t]
   \centering
   \includegraphics[width=0.47\textwidth]{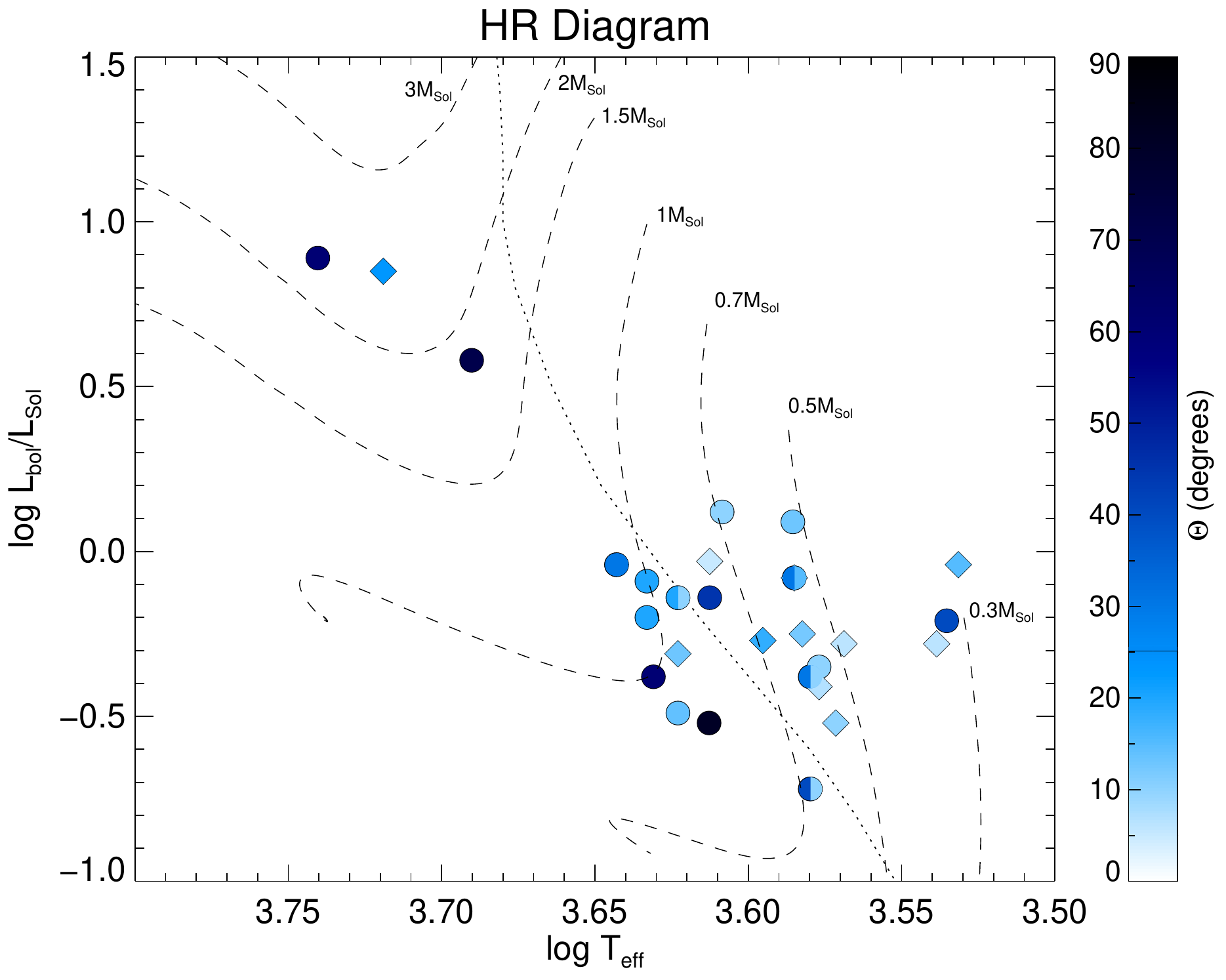}
   \caption{HR diagram with the stars in our sample (filled diamonds) and other stars from 
   the literature (filled circles). Colours represent the stars' magnetic obliquities. Two 
   colours are used when two different values were found at different epochs. Dashed lines 
   represent mass tracks from pre-main sequence evolutionary models of \citet{tognelli11}, while 
   the dotted line represents the limit where a radiative core begins to develop, according 
   to these models (stars to the right of this line should be fully convective). 
   } 
   \label{fig:hrd}
\end{figure}

In order to further investigate this hypothesis, we plotted an HR diagram using different 
colours to represent the magnetic obliquity found for each star (Fig. \ref{fig:hrd}). 
In a few cases, two different observations using spectropolarimetry resulted in different values 
of the magnetic obliquity (with differences of up to 30$^{\circ}$ between the two epochs). Both 
values are represented in the figure by a two-coloured symbol. We used different symbols to 
differentiate the sample from our study (filled diamonds) from those taken from the literature 
(filled circles). Theoretical mass tracks from \citet{tognelli11} are also plotted for reference. 
The dotted line represents the evolutionary phase in which a radiative core is expected to 
begin to develop, according to \citet{tognelli11} \citep[see also][]{gregory12}, meaning that 
stars to the right of this line should be fully convective. 
This figure seems to suggest that fully convective stars usually have low magnetic obliquities 
(though there are a few cases where it is as high as 40$^{\circ}$), while the stars with large 
magnetic obliquities ($\gtrsim 60^{\circ}$) are those that have developed at least a small 
radiative core. The average magnetic obliquity among the stars in the fully convective part of 
the HR diagram is 17$^{\circ}$ (with an rms dispersion of 12$^{\circ}$), while in the partly 
radiative part of the HR diagram it is slightly larger, 35$^{\circ}$ (with an rms deviation of 
22$^{\circ}$). This could be complementary with other studies of magnetic fields in young stars 
that suggest that, as a radiative core develops, stellar magnetic fields tend to become more 
complex and less intense than in fully convective stars \citep{gregory12,folsom16}. If this is 
indeed the case, it may explain why our sample shows a tendency towards low magnetic obliquities, 
as most of our stars are fully convective. 

The apparent evolution of the magnetic field configuration across the HR diagram \citep[in terms 
of the strength of the field and its complexity, as demonstrated by][as well as in terms of the 
magnetic obliquity, as shown here]{folsom16, villebrun19} points to a dynamic origin of stellar 
magnetic fields, likely as the consequence of a dynamo effect. This study and previous ones such 
as \citet{folsom16} also seem to suggest that there are intrinsic differences between magnetic 
fields of stars that are fully convective and those that have begun to develop a radiative core. 

\section{Summary and conclusions}\label{sec:conc}

We have analysed the variability of the HeI$\lambda$5876 emission line profile in a sample of 10 
CTTSs and 1 WTTS, measured with high-resolution spectroscopy over a period of up to 10 days. 
Most observations show a simple profile, dominated by a narrow component (NC) that is believed to 
originate in accretion shocks near the stellar surface. For four stars, the HeI$\lambda$5876 line 
profile is seen to be composed of a combination of narrow and broad component (BC), the latter 
being believed to have multiple origins, likely formed in both hot winds and accretion columns 
\citep{beristain01}. 
For these four stars, each individual observation was fitted with a combination of two Gaussians, 
one broader than the other, in order to separately investigate the origin of the two components. 
Both appear to be variable in their own way on the time-scale spanned by our observations. 
With the resolution of our spectra we can clearly see that the NC is not truly Gaussian, but 
this approximation does seem to be valid when analysing its radial velocity. 
This can be illustrated by the fact that, even though the star 
T Tau's He~I line profile is clearly dominated by the BC, its stellar rotation period of 2.8 days 
is recovered from the radial velocity measurements of the NC, while this period is not clear in a 
radial velocity analysis of the composite profile. This shows that an analysis of the NC of the 
He~I line is still possible even for stars whose He~I line profile is clearly dominated by the BC. 

The NC of the HeI$\lambda$5876 emission line is shown to be redshifted by an average of 
6.4~km~s$^{-1}$ in our sample, with a standard deviation of 1.4~km~s$^{-1}$. By taking the median 
redshifts for each star and de-projecting them from our line of sight, we find that the material 
responsible for this emission is traveling at an average of 9~km~s$^{-1}$ in our sample (with a 
standard deviation of 4.3~km~s$^{-1}$). This velocity is consistent with material tracing the 
post-shock region of accretion shocks, after considerable deceleration of the free-falling 
material has occurred. 

By measuring the amplitude of the radial velocity variability ($\Delta V_{rad}$) of the NC of the 
HeI$\lambda$5876 emission line, along with the stars' projected rotational velocities ($v\sin i$), 
we were able to estimate these stars' magnetic obliquities ($\Theta$ - the angle between the axis 
of their magnetic field and rotation axis). We find an average magnetic obliquity of 
$11.4^{\circ}$ in our sample, with an rms dispersion of $5.4^{\circ}$. The magnetic axis thus 
seems close to being aligned with the stellar rotation axis in our sample. This is not entirely 
in agreement with other studies of magnetic field configurations \citep[e.g.,][]{johnstone14}, 
which find several cases of misalignments larger than 20$^{\circ}$. This difference may simply be 
due to an issue of low number statistics. However, there is also the possibility that our sample 
may be subject to a selection bias. With the exception of T Tau, the stars in this sample were 
chosen on the basis that their He~I line profiles are dominated by the NC. The star T Tau, whose 
He~I line profile is clearly dominated by the BC (see Fig. \ref{fig:hei_profiles}), is the star 
that presents the largest magnetic obliquity in our sample (of $23^{\circ} \pm 10^{\circ}$). If 
the mechanism that originates the BC of the HeI$\lambda$5876 line is somehow connected to larger 
magnetic obliquities, then this would result in a selection bias in our sample. However, to 
confirm this hypothesis, a larger study of magnetic obliquities, including stars with more 
diverse He~I profiles, would be needed. 

We find tentative evidence for a trend between the position of a star on the HR diagram and its 
magnetic obliquity. This result is based on a small sample and should be taken with care, but the 
possibility that a star's inner structure can play a strong role in determining its magnetic 
obliquity would be consistent with other studies that show an apparent evolution of a star's 
magnetic field as it evolves across the HR diagram \citep[e.g.,][]{gregory12,folsom16,villebrun19}. 
In particular, there may be a considerable difference between the magnetic field configurations 
of stars that are fully convective and those that are partially radiative, a possibility that 
merits further investigation as it is unfortunately still subject to low number statistics. 
If this is indeed the case, it supports the idea that magnetic fields in T Tauri stars are 
generated by a dynamo, rather than by fossil fields, and that the existence of a boundary 
between radiative and convective zones plays an important role in determining the geometry of 
that magnetic field. It could also be responsible for our sample's bias towards low magnetic 
obliquities, since most of the stars in our sample are in the fully convective portion of the 
HR diagram. 

Besides our main results, we also find some interesting aspects of individual sources. A 
joint analysis of the variability of the He~I, H$\alpha$ and H$\beta$ lines show, in at least 
three sources (DK~Tau, DN~Tau and IP~Tau), evidence for a spatial association between the 
accretion shocks on the stellar surface and the accretion columns, which would be expected in 
a scenario in which disc-locking takes place. Meanwhile another source, DF~Tau, shows evidence 
for a variable wind that seems to be spatially associated with the accretion shock. This could 
be a stellar wind that is launched very close to the accretion shock, or it could 
result from a non-uniform disc wind being launched from close to the truncation radius, in a 
part of the inner disc where an accretion column is forming. Since this variability seems to 
be consistent with having the same period as the stellar rotation, this would also be close 
to the co-rotation radius. 

We find that the star IW~Tau, previously reported as a non-accreting, weak-line T Tauri star, 
in fact presents redshifted He~I emission, which means that this emission must come from 
matter falling onto the star. It seems therefore that this star is still weakly 
accreting. An analysis of the flux of several emission lines that are known to be linked with 
accretion leads us to estimate a mass accretion rate of 
$\dot{M}_{acc} = 3 \times 10^{-10}~\mathrm{M}_{\odot}~\mathrm{yr}^{-1}$ for this star. 
We should note that this value is likely overestimated, since at such low limits of accretion, 
these lines are likely contaminated with a strong contribution from chromospheric activity. 

Finally, we recover the stellar rotation period of 2.8 days for the star T~Tau in both the 
radial velocity analysis of the NC of the He~I line, as well as in a portion of a 2-dimensional 
periodogram of this line. However we find a slightly larger period, of 3.1 days, in the wings 
of this line and in the radial velocity analysis of its BC, which is redshifted. Since the BC 
in this case is believed to originate from the accretion columns, while the NC originates at 
the accretion shock near the stellar surface, this may be an indication that there is 
differential rotation throughout the accretion columns of this system. Although, seeing as the 
two periods are very similar, this is not enough evidence to conclude this with certainty and 
further investigation would be needed to support this claim. 

\section*{Acknowledgements}
This paper was based on observations made at Observatoire de Haute Provence (CNRS), France. 
This project has received funding from the European Research Council (ERC) under the European 
Union's Horizon 2020 research and innovation programme (grant agreement No 742095, \textit{SPIDI}: 
Star-Planets-Inner Disk-Interactions, spidi-eu.org; and grant agreement No 743029, EASY: 
Ejection-Accretion-Structures-in-YSOs). 
The work also received support from the CAPES/COFECUB collaboration under grant number 
88887.160792/2017-00. The authors thank Noemi Roggero for the K2 light curves 
and Antonella Natta for very helpful discussions. 

\vspace{-0.5cm}

\section*{Data availability}
The data underlying this article can be accessed through the SOPHIE archive at 
\url{http://atlas.obs-hp.fr/sophie/}.

\vspace{-0.5cm}

\bibliographystyle{mnras}  
\bibliography{references.bib}

\vspace{-0.5cm}

\begin{appendix}

\section{Additional figures}

The following pages present figures that are discussed in the text (Figs. \ref{fig:he1veil1} 
through \ref{fig:k2lcs}).

\begin{figure*}%[p] 
   \centering
   \includegraphics[width=0.24\textwidth]{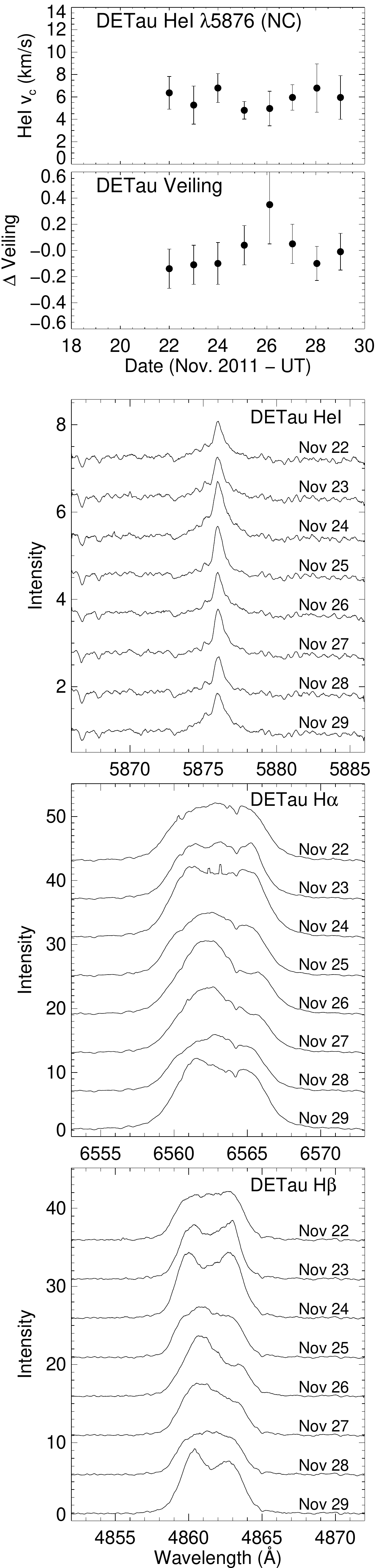}
   \includegraphics[width=0.24\textwidth]{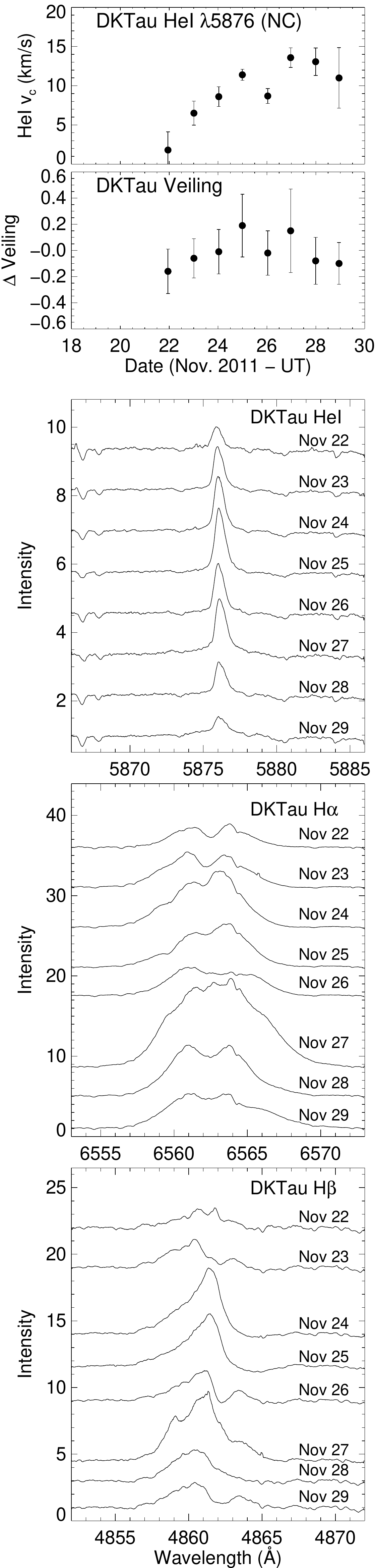}
   \includegraphics[width=0.24\textwidth]{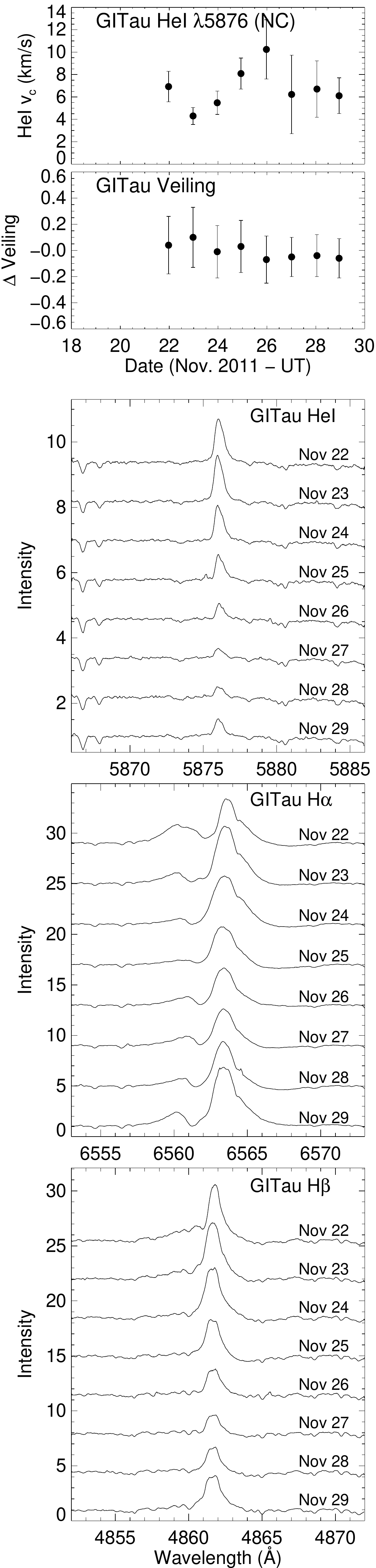}
   \includegraphics[width=0.24\textwidth]{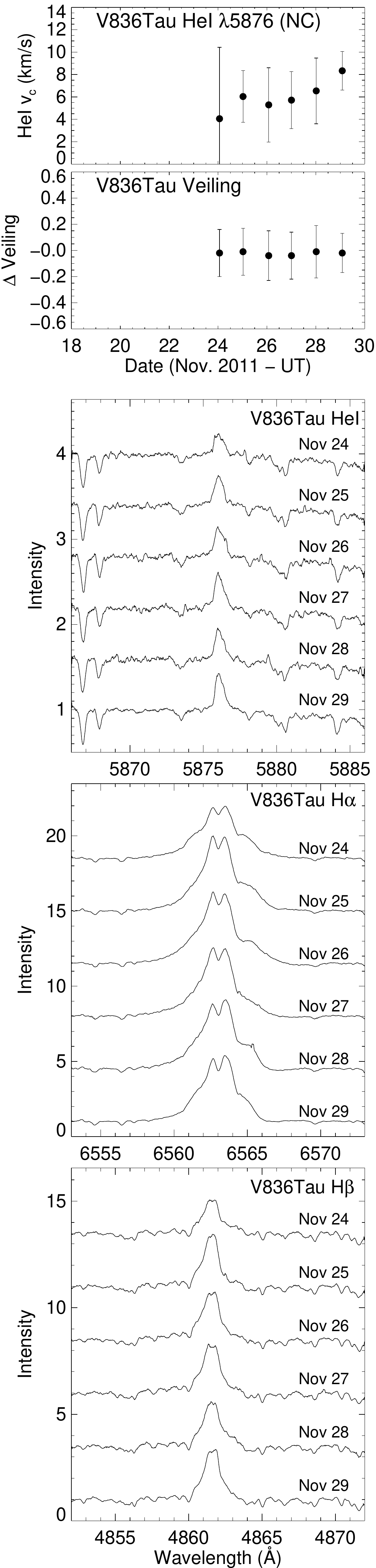}
   \caption{Variability of the radial velocity of the HeI$\lambda5876$ narrow component, veiling, 
   the HeI$\lambda5876$ profile, the H$\alpha$ profile and the H$\beta$ profile for the stars in 
   our sample that were observed for only one rotation cycle. 
   }
   \label{fig:he1veil1}
\end{figure*}

\begin{figure*}%[p] 
   \centering
   \includegraphics[width=0.48\textwidth]{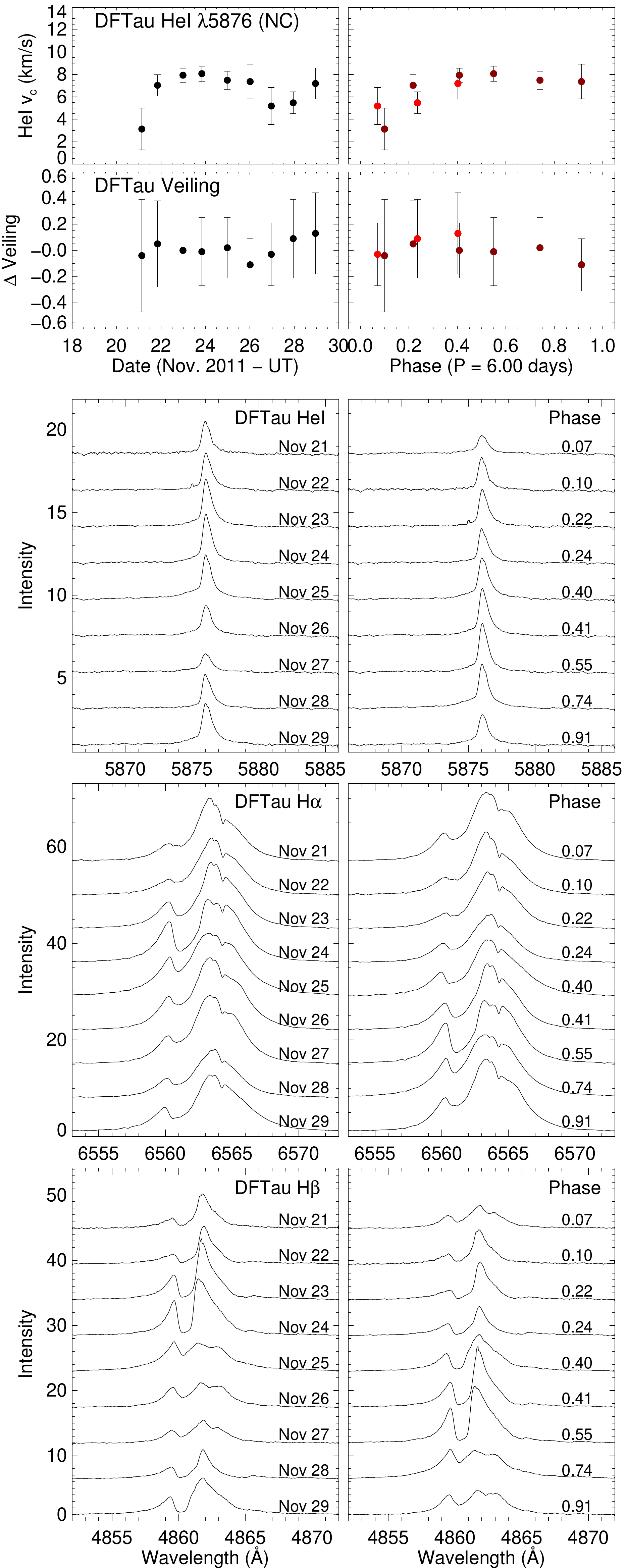}
   \includegraphics[width=0.48\textwidth]{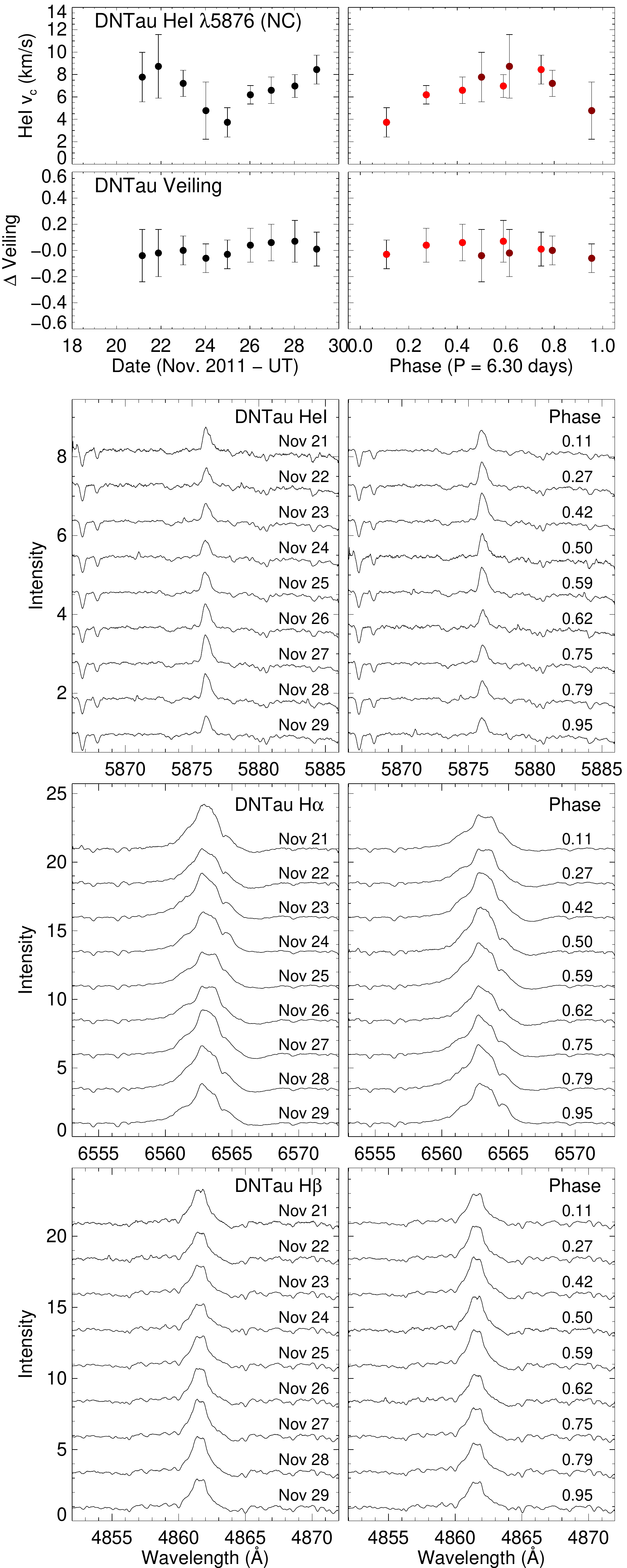}
   \caption{Variability of the radial velocity of the HeI$\lambda5876$ narrow component, veiling, 
   the HeI$\lambda5876$ profile, the H$\alpha$ profile and the H$\beta$ profile for the rest of 
   the stars in our sample, shown in full (\textit{left panels}) and folded in phase using the 
   period believed to be the stellar rotation period (\textit{right panels}).
   Different colors represent different rotation cycles. 
   The initial dates for the phase calculation are shifted by the amount necessary for 
   the hotspot to be in full view around phase $\phi=0.5$. Thus, in a few cases, the phase plots 
   show three colours even though the star was only observed over two rotation periods. 
   }\ContinuedFloat 
   \label{fig:he1veil2}
\end{figure*}

\begin{figure*}%[p] 
   \centering
   \includegraphics[width=0.48\textwidth]{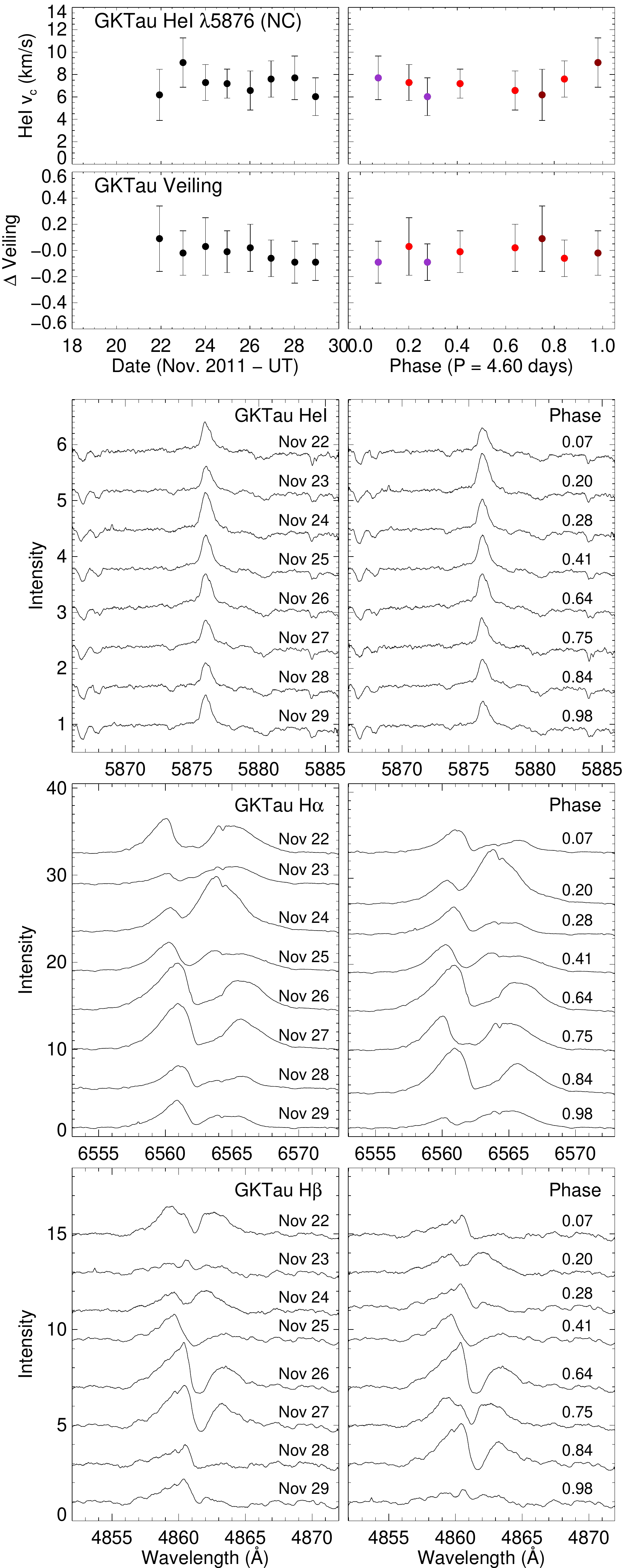}
   \includegraphics[width=0.48\textwidth]{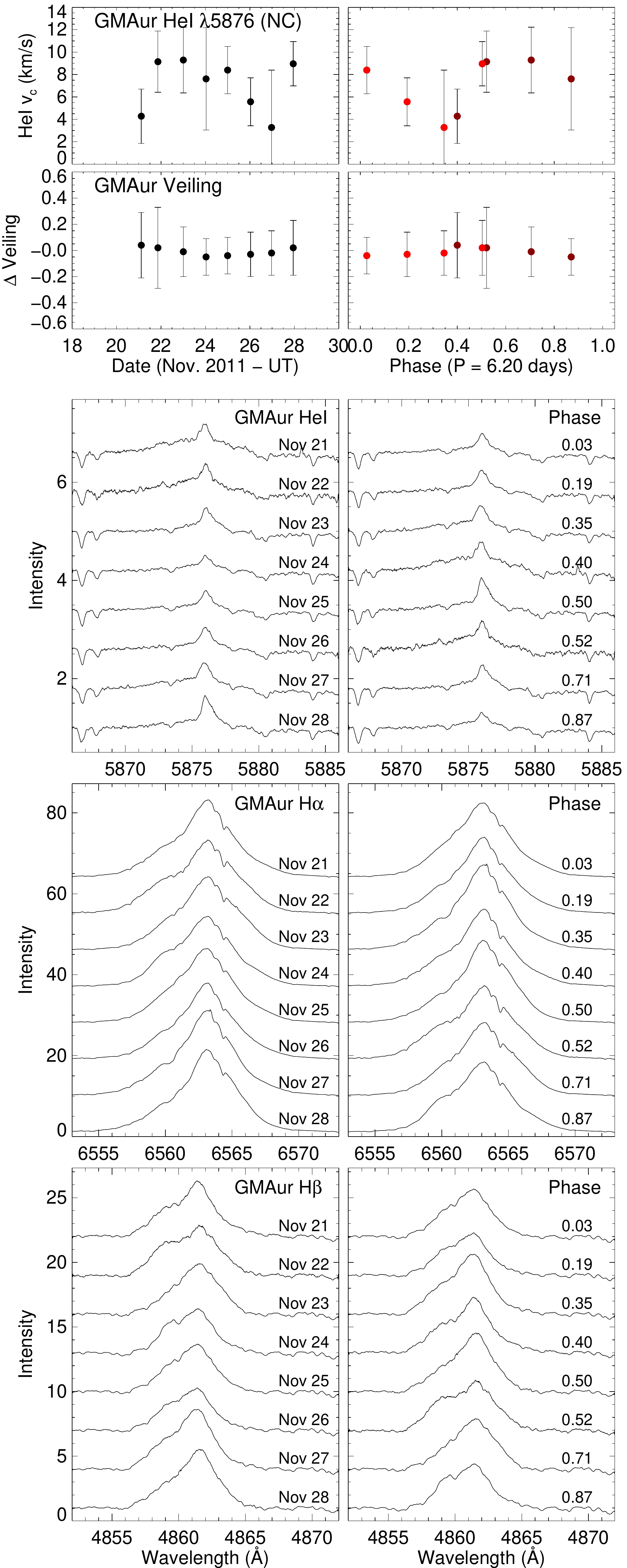}
   \caption{Continued.  
   }\ContinuedFloat 
\end{figure*}

\begin{figure*}%[p] 
   \centering
   \includegraphics[width=0.48\textwidth]{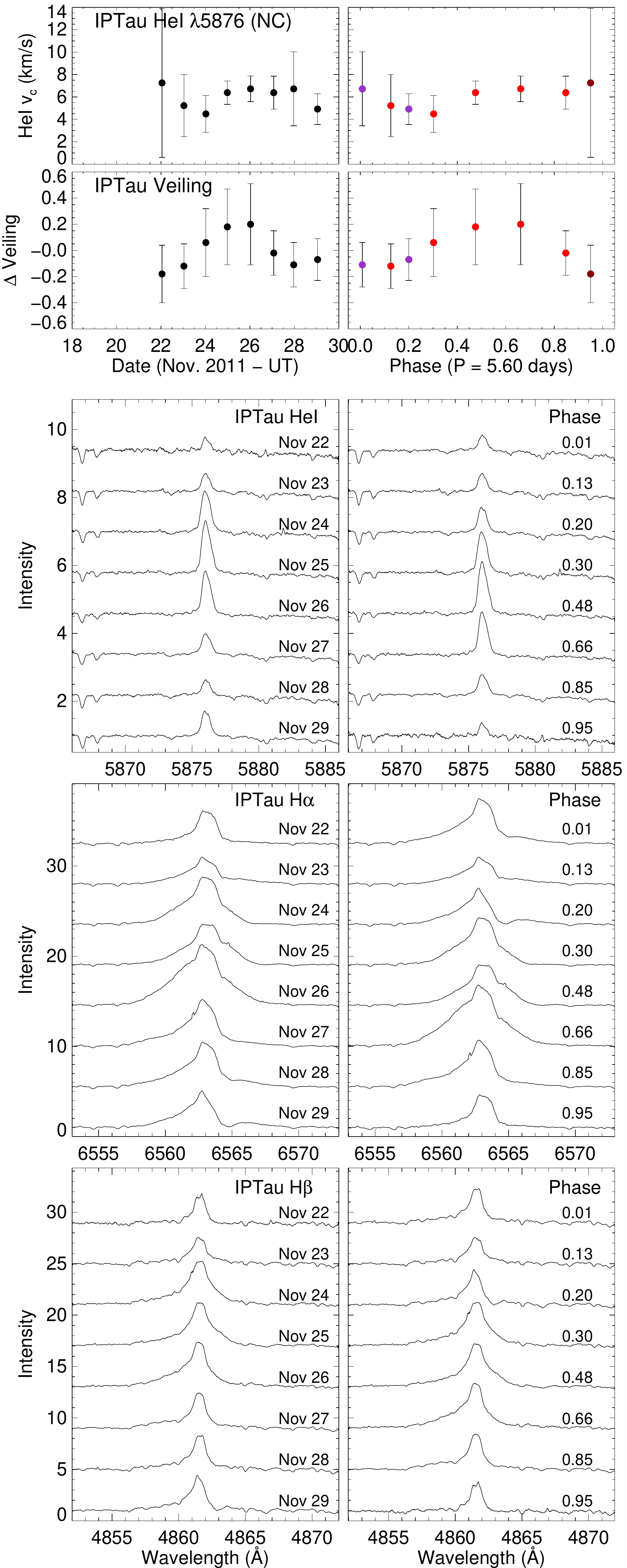}
   \includegraphics[width=0.48\textwidth]{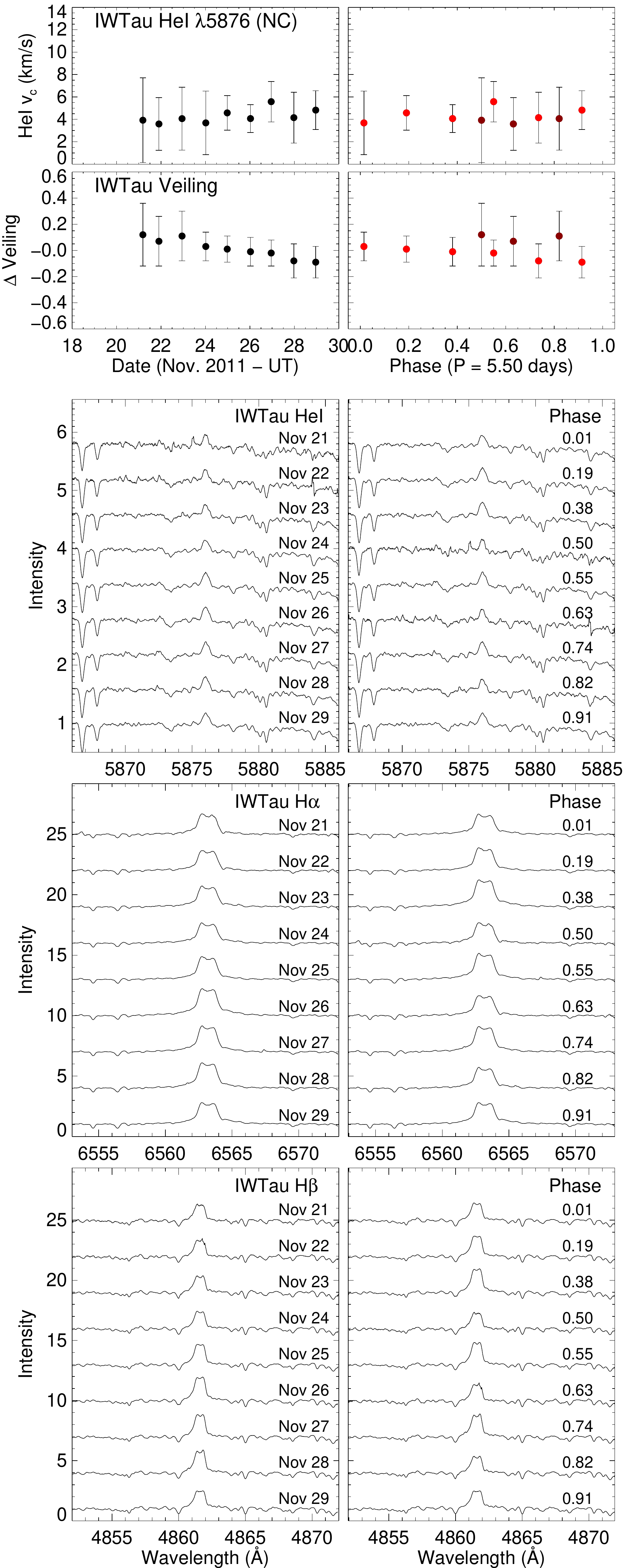}
   \caption{Continued.  
   }\ContinuedFloat 
\end{figure*}

\begin{figure}%[p] 
   \centering
   \includegraphics[width=0.47\textwidth]{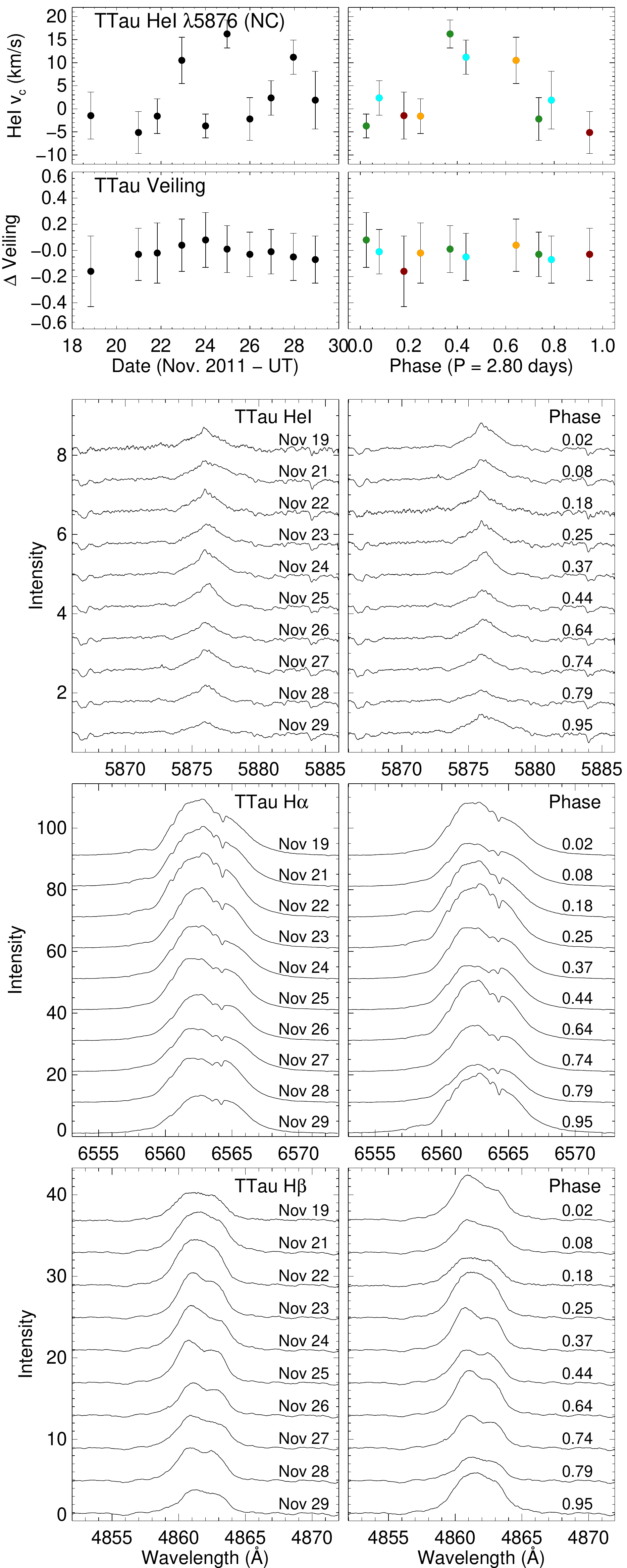}
   \caption{Continued.  
   }
\end{figure}

\begin{figure}%[t]
  \centering
  \includegraphics[width=0.47\textwidth]{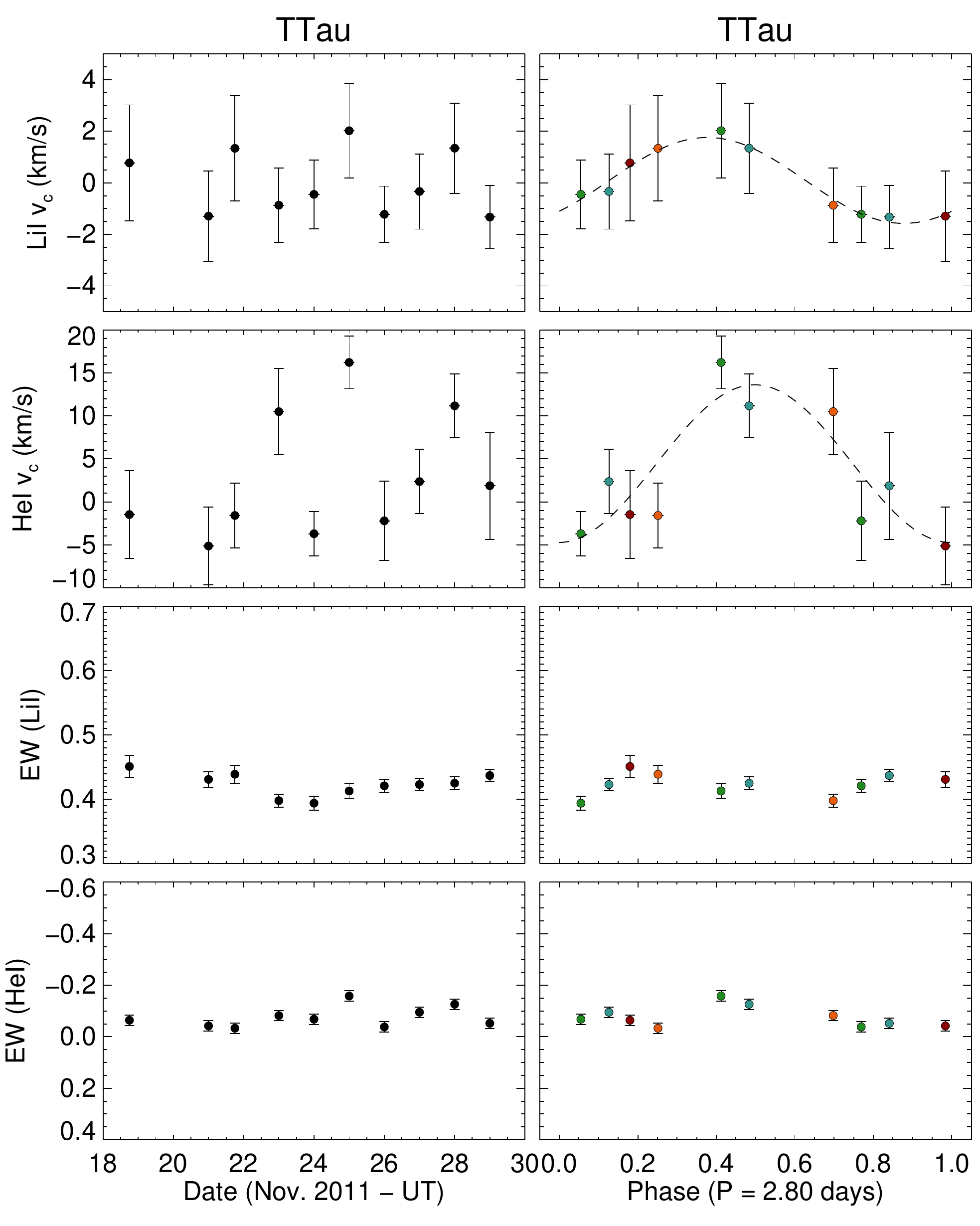}
  \caption{Plots showing the variability of the radial velocity of the LiI$\lambda6708$ line, 
  the radial velocity of the HeI$\lambda5876$ line, the equivalent width of the LiI$\lambda6708$ 
  line and the equivalent width of the HeI$\lambda5876$ line. 
  \textit{Right panels} show the same plots as the \textit{left panels}, but folded in phase with 
  the stellar rotation period. Different colors represent different rotation cycles. 
  }\ContinuedFloat 
  \label{fig:li_he_ew}
\end{figure}

\begin{figure*}%[t]
  \centering
  \includegraphics[width=0.49\textwidth]{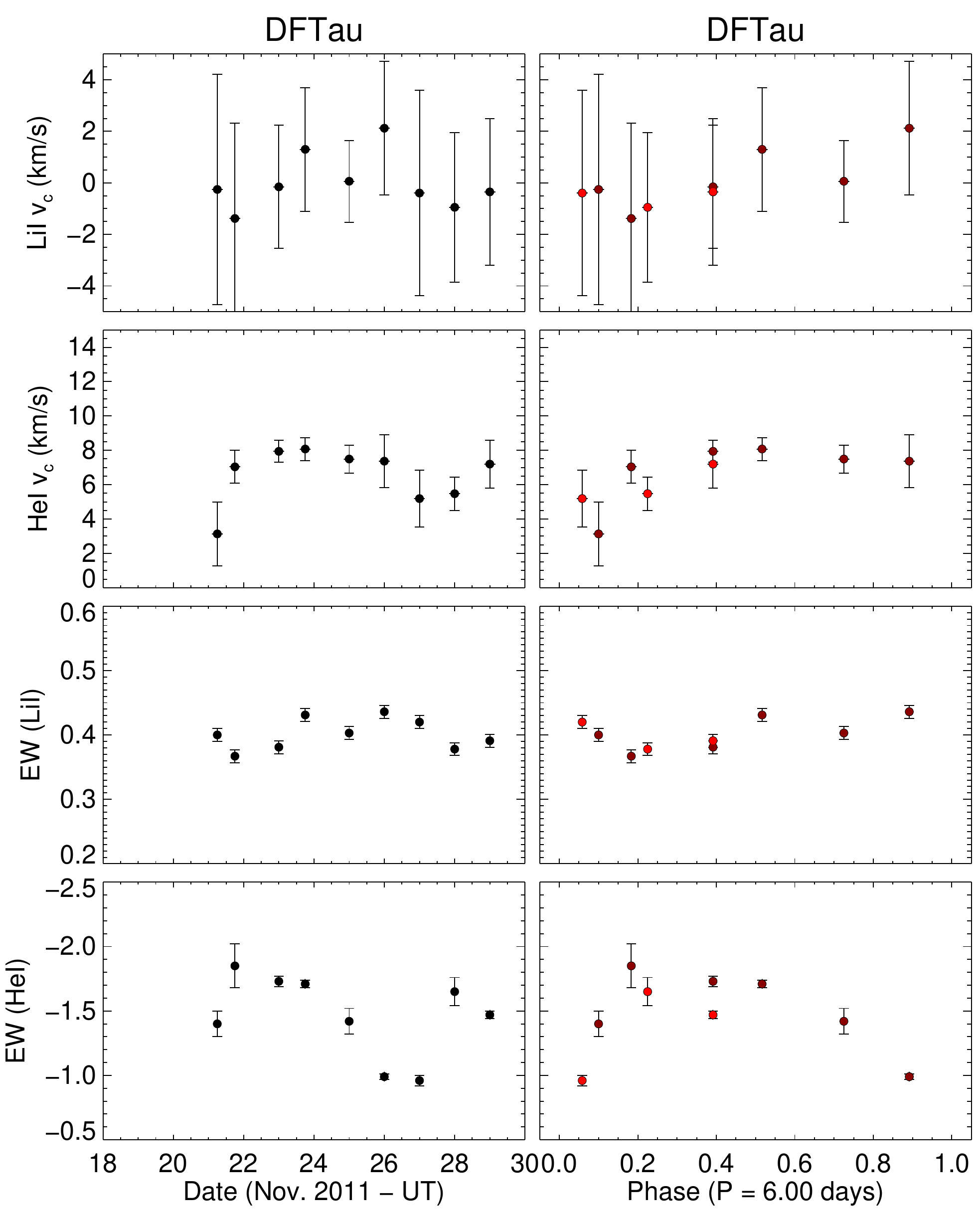}
  \includegraphics[width=0.49\textwidth]{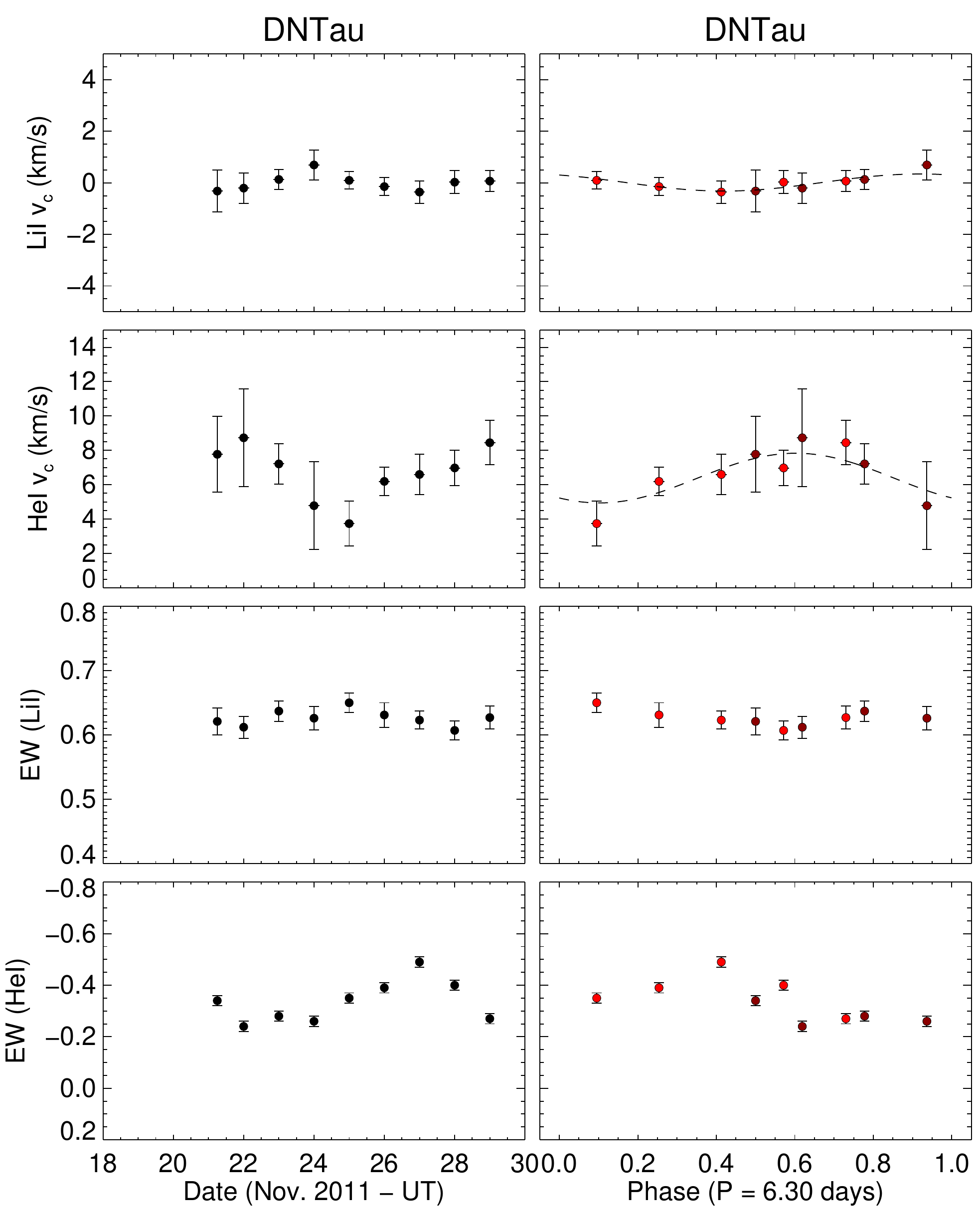} \newline \newline
  \includegraphics[width=0.49\textwidth]{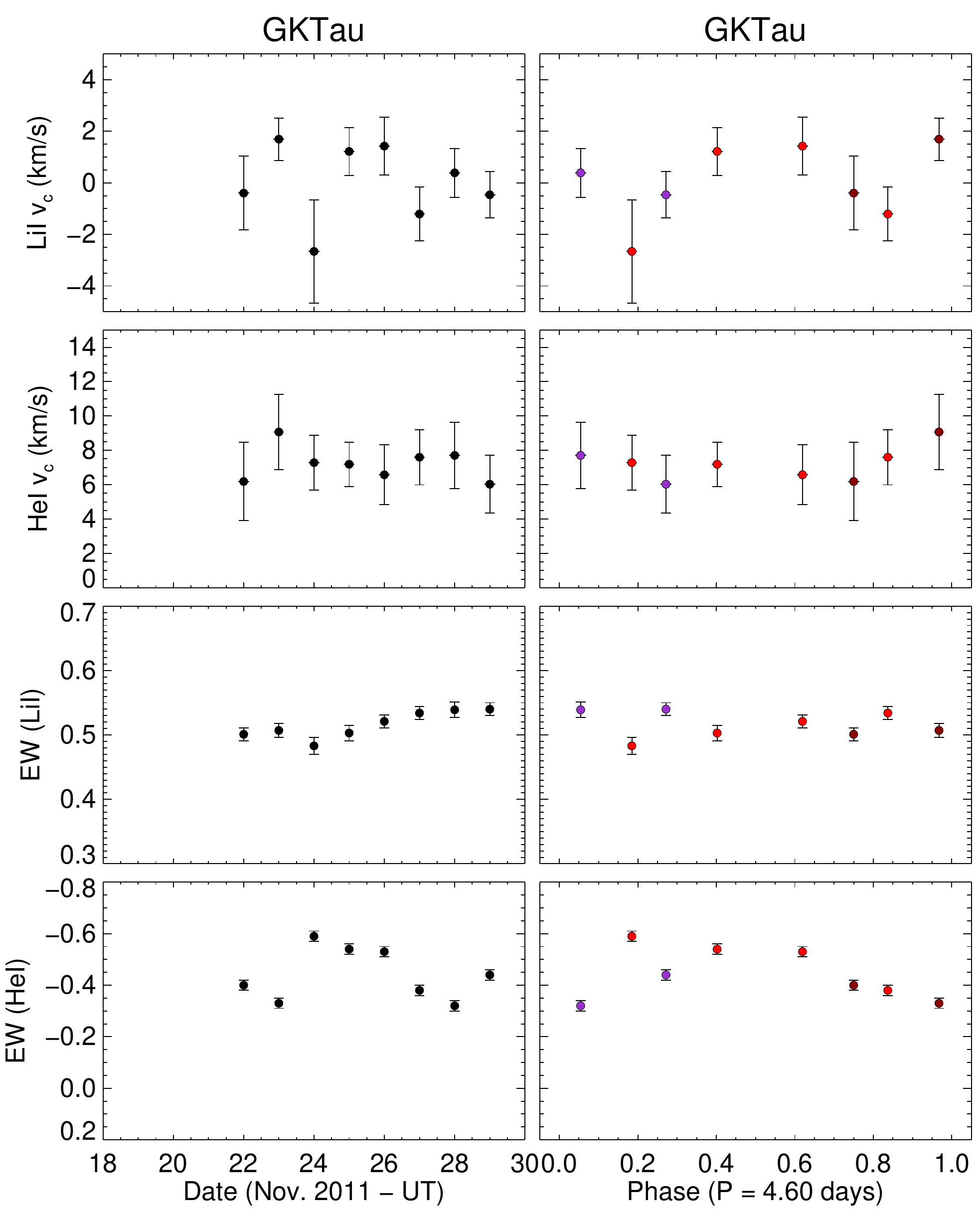}
  \includegraphics[width=0.49\textwidth]{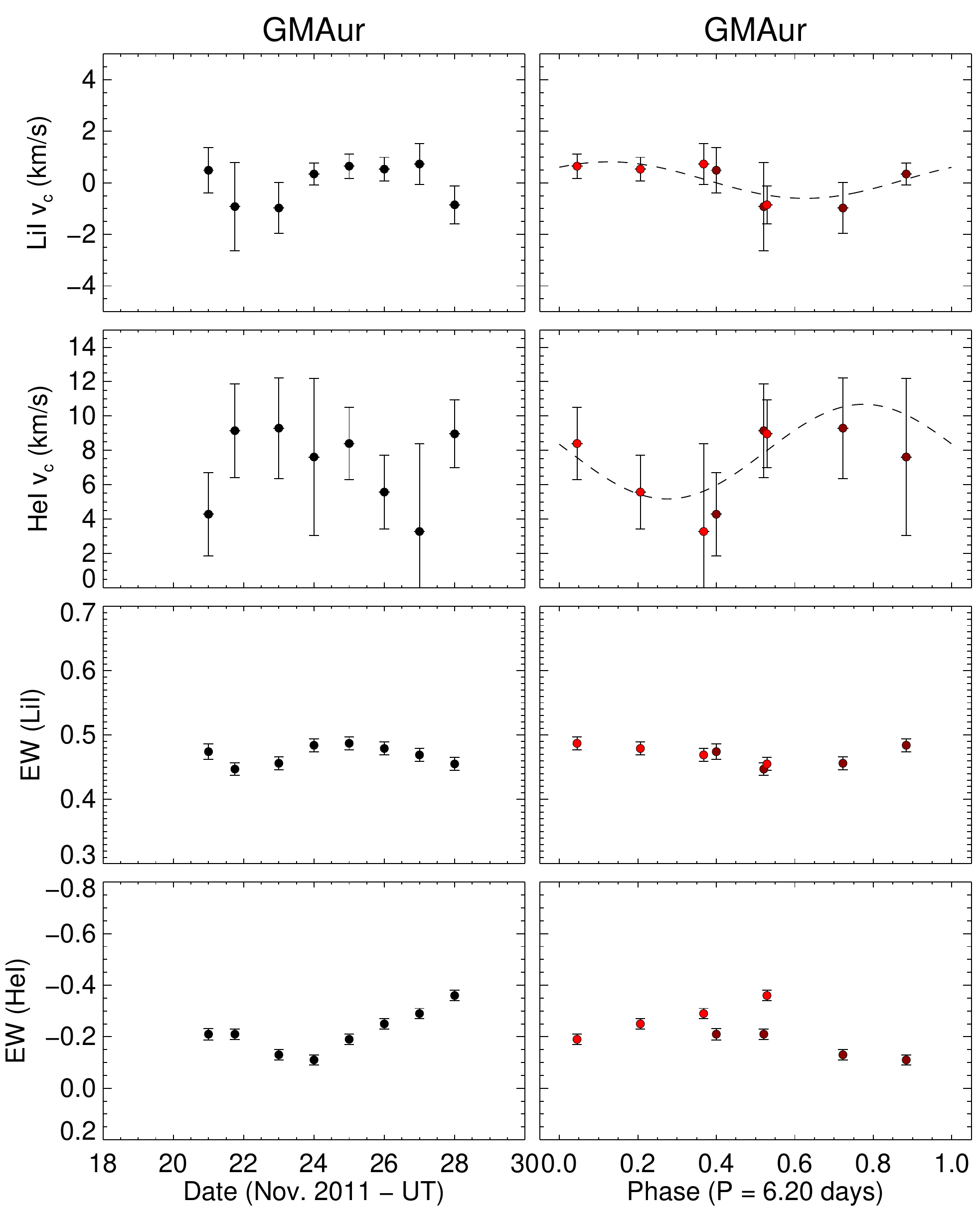}
  \caption{Continued. 
  }\ContinuedFloat 
\end{figure*}

\begin{figure*}%[t]
  \centering
  \includegraphics[width=0.49\textwidth]{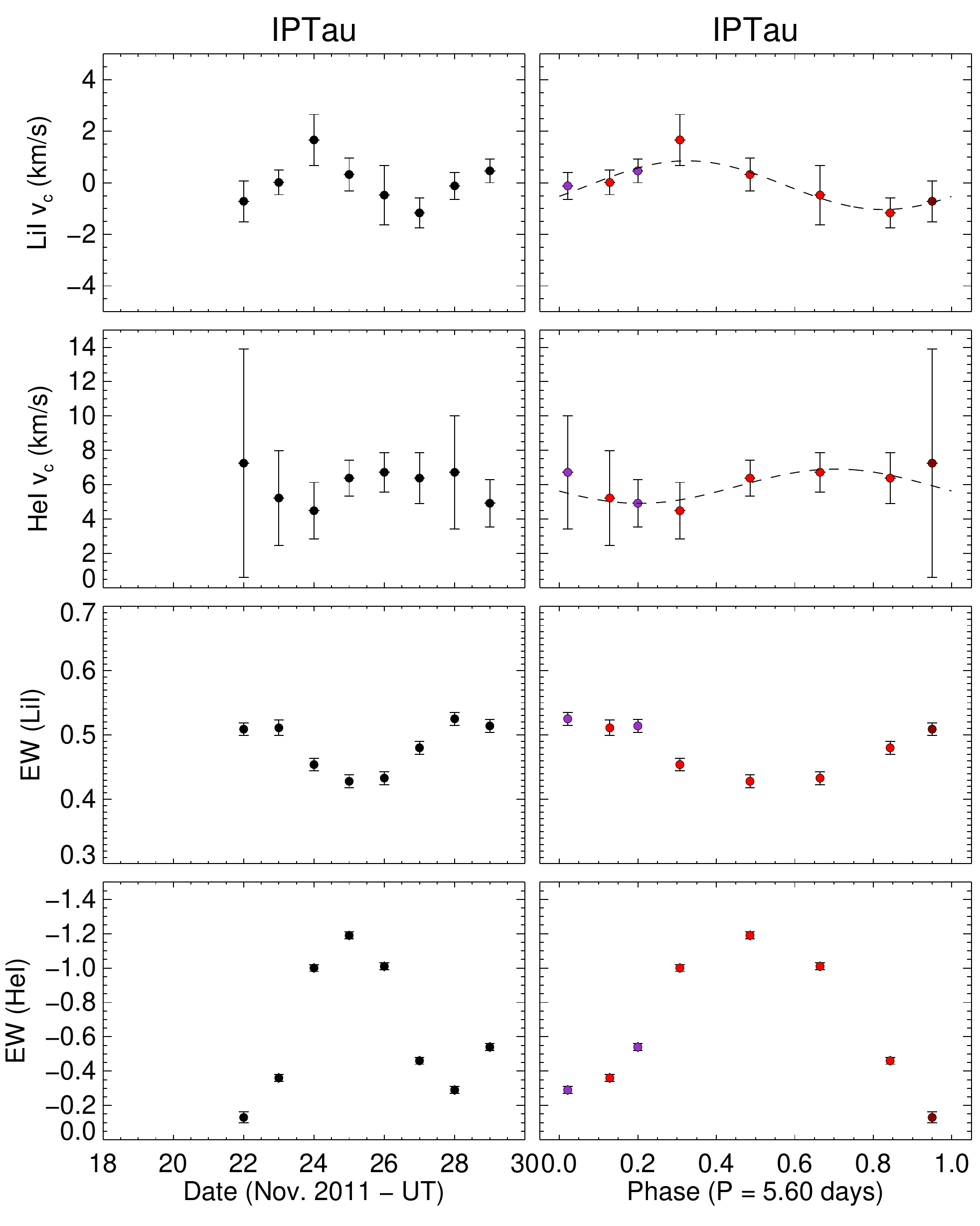}
  \includegraphics[width=0.49\textwidth]{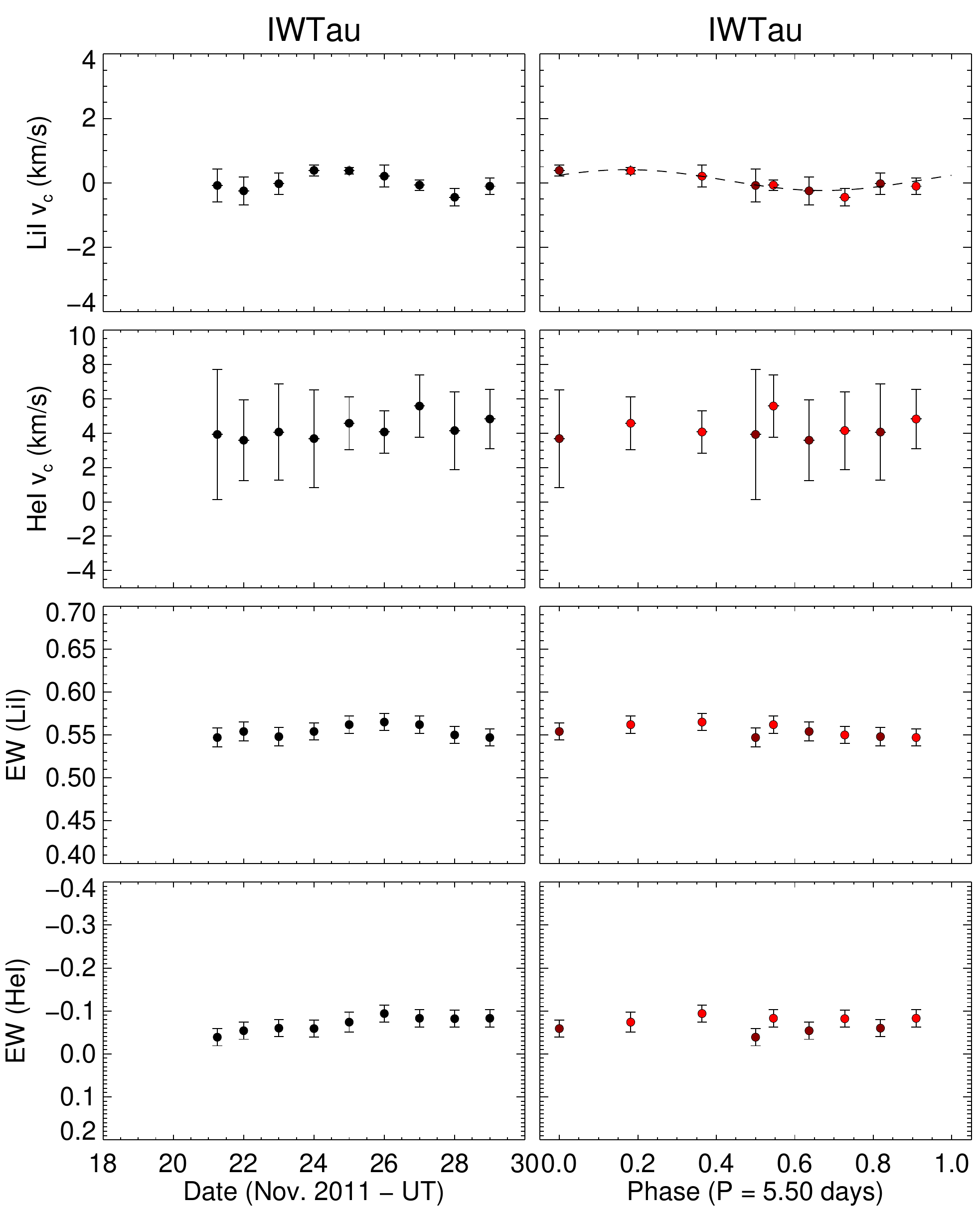}
  \caption{Continued.  
  }\ContinuedFloat 
\end{figure*}

\begin{figure*}%[t]
  \centering
  \includegraphics[width=0.245\textwidth]{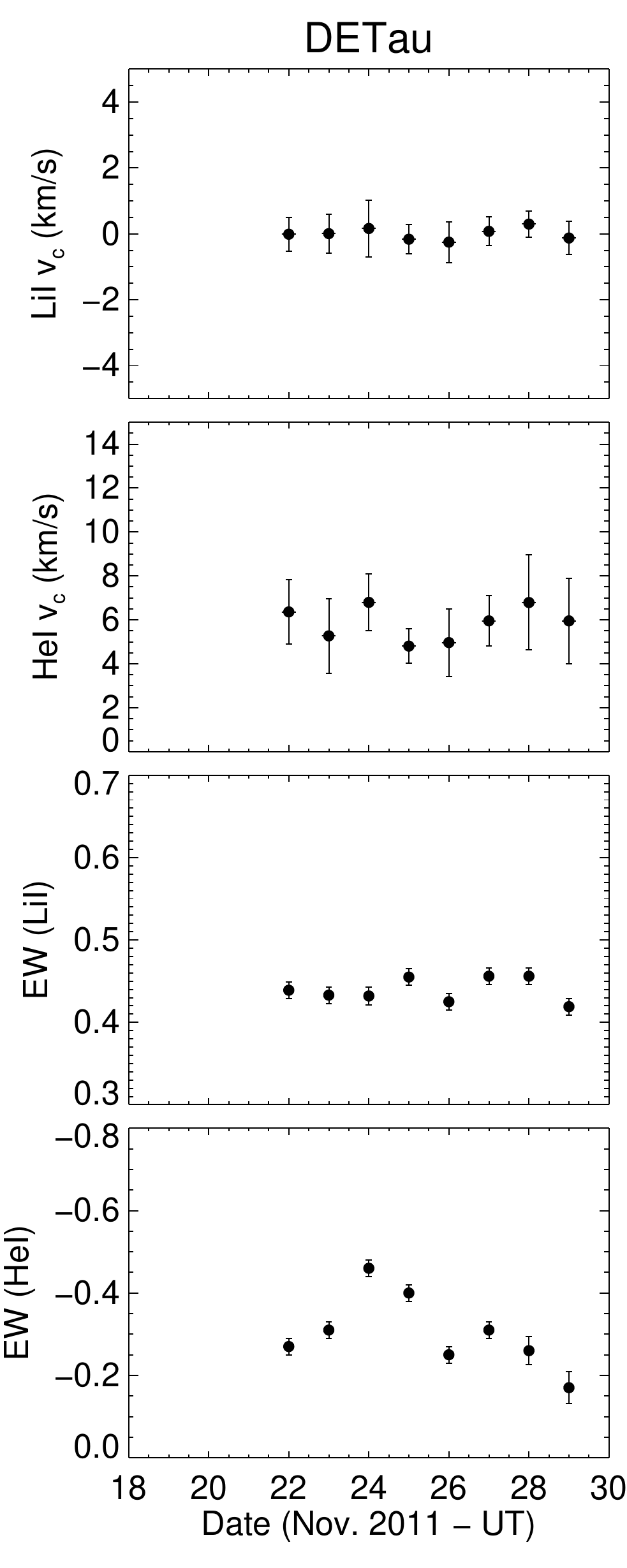}
  \includegraphics[width=0.245\textwidth]{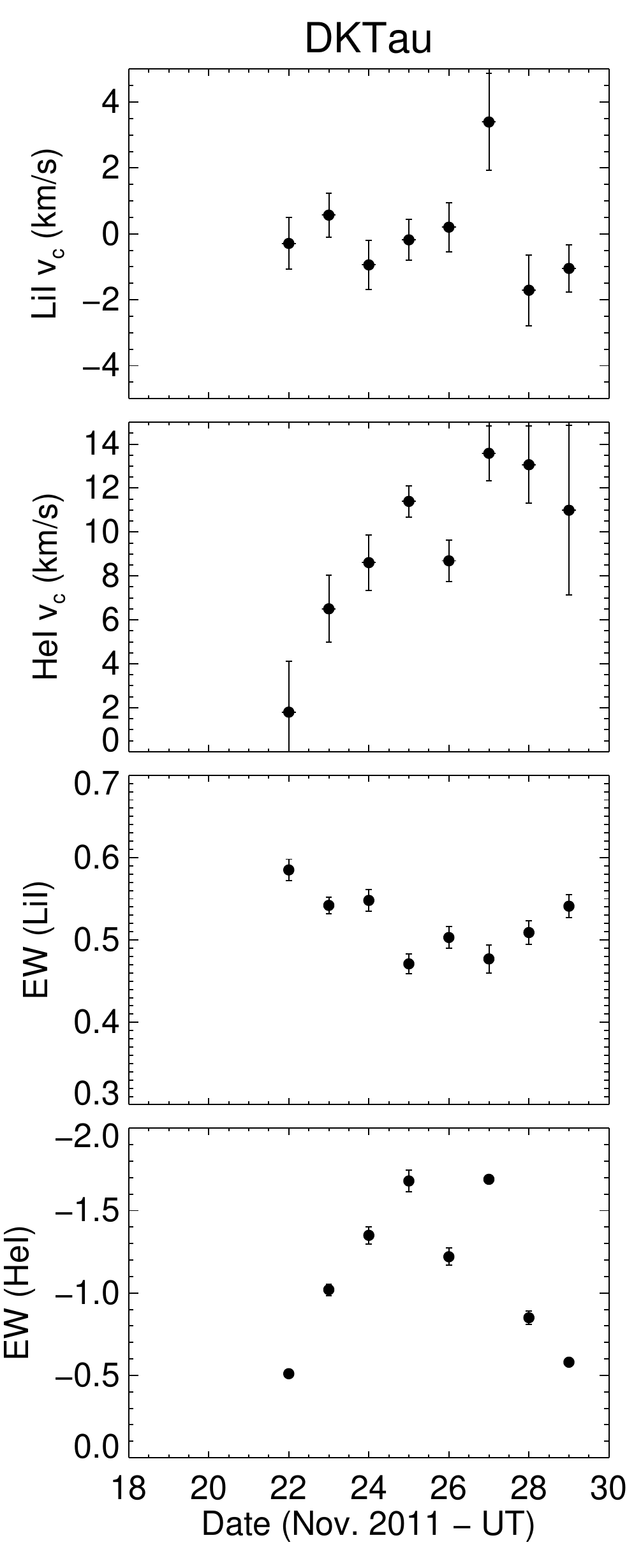}
  \includegraphics[width=0.245\textwidth]{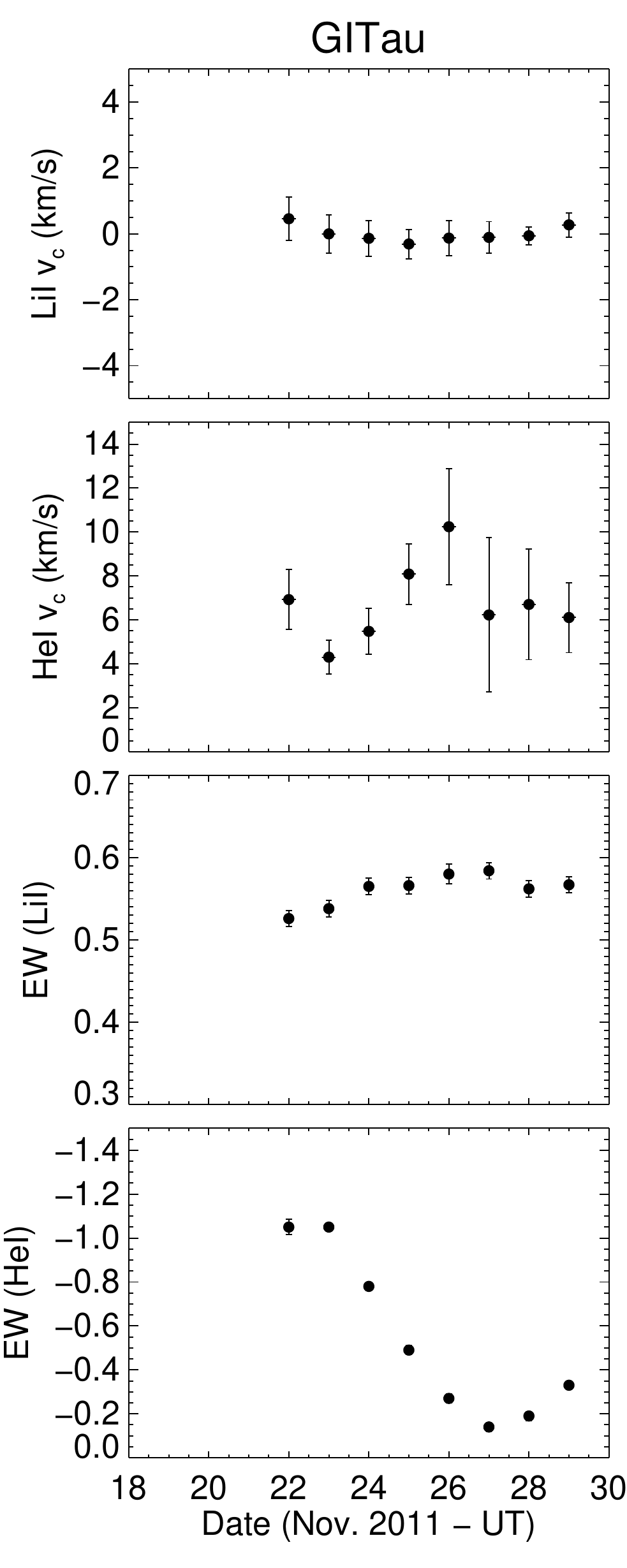}
  \includegraphics[width=0.245\textwidth]{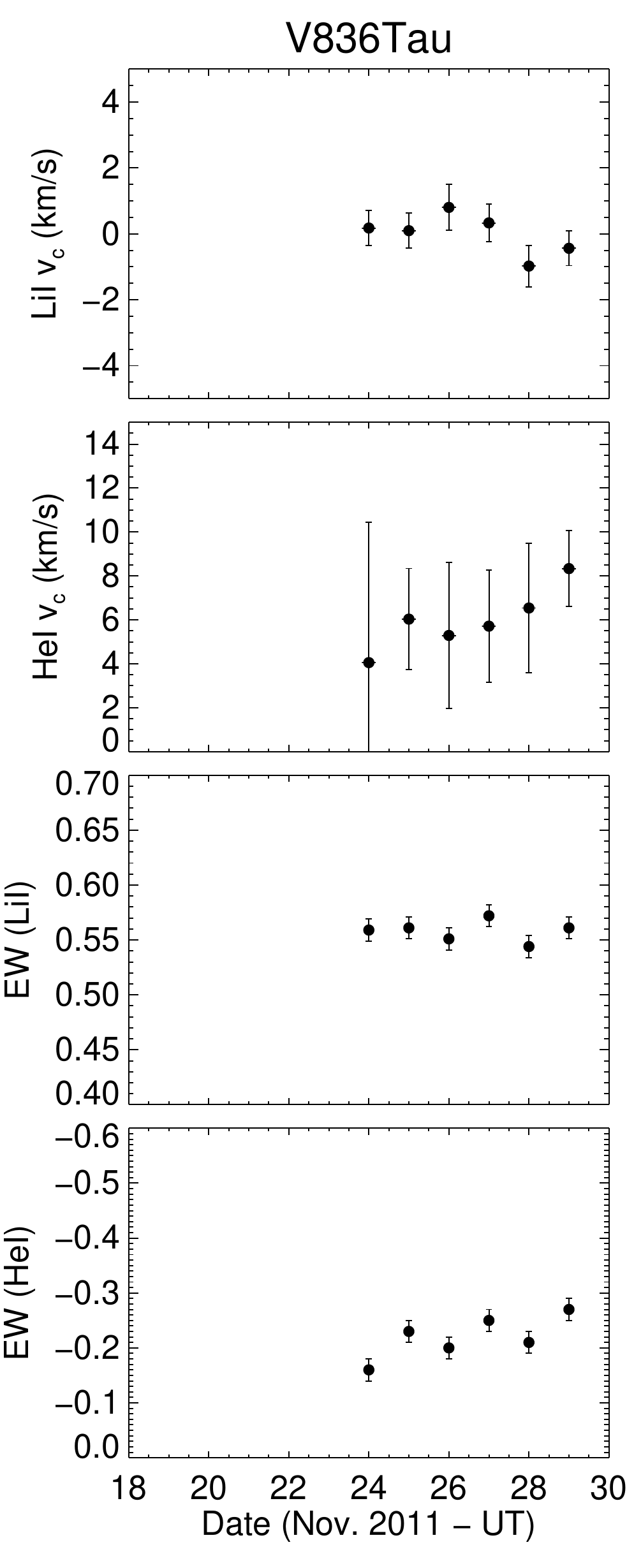}
  \caption{Continued. This time only one plot is shown for each star, since only one rotation 
  period was measured. 
   }
\end{figure*}

\begin{figure*}%[p] 
   \centering
   \includegraphics[width=0.48\textwidth]{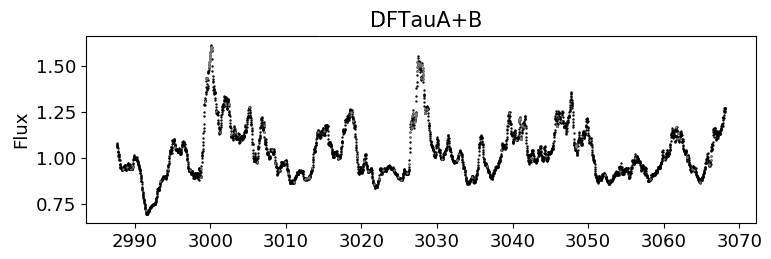}
   \includegraphics[width=0.48\textwidth]{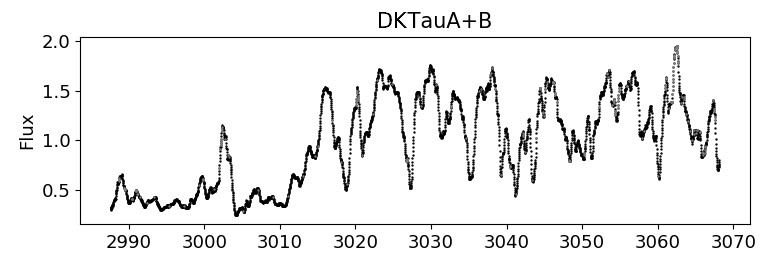}
   \includegraphics[width=0.48\textwidth]{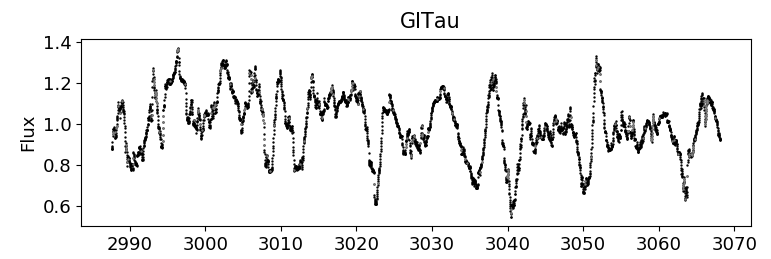}
   \includegraphics[width=0.48\textwidth]{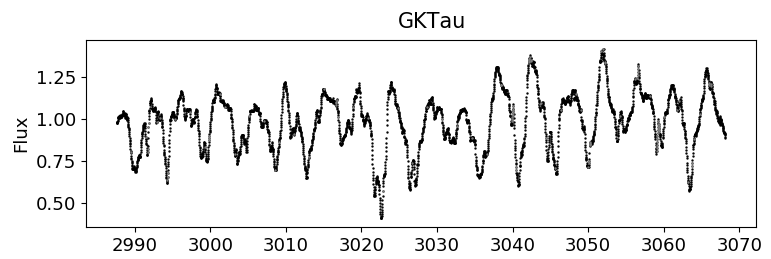}
   \includegraphics[width=0.48\textwidth]{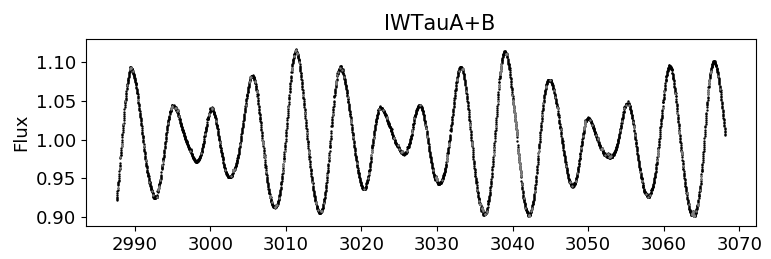}
   \includegraphics[width=0.48\textwidth]{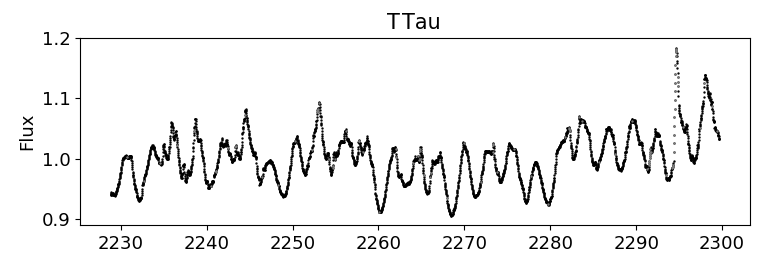}
   \caption{Light curves of the stars in our sample that were observed by K2.
   }
   \label{fig:k2lcs}
\end{figure*}

\section{Additional tables} 

Table \ref{tab:eqw} shows the equivalent widths of the He~I, H$\alpha$ and H$\beta$ 
lines in this study. In order to demonstrate the variability of these emission lines, 
the average and standard deviation over all nights are given.
Table \ref{tab:magob_otr} shows information on the magnetic obliquities of other T Tauri stars 
taken from the literature.  \newpage

\begin{table}
\caption{Equivalent widths of H$\alpha$, H$\beta$ and HeI$\lambda$5876 for our sample.}
\label{tab:eqw}
\centering
\begin{tabular}{l c c c c c c}
\hline
Star    & \multicolumn{2}{ c }{H$\alpha$} & \multicolumn{2}{ c }{H$\beta$} & \multicolumn{2}{ c }{He~I} \\
        & <EW>  & $\sigma$ & <EW>  & $\sigma$ & <EW>  & $\sigma$ \\
\hline
DETau   &   -65  &     7   &  -30  &      4   &  -1.0 &    0.2   \\
DFTau   &   -64  &    11   &  -17  &      6   &  -2.5 &    0.7   \\
DKTau   &   -35  &    19   & -9.5  &    4.7   &  -1.1 &    0.4   \\
DNTau   &  -7.6  &   1.0   & -3.3  &    0.5   & -0.34 &   0.08   \\
GITau   & -10.3  &   3.2   & -5.0  &    2.6   & -0.54 &   0.35   \\
GKTau   &   -21  &     8   & -4.1  &    2.1   & -0.44 &   0.10   \\
GMAur   &  -106  &     5   &  -16  &      2   &  -1.1 &    0.4   \\
IPTau   &   -16  &     6   & -7.9  &    3.1   & -0.62 &   0.37   \\
IWTau   &  -3.8  &   0.6   & -1.7  &    0.3   & -0.07 &   0.02   \\
TTau    &   -98  &    13   &  -17  &      4   &  -1.0 &    0.3   \\
V826Tau &  -1.5  &   0.2   & -0.27 &   0.16   & -0.07 &   0.01   \\
V836Tau & -14.2  &   1.4   & -3.4  &    0.4   & -0.22 &   0.04   \\
\hline
\end{tabular}

 \textbf{Notes:} We show the mean equivalent width (<EW>) of  each emission line over 
 the several days observations, along with the standard deviation ($\sigma$), in order 
 to illustrate the variability of these lines. All values are given in \AA. 
 
\end{table}

\begin{table*}
\caption{Magnetic obliquities of other CTTSs from the literature}
\label{tab:magob_otr}
\centering
\begin{tabular}{l c c c c c}
\hline
Star & Spectral type & Mass ($M_{\odot}$) & $\Theta (^{\circ})$ & Source of $\Theta$ & References  \\ 
\hline
AA Tau            & M0.6 & 0.57 &  10  &  ZDI  & \citep{johnstone14,donati10b} \\ 
BP Tau (Feb 2006) & M0.5 & 0.62 &  10  &  ZDI  & \citep{johnstone14,donati08}  \\ 
BP Tau (Dec 2006) &  "   &  "   &  30  &  ZDI  & \citep{johnstone14,donati08}  \\ 
CI Tau            & K5.5 & 0.90 &  20  &  ZDI  & \citep{donati20}  \\             
CR Cha            & K2   & 2.00 &  70  &  ZDI  & \citep{johnstone14,hussain09} \\ 
CV Cha            & G8   & 1.05 &  60  &  ZDI  & \citep{johnstone14,hussain09} \\ 
DN Tau (2010)     & M0.3 & 0.55 &  30  &  ZDI  & \citep{donati13}  \\             
DN Tau (2012)     &  "   &  "   &  15  &  ZDI  & \citep{donati13}  \\             
GQ Lup (2009)     & K5.0 & 0.89 &  30  &  ZDI  & \citep{johnstone14,donati12}  \\ 
GQ Lup (2011)     &  "   &  "   &  30  &  ZDI  & \citep{johnstone14,donati12}  \\ 
LkCa 15           & K5.5 & 0.89 &  20  &  ZDI  & \citep{donati19}  \\             
TW Hya (2008)     & M0.5 & 0.69 &  40  &  ZDI  & \citep{johnstone14,donati11b} \\ 
TW Hya (2010)     &  "   &  "   &  10  &  ZDI  & \citep{johnstone14,donati11b} \\ 
V2129 Oph (2005)  & K6   & 1.35 &  20  &  ZDI  & \citep{johnstone14,donati07}  \\ 
V2129 Oph (2009)  &  "   &  "   &  10  &  ZDI  & \citep{johnstone14,donati11a} \\ 
V2247 Oph         & M1   & 0.36 &  40  &  ZDI  & \citep{johnstone14,donati10a} \\ 
V4046 Sgr A       & K5   & 0.95 &  60  &  ZDI  & \citep{johnstone14,donati11c} \\ 
V4046 Sgr B       & K5   & 0.85 &  80  &  ZDI  & \citep{johnstone14,donati11c} \\ 
DR Tau            & K6   & 0.88 &  14  &  $\Delta V_r (HeI)$ & \citep{petrov11}          \\     
EX Lup            & M0   & 0.60 &  13  &  $\Delta V_r (HeI)$ & \citep{sicilia-aguilar15}  \\     
RU Lup            & K7   & 0.65 &  10  &  $\Delta V_r (HeI)$ & \citep{gahm13, alcala17}   \\     
RW Aur A          & K6.5 & 1.13 &  45  &  Modeling spectral variability &  \citep{petrov01} \\  
\hline
\end{tabular}

 \textbf{Notes:} To maintain consistency with this work, for the star EX~Lup 
 we took the value of $\Theta$ derived from the amplitude of the radial velocity variability 
 of the HeI$\lambda5876$ line. Masses and spectral types are from \citet{herczeg14}, where 
 available. For the stars not included in that study, masses were re-derived using the same mass tracks they used \citep[the PISA tracks,][]{tognelli11}, in order to maintain consistency among the sample. 
\label{lastpage}
\end{table*}

\end{appendix}

%\AtEndDocument{\includepdf[pages=1-7,offset=0 0]{onlinematerial.pdf}}
\begin{figure*} 
 \vspace{-1cm}
 \centering
 \includegraphics[page=1,trim=1.25cm 2.5cm 1.5cm 1.5cm,clip,width=\textwidth]{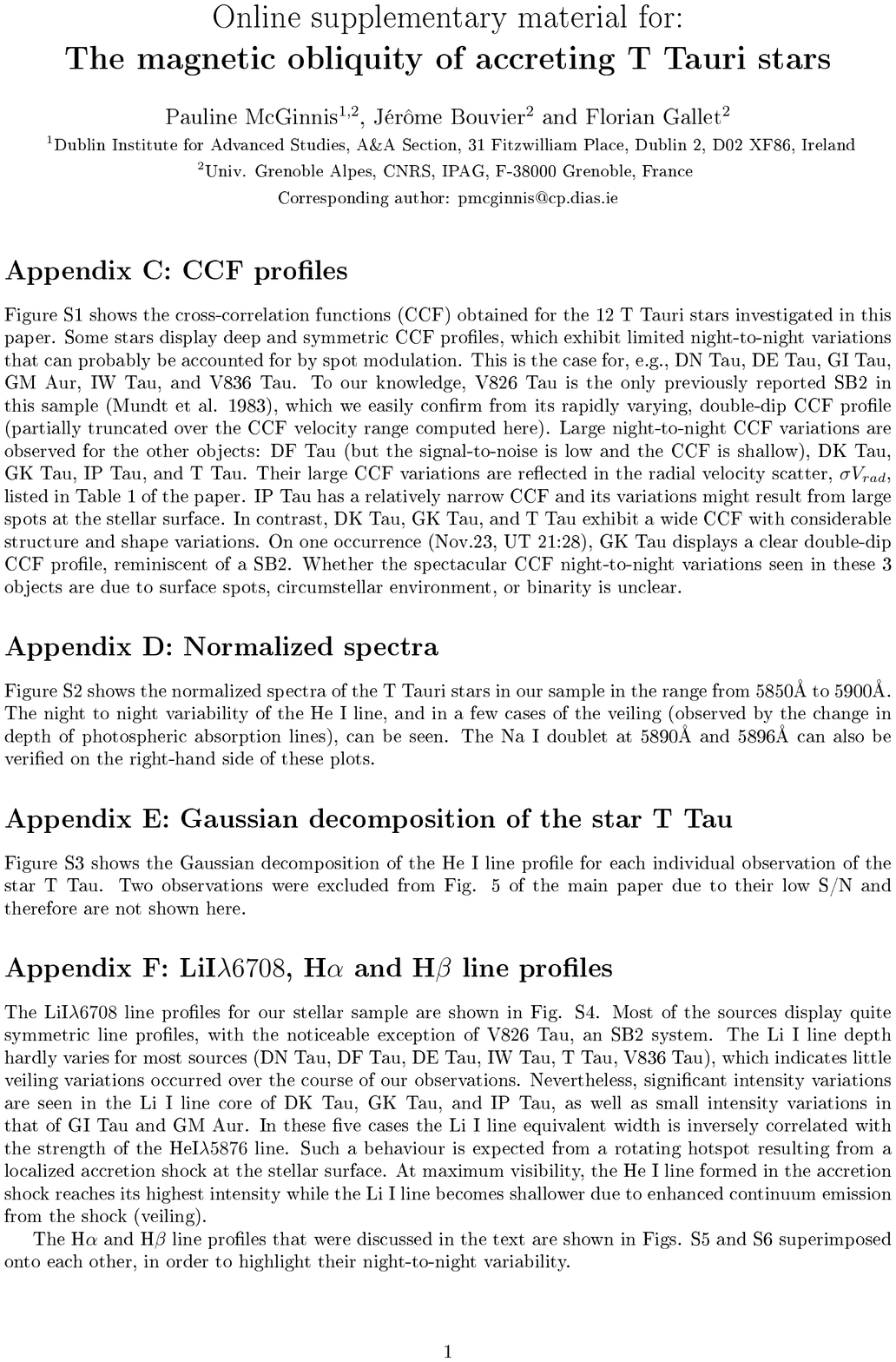}
\end{figure*}

\begin{figure*} 
 \vspace{-0.8cm}
 \centering
 \includegraphics[page=2,trim=1.25cm 2.5cm 1.5cm 1.5cm,clip,width=\textwidth]{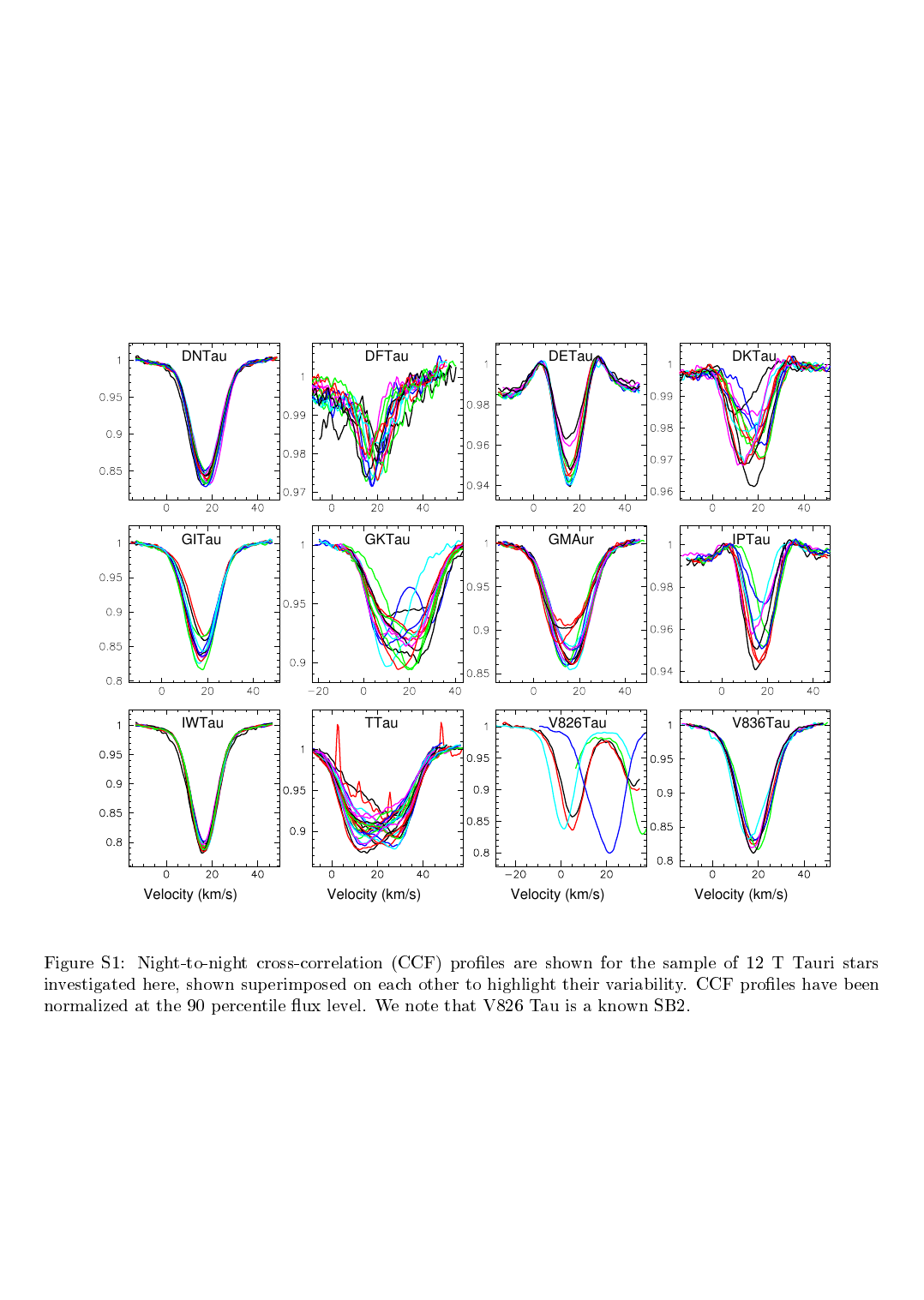}
\end{figure*}

\begin{figure*} 
 \vspace{-0.8cm}
 \centering
 \includegraphics[page=3,trim=1.25cm 2.5cm 1.5cm 1.5cm,clip,width=\textwidth]{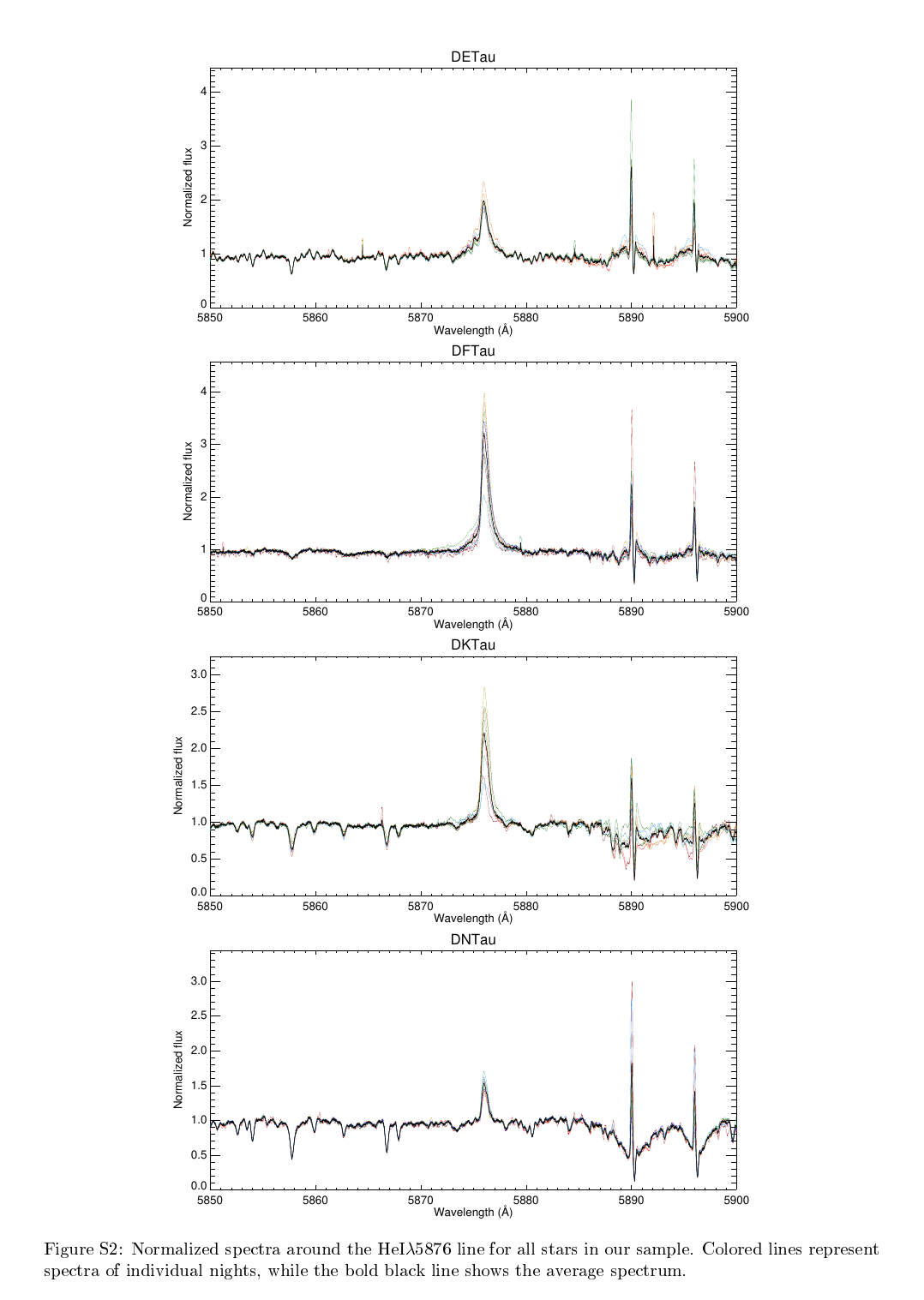}
\end{figure*}

\begin{figure*} 
 \vspace{-0.8cm}
 \centering
 \includegraphics[page=4,trim=1.25cm 2.5cm 1.5cm 1.5cm,clip,width=\textwidth]{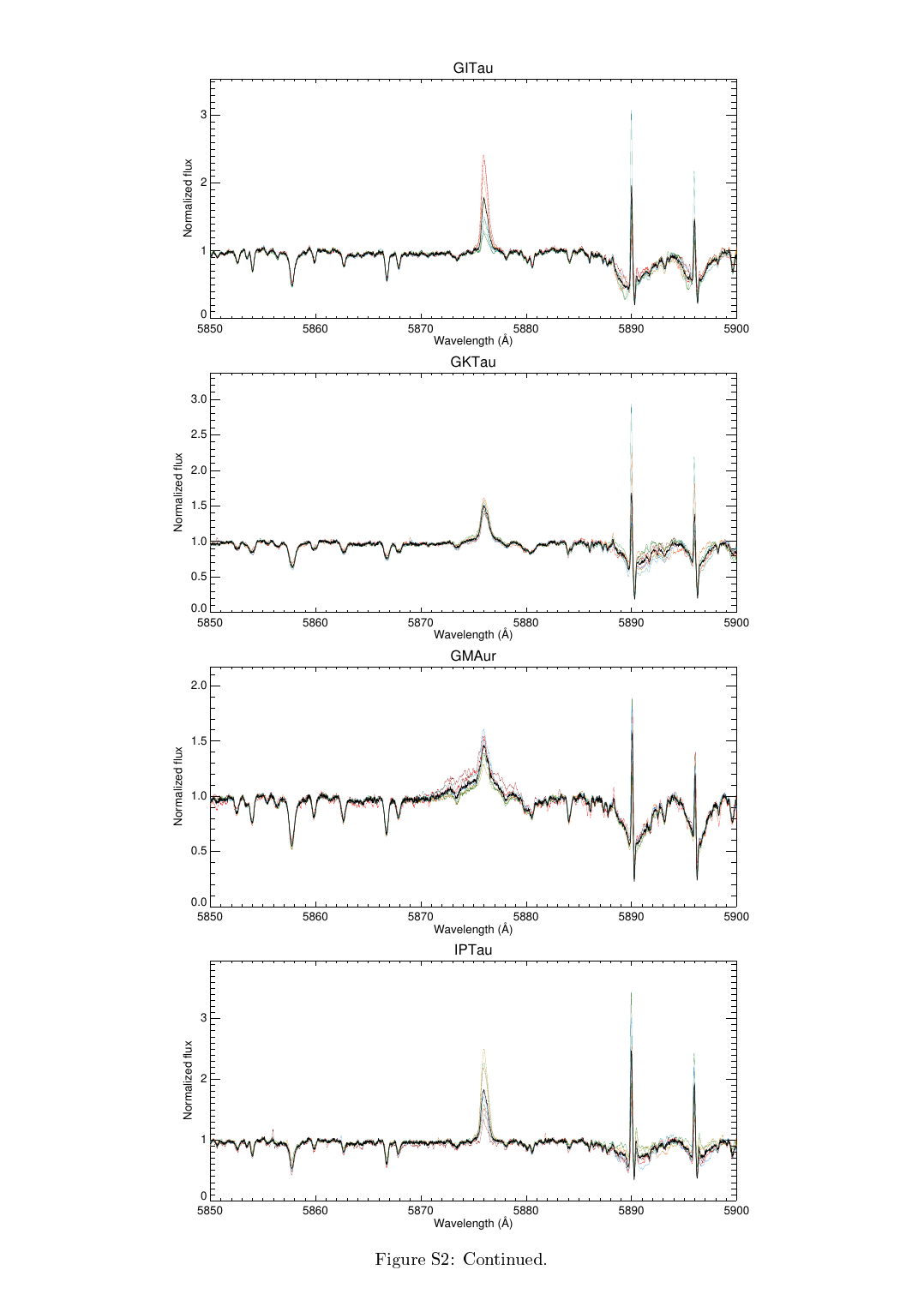}
\end{figure*}

\begin{figure*} 
 \vspace{-0.8cm}
 \centering
 \includegraphics[page=5,trim=1.25cm 2.5cm 1.5cm 1.5cm,clip,width=\textwidth]{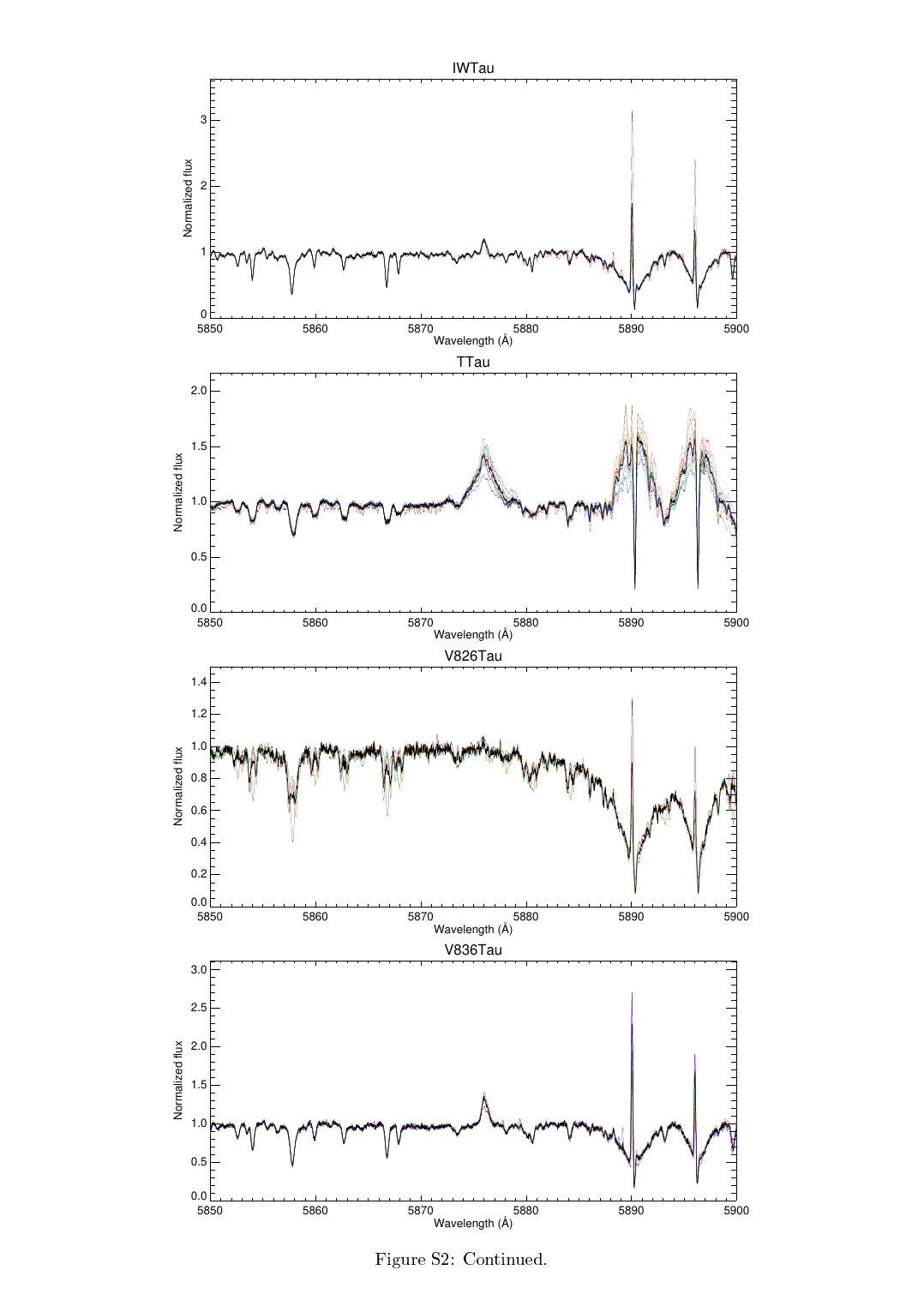}
\end{figure*}

\begin{figure*} 
 \vspace{-0.8cm}
 \centering
 \includegraphics[page=6,trim=1.25cm 2.5cm 1.5cm 1.5cm,clip,width=\textwidth]{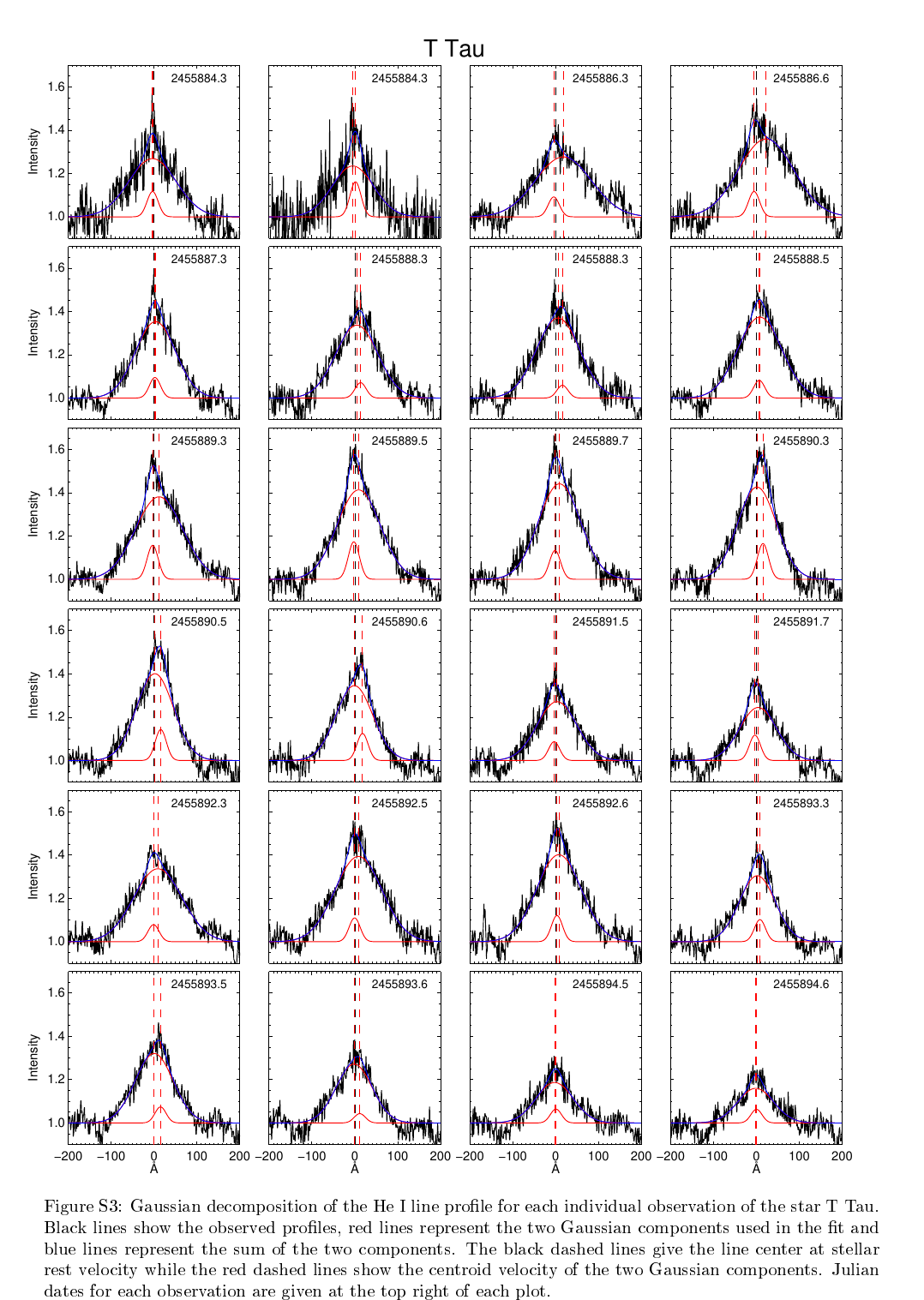}
\end{figure*}

\begin{figure*} 
 \vspace{-0.8cm}
 \centering
 \includegraphics[page=7,trim=1.25cm 2.5cm 1.5cm 1.5cm,clip,width=\textwidth]{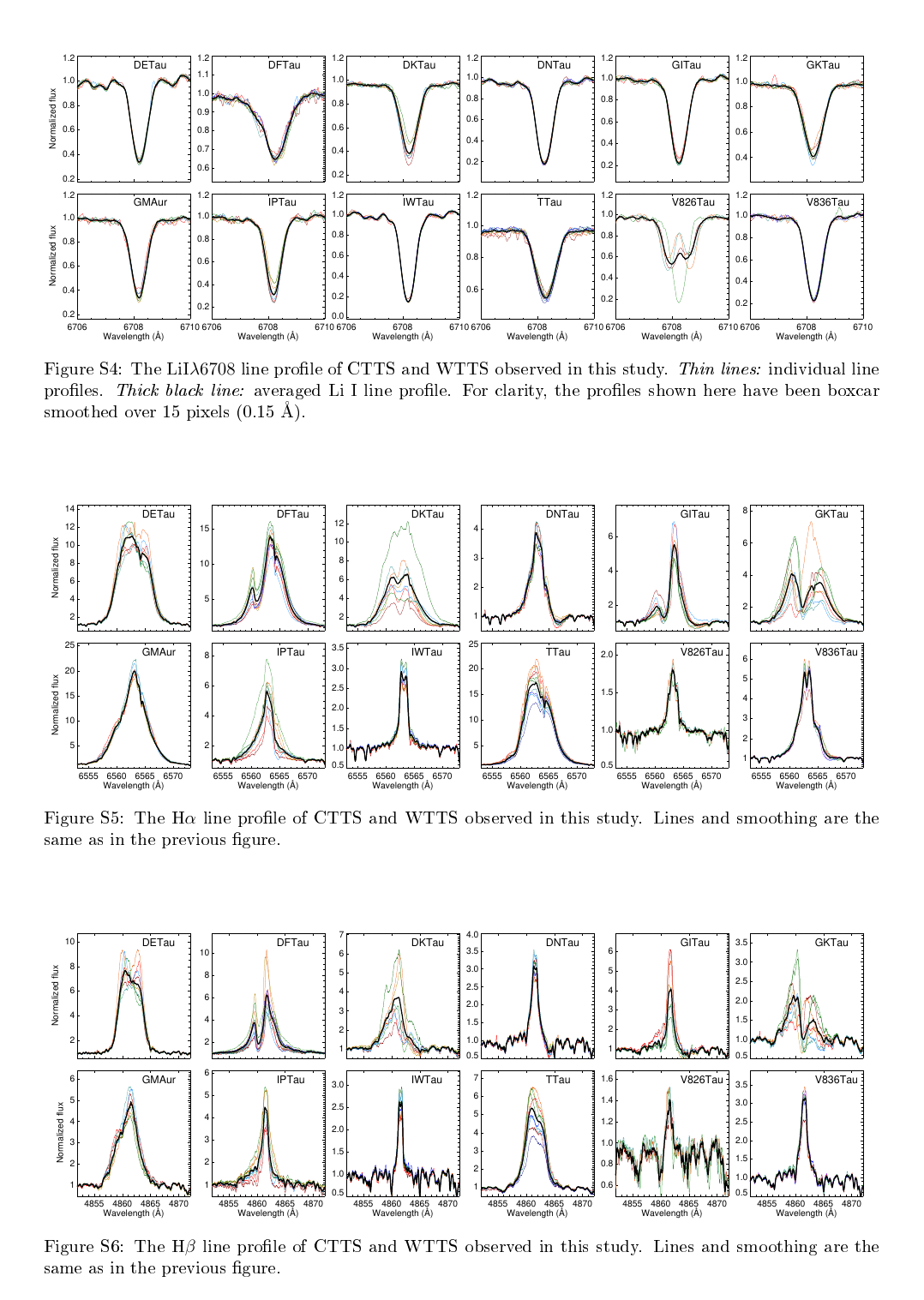}
\end{figure*}

\bsp	% typesetting comment
\end{document}